\documentstyle[epsfig,twocolumn,aps,rmp]{revtex}

\newcommand{\noprintlabel}{}

\def\fig{.}
\def\note #1]{{\bf #1]}}
\def\etal{{\it et al.}}
\def\eg{{\it e.g.}}
\def\cf{{\it cf.}}
\def\ie{{\it i.e.}}
\def\etc{{\it etc.}}

\def\mHz{\,{\rm mHz}}
\def\muHz{\,\mu{\rm Hz}}
\def\nHz{\, {\rm nHz}}
\def\cm{\, {\rm cm}}
\def\km{\, {\rm km}}
\def\g{\, {\rm g}}
\def\m{\, {\rm m}}
\def\s{\, {\rm s}}
\def\K{\, {\rm K}}

\def\erg{\, {\rm erg}}
\def\dd{{\rm d}}
\def\nablarad{\nabla_{\rm rad}}
\def\nablaad{\nabla_{\rm ad}}
\def\Msun{M_\odot}
\def\Rsun{R_\odot}
\def\Lsun{L_\odot}
\def\Xs{X_{\rm s}}
\def\Zs{Z_{\rm s}}
\def\alphac{\alpha_{\rm c}}
\def\clight{\tilde c}
\def\hyd{\,{\rm {}^1 H}}
\def\helfour{\,{\rm {}^4 He}}
\def\pres{p}
\def\force{{\cal F}}
\def\rt{r_{\rm t}}
\def\omegac{\omega_{\rm c}}
\def\kB{k_{\rm B}}

\def\CH{{\cal H}}

\def\Hone{\CH_1}
\def\Htwo{\CH_2}
\def\Fsurf{F_{\rm surf}}
\def\deltar{\delta_r}
\def\omegai{\omega_{\rm i}}
\def\PP{{\cal P}}
\def\Ib{I_{\rm b}}
\def\Ir{I_{\rm r}}
\def\VD{V_{\rm D}}
\def\Ombar{\bar\Omega}
\def\CJ{{\cal J}}
\def\CC{{\cal C}}
\def\CK{{\cal K}}
\def\CT{{\cal T}}
\def\Ye{Y_{\rm e}}
\def\cw{c_{\rm w}}

\def\gwig{{\leavevmode\kern0.3em\raise.3ex\hbox{$>$}
\kern-0.8em\lower.7ex \hbox{$\sim$}\kern0.3em}}
\def\lwig{{\leavevmode\kern0.3em\raise.3ex\hbox{$<$}
\kern-0.8em\lower.7ex \hbox{$\sim$}\kern0.3em}}

\newcommand{\pmbol}[1]{\setbox0=\hbox{#1}%
 \kern-0.018em\copy0\kern-\wd0
 \kern0.036em\copy0\kern-\wd0
 \kern-0.018em\raise0.028em\box0} % Poorman's bold.
\newcommand{\del}{\nabla} 
\newcommand{\bolda}{{\bf a}}  % bold a
\newcommand{\boldr}{{\bf r}}  % bold r
\newcommand{\boldg}{{\bf g}}  % bold g
\newcommand{\boldk}{{\bf k}}  % bold k
\newcommand{\boldn}{{\bf n}}  % bold n
\newcommand{\boldv}{{\bf v}}  % bold v
\newcommand{\boldOmega}{\Omega\kern-0.68em\Omega\kern-0.68em\Omega}
\newcommand{\scalarprod}{\cdot}
\newcommand{\vectordel}{\pmbol{$\nabla$}}       % So-so vector del
\newcommand{\diverge}{{\rm div\,}}        % Divergence
\newcommand{\divh}{{\rm div_h\,}}        % Horizontal divergence
\newcommand{\wpard}[2]{{\partial #1 \over \partial #2} } % Partials without prens
\newcommand{\bfdelta}{\pmb{$\delta$}}  % bold delta
\newcommand{\bfdelr}{\bfdelta\boldr}  % bold delta r
\newcommand{\boldxi}{\xi\kern-0.45em\xi\kern-0.45em\xi}
\newcommand{\boldxih}{\boldxi_{\rm h}}
\newcommand{\nablah}{\nabla_{\rm h}}
\newcommand{\xih}{\xi_{\rm h}}
\newcommand{\boldkh}{\boldk_{\rm h}}
\newcommand{\kh}{k_{\rm h}}

\newcommand{\be}{\begin{equation}}
\newcommand{\bea}{\begin{eqnarray}}
\newcommand{\ee}
{\end{equation}}
\newcommand{\eea}
{\end{eqnarray}}
\newcommand{\eel}[1]
{\label{#1}
\end{equation}
\ifx\noprintlabel\undefined{\scriptsize\bf [#1]}\par\noindent\fi}

\newcommand{\plabel}[1]
{\label{#1}
\ifx\noprintlabel\undefined{\par\noindent\scriptsize\bf [#1]}\par\noindent\fi}

\newcommand{\eenl}[1]
{\end{equation}
\ifx\noprintlabel\undefined{\scriptsize\bf [#1]}\par\noindent\fi}

\newcommand{\eeal}[1]
{\label{#1}
\end{eqnarray}
\ifx\noprintlabel\undefined{\scriptsize\bf [#1]}\par\noindent\fi}
\newcommand{\eeanl}[1]
{\end{eqnarray}
\ifx\noprintlabel\undefined{\scriptsize\bf [#1]}\par\noindent\fi}

\newcommand{\figlab}[1]{\label{#1}
\ifx\noprintlabel\undefined{\scriptsize\bf [#1]}\fi}

\newcommand{\inclfig}[3]
{\ifx\nofigures\undefined{\leavevmode\epsfxsize=#1\epsfbox{#2}}\else
{\bf [#3]}\fi}

\newcommand{\incltwofig}[4]
{\ifx\nofigures\undefined{\leavevmode\epsfxsize=#1\epsfbox{#2}
\epsfxsize=#1\epsfbox{#3}}\else
{\bf [#4]}\fi}

\newcommand{\Eq}[1]{Eq.~(\ref{#1})}

\newcommand{\incltit}[1]
{\ifx\notitles\undefined{``#1,''}\fi}

\newcommand{\incltitnc}[1]   % Version without comma
{\ifx\notitles\undefined{``#1''}\fi}

\newcommand{\incllast}[1]
{\ifx\nolastpage\undefined{--#1}\else{}\fi}

\begin{document}
\title{Helioseismology}
\author{J{\o}rgen Christensen-Dalsgaard}
\address{High Altitude Observatory, National Center for Atmospheric Research,
Boulder, CO, USA, \\
Teoretisk Astrofysik Center, Danmarks Grundforskningsfond, and \\
Institut for Fysik og Astronomi, Aarhus Universitet, DK 8000 Aarhus C,
Denmark\thanks{Permanent address. Electronic address: jcd@phys.au.dk}\\
\\
{\rm (Submitted: 9 November 2001. Revised: 4 April 2002.
Scheduled for publication in {\it Rev. Mod. Phys.}, Oct. 2002)}}
\maketitle

\begin{abstract}
Oscillations detected on the solar surface provide a unique possibility
for investigations of the interior properties of a star.
Through major observational efforts, including extensive observations
from space, as well as development of sophisticated tools for the
analysis and interpretation of the data, we have been able to infer
the large-scale structure and rotation of the solar interior with substantial
accuracy, and we are beginning to get information about the complex
subsurface structure and dynamics of sunspot regions, which dominate
the magnetic activity in the solar atmosphere and beyond.
The results provide a detailed test of the modeling of stellar
structure and evolution, and hence of the physical properties
of matter assumed in the models.
In this way the basis for using stellar modeling in other branches
of science is very substantially strengthened; an important example
is the use of observations of solar neutrinos to constrain the properties
of the neutrino.
\end{abstract}

\tableofcontents

\section{Introduction}

\plabel{sec:introduc}

By the standards of astrophysics, stars are relatively well understood.
Modelling of stellar evolution has explained, or at least accounted for,
many of the observed properties of stars.
Stellar models are computed on the basis of the assumed physical
conditions in stellar interiors, including the thermodynamical
properties of stellar matter, the interaction between matter and
radiation and the nuclear reactions that power the stars.
By following the changes in structure as the stars evolve
through the fusion of lighter elements into heavier, starting
with hydrogen being turned into helium, the models predict how
the observable properties of the stars change as they age.
These predictions can then be compared to observations.
Important examples are the distributions of stars in terms of
surface temperature and luminosity, particularly for stellar clusters
where the stars, having presumably been formed in the same interstellar
cloud, can be assumed to share the same age and original composition.
These distributions are generally in reasonable agreement with the models;
the comparison between observations and models furthermore provides
estimates of the ages of the clusters, of considerable interest to
the understanding of the evolution of the Galaxy.
Additional tests, generally quite satisfactory,
are provided in the relatively few cases where
stellar masses can be determined with reasonable accuracy from
the motion of stars in binary systems.
Such successes give some confidence in the use of stellar models
in other areas of astrophysics.
These include studies of element synthesis in late stages of stellar 
evolution, the use of supernova explosions as `standard candles' in
cosmology, and estimates of the primordial element composition
from stellar observations.

An important aspect of stellar astrophysics is the use of stars
as physics laboratories. 
Since the basic properties of stars and their modeling are
presumed to be relatively well established, one may hope to use
more detailed observations to provide information about the 
physics of stellar interiors, to the extent that it is reflected
in observable properties.
This is of obvious interest: conditions in the interiors of stars
are generally far more extreme, in terms of temperature and density,
than achievable under controlled circumstances in terrestrial laboratories.
Thus sufficiently detailed
stellar data might offer the hope of providing information on the
properties of matter under these conditions.

Yet in reality there is little reason to be complacent about the
status of stellar astrophysics.
Most observations relevant to stellar interiors provide 
only limited constraints on the detailed properties of the stars.
Where more extensive information is becoming available, such
as determinations of detailed surface abundances, the models often
fail to explain it.
Furthermore, the models are in fact extremely simple, compared
to the potential complexities of stellar interiors.
In particular, convection, which dominates
energy transport in parts of most stars, is treated very crudely
while other potential hydrodynamical instabilities are generally neglected.
Also stellar rotation is rarely taken into account, yet could
have important effects on the evolution.
These limitations could have profound effects on, for example,
the modeling of late stages of stellar evolution, which depend
sensitively on the composition profile established during the life of the star.

The Sun offers an example of a star that can be studied in very great detail.
%Its mass is known with high accuracy from the motion of the planets,
%its radius and total luminosity can be determined directly,
%given that its distance is precisely known,
%and dating of meteorites provides a precise measure of the age of the Sun.
Furthermore, it is a relatively simple star:
it is in the middle of its life, with approximately half the original
central abundance of hydrogen having been used, and, compared to
some other stars, the physical conditions in the solar interior are
relatively benign.
Thus in principle the Sun
provides an ideal case for testing the theory of stellar evolution.

In practice, the success of such tests was for a long time somewhat doubtful.
Solar modeling depends on two unknown parameters: the initial helium
abundance and a parameter characterizing the efficacy of convective
energy transport near the solar surface.
These parameters can be adjusted to provide a model of solar mass,
matching the solar radius and luminosity at the age of the Sun.
Given this calibration, however, the measured surface properties of
the Sun provide no independent test of the model.
Furthermore, two potentially severe problems with solar models
have been widely considered.
One, the so-called faint early Sun problem, resulted from the realization
that solar models predicted that the initial luminosity of the Sun,
at the start of hydrogen fusion, was approximately 70 per cent of the
present value, yet geological evidence indicated that there had
been no major change in the climate of the Earth over the past 3.5 Gyr
({\eg}, Sagan and Mullen, 1972).%
\footnote{The change in luminosity was noted by Schwarz\-schild (1958)
who speculated about possible geological consequences.}
This change in luminosity is a fundamental effect of the conversion
of hydrogen to helium and the resulting change in solar structure;
thus the attempts to eliminate it resorted to rather drastic measures,
such as suggestions for changes to the gravitational constant.
As noted by Sagan and Mullen, a far more likely explanation is a readjustment
of conditions in the Earth's atmosphere to compensate for the
change in luminosity.
A more serious concern was the fact that attempts to detect the
neutrinos created by the fusion reactions in the solar core found
values far below the predictions.
This evidently raised doubts about the computations of solar models,
and hence on the general understanding of stellar evolution, and
led to a number of suggestions for changing the models such as 
to bring them into agreement with the neutrino measurements.

The last three decades have seen a tremendous growth in our information
about the solar interior, through the detection and extensive observation
of oscillations of the solar surface.
Analyses of these oscillations, appropriately termed helioseismology,
have resulted in extremely precise and detailed information about
the properties of the solar interior, rivaling or in some respects
exceeding our knowledge about the interior of the Earth.

\section{Early history of helioseismology} \plabel{sec:history}

The development of helioseismology has to a large extent been driven
by observations.
Hence in the following I provide an overview of the evolution of
observations of solar oscillations.
Discussions of the development of helioseismic inferences
follow in later sections.

It is possible that the first indications of solar oscillations were
detected by Plaskett (1916), who observed fluctuations in the solar surface
Doppler velocity in measurements of the solar rotation rate.
It was not clear, however, whether the fluctuations were truly solar
or whether they were induced by effects in the Earth's atmosphere.
The solar origin of these fluctuations was established by 
Hart (1954, 1956).
%he also investigated some of their properties, including the variation
%with height in the solar atmosphere.
However, the first definite observations of oscillations
of the solar surface were made by Leighton {\etal}\ (1962).
%with a preliminary report by Leighton (1961).
They detected roughly periodic oscillations in local Doppler velocity
with periods of around 300 s and a lifetime of at most a few periods.
Strikingly, they noted the potential for using the observed period to
probe the properties of the solar atmosphere.
A confirmation of the initial detection of the oscillations was made by 
Evans and Michard (1962).
The observations by Leighton {\etal}\ (1962) also
led to the detection of convective motion on supergranular scales.
As discussed in section~\ref{sec:timedist}, 
the study of solar oscillations and supergranulation has recently
come together again.

Early observations of the five-minute oscillations were of short
duration and limited spatial extent. 
With only such information,
the oscillations were generally interpreted as local phenomena
in the solar atmosphere, of limited spatial and temporal coherence,
possibly waves induced by penetrating convection
({\eg}, Bahng and Schwarzschild, 1963).
However, attempts at determining their structure were made by
several authors, including Frazier (1968);
through observations and Fourier transforms
of the oscillations as a function of position and time,
he could make power spectra as a function of wavenumber and frequency,
showing some localization of power.
Such observations indicated a less superficial nature of the oscillations,
and inspired major theoretical advances in the understanding of their nature:
Ulrich (1970) and Leibacher and Stein (1971) proposed that the observations
resulted from standing acoustic waves in the solar interior.
Such calculations were further developed by Wolff (1972)
and Ando and Osaki (1975), 
who found that oscillations in the relevant frequency and wavenumber
range may be linearly unstable.
However, the definite breakthrough were the observations
by Deubner (1975) which for the first time identified ridges
in the wavenumber-frequency diagram, reflecting the modal structure
of the oscillations.
Similar observations were reported by Rhodes {\etal}\ (1977),
who furthermore compared the frequencies with computed models to
obtain constraints on the properties of the solar convection zone.

The year of 1975 was indeed the {\it annus mirabilis}  of helioseismology.
An important event was the announcement by H. A. Hill of the detection
of oscillations in the apparent solar diameter
(see Hill {\etal}, 1976; Brown {\etal}, 1978).
This was the first suggestion of truly global oscillations of the 
Sun and immediately indicated the possibility of using such
data to investigate the properties of the solar interior
({\eg}, Scuflaire {\etal}, 1975; Christensen-Dalsgaard and Gough, 1976;
Iben and Mahaffy, 1976; Rouse, 1977).
Simultaneously, Brookes {\etal}\ (1976) and
Severny {\etal}\ (1976) announced independent detections
of a solar oscillation with a period of 160 min, with similarly
interesting diagnostic potentials.
Even though these detections have since been found to be of likely 
non-solar origin, they played a very important role as inspiration
for the development of helioseismology.

(For the present author, the announcement by Hill was particularly
significant.
It took place at a conference in Cambridge in the Summer of 1975.
I was engaged, with Douglas Gough, in modeling solar structure and
oscillations, as part of an investigation of mixing induced by
oscillations as a possible explanation of the solar neutrino problem.
As a result, we had available solar models and programmes for
computing their frequencies.
Hill presented an observed spectrum and I was able, the following day,
to compare it with frequencies computed for a model;
the agreement was quite striking.
It has since transpired that the observations had little to do with
global oscillations of the Sun; and the model was surely far too crude
for such a comparison.
Even so, the event was a major personal turning point, directing my 
scientific efforts towards helioseismology.)

The next major observational step was the identification by
Claverie {\etal}\ (1979)
of modal structure of five-minute oscillations in Doppler-velocity
observations in light integrated over the solar disk.
Such observations are sensitive only to oscillations of the lowest
spherical-harmonic degree, and hence these were the first 
confirmed detection of truly global modes of oscillations.
The frequency pattern, with regularly spaced peaks, matched theoretical
predictions based on the asymptotic theory of acoustic modes
of high radial order
(Christensen-Dalsgaard and Gough, 1980a; see also Section \ref{sec:pasymp}).
Further observations, with much higher frequency resolution,
were made from the Geographical South Pole 
during the austral summer 1979--80
(Grec {\etal}, 1980);
these resolved the individual multiplets in the low-degree spectrum
and allowed a comparison between the frequency data, including also the
so-called small frequency separation, and solar models.
The structure of the frequency spectrum was analyzed asymptotically
by Tassoul (1980).
It was pointed out by Gough (1982) that the small separation 
was related to the curvature of sound speed in the solar core;
thus it would, for example, provide evidence for mixing 
of material in the core (see Section~\ref{sec:pasymp}).

The existence of oscillations in the five-minute range, both a low
degree as detected by Claverie {\etal}\ (1979) and at high wavenumbers
as found by Deubner (1975), strongly suggested a common cause
({\eg}, Christensen-Dalsgaard and Gough, 1982).
The gap between these observations was filled by Duvall and Harvey (1983),
who made detailed observations at intermediate degree.
This also allowed a definite identification of the order of the modes,
even at low degree, by establishing the connection with the high-degree
modes for which the order could be directly determined.
By providing a full range of modes these and subsequent 
observations opened the possibilities for detailed inferences of
properties of the solar interior, such as 
the internal solar rotation (Duvall {\etal}, 1984) and
the sound speed (Christensen-Dalsgaard {\etal}, 1985).
%The former analysis, in particular, established that the interior
%of the Sun, even quite close to the core, rotates at approximately
%the same speed as the surface; no evidence was found for a
%rapidly rotating core.

\section{Overall properties of the Sun}

\plabel{sec:sunprop}

The Sun is unique amongst stars in that its properties are
known with high precision. 
The product $G \Msun$, where $G$ is the gravitational constant and
$\Msun$ is the mass of the Sun, is known with very high accuracy
from planetary motion.
Thus the factor limiting the accuracy of $\Msun$ is the value of $G$;
the commonly used value is $\Msun = 1.989 \times 10^{33} \g$.
The solar radius $\Rsun$ follows from the apparent diameter and the distance to 
the Sun.
%Difficulties arise from the definition of the point in the solar limb intensity
%profile at which to determine the apparent diameter, 
%and the physical height in the solar atmosphere to which this corresponds.
Most recent computations of solar models have used
$\Rsun = 6.9599 \times 10^{10} \cm$
(Auwers, 1891).%
\footnote{Brown \& Christensen-Dalsgaard (1998) obtained the value
of $(695.508 \pm 0.026)$~Mm from a careful analysis of 
daily timings at noon of solar transits with a telescope fixed in the direction
of the meridian,
combined with modeling of the limb intensity; this value refers
to the solar photosphere, defined as the point where the temperature
equals the effective temperature.
This value has not yet been used for detailed solar modeling, however.}
The solar luminosity $\Lsun$ 
is determined from satellite irradiance measurements,
suitably averaged over the variation of around 
0.1~\% during the solar cycle 
({\eg}, Willson and Hudson, 1991; Pap and Fr\"ohlich, 1999);
a commonly used value is $\Lsun = 3.846 \times 10^{33} \erg \s^{-1}$.
Finally, the age of the Sun is obtained from age determinations
for meteorites, combined with modeling of the formation history
of the solar system ({\eg}, Guenther, 1989; Wasserburg, in
Bahcall and Pinsonneault, 1995).
Based on a careful analysis, Wasserburg estimated the
age as $t_\odot = (4.566 \pm 0.005) \times 10^9$~yr.

The composition of stellar matter is traditionally characterized
by the relative abundances by mass $X$, $Y$ and $Z$ of hydrogen,
helium and `heavy elements' ({\ie}, elements heavier than helium).
The solar surface composition can in principle be determined from
spectroscopic analysis.
In practice, the principle works for most elements heavier than helium;
for elements with lines in the solar photospheric spectrum,
abundances can be determined with reasonable precision,
although often limited by uncertainties in the relevant basic
atomic parameters and in the modeling of the solar atmosphere
({\eg}, Asplund {\etal}, 2000b),
as well as by blending with weak lines 
({\eg}, Allende Prieto {\etal}, 2001).
The relative abundances so obtained are generally in good agreement
with solar-system abundances as inferred from meteorites
({\eg}, Anders and Grevesse, 1989; Grevesse and Sauval, 1998).
A striking exception is the abundance of lithium, which is lower
by about a factor 150, relative to silicon,
in the Sun than in meteorites.
There have been suggestions that the beryllium abundance is lower also,
but the most recent determinations seems to indicate that the 
solar beryllium abundance is similar to the meteoritic value
({\eg}, Balachandran and Bell, 1998).
As discussed in Section \ref{sec:helsun} these observations are of great 
interest in connection with investigations of solar internal structure
and dynamics.

The noble gases, including helium,
do not have lines in the photospheric spectrum
as a result of the large excitation energies of the relevant
atomic transitions.
It is true that helium can be detected in the solar spectrum,
but only through lines formed high in the solar atmosphere where
conditions are complex and uncertain and a reliable abundance
determination is therefore not possible.
As a result, the solar helium abundance is not known from
`classical' observations.
Typically, the initial abundance $Y_0$ by mass is used as a free
parameter in solar-model calculations.
On the other hand, spectroscopic data do provide a measure of
the ratio $\Zs/\Xs$ of the present surface abundances
heavy elements and hydrogen;
commonly used values are 0.0245 (Grevesse and Noels, 1993)
and 0.023 (Grevesse and Sauval, 1998).

Solar surface rotation can be determined by following the motion of features
on the solar surface ({\eg}, sunspots) as they move across the solar disk,
or through Doppler measurements.
The angular velocity $\Omega$ obtained from Doppler measurements,
as a function of co-latitude $\theta$, can be fitted
by the following relation
\be
{\Omega \over 2 \pi} = 451.5 \nHz - 65.3 \nHz \, \cos^2 \theta
- 66.7 \nHz \, \cos^4 \theta 
\eel{eqn:surfrot}
(Ulrich {\etal}, 1988),
although there are significant departures from this relation,
as well as variations with time (see also Section \ref{sec:change}).

\section{Solar structure and evolution}

\subsection{`Standard' solar models} \plabel{sec:standardsol}

As a background for the discussion of the helioseismically inferred
information about the solar internal structure, it is useful briefly
to summarize the principles of computation of `standard' solar models.%
\footnote{
Further discussion of such models, and detailed results, have been
provided by, for example, Bahcall and Pinsonneault (1992, 1995),
Christensen-Dalsgaard {\etal}\ (1996), Brun {\etal}\ (1998),
and Bahcall {\etal}\ (2001).}
Such models are assumed to be spherically symmetric, ignoring
effects of rotation and magnetic field. 
In that case, the basic equations of stellar structure can be written
\begin{mathletters}
\label{eqn:solstruc}
\begin{eqnarray}
{\dd p \over \dd r} & = &- {G m \rho \over r^2} \; , \label{eqn:hydrost} \\
{\dd m \over \dd r} & = & 4 \pi r^2 \rho \; , \label{eqn:mass} \\
{\dd T \over \dd r} & = & \nabla {T \over p} {\dd p \over \dd r} \; ,
\label{eqn:tgrad} \\
{\dd L \over \dd r} & = & 4 \pi r^2 \left[ \rho \epsilon -
\rho {\dd \over \dd t }\left( {u \over \rho }\right)
 + {p \over \rho }{\dd \rho \over \dd t }\right] \; .
\label{eqn:energy}
\end{eqnarray}
\end{mathletters}
Here $r$ is distance to the center, $p$ is pressure, $m$ is the mass
of the sphere interior to $r$, 
$\rho$ is density, $T$ is temperature,
$L$ is the flow of energy per unit time through the sphere of radius $r$,
$\epsilon$ is the rate of nuclear energy generation per unit mass
and time, and $u$ is the internal energy per unit volume.%
\footnote{During most of the evolution of the Sun, the last two terms
in \Eq{eqn:energy} are very small compared to the nuclear term.}
Also, the temperature gradient has been characterized by
$\nabla = \dd \ln T/ \dd \ln p$ and is determined by the mode of
energy transport.
Where energy is transported by radiation, $\nabla = \nablarad$,
where the radiative gradient is given by
\be
\nablarad =
{3 \over 16 \pi a \clight  G} {\kappa p \over T^4}{L(r) \over m (r)} \; ;
\eel{eqn:nablarad}
here $\clight$ is the speed of light, $a$ is the radiation density constant
and $\kappa$ is the opacity,
defined such that $1/(\kappa \rho)$ is the mean free path of a photon.
In regions where $\nablarad$ exceeds the adiabatic gradient 
$\nablaad = (\partial \ln T / \partial \ln p)_s$,
the derivative being taken at constant specific entropy $s$,
the layer becomes unstable to convection.
In that case energy transport is predominantly by convective motion;
as discussed below,
the detailed description of convection is highly uncertain.

Energy generation in the Sun results from the fusion of
hydrogen into helium.
The net reaction can be written as
\be
4 \hyd \rightarrow \helfour + 2 {\rm e}^+ + 2 \nu_{\rm e} \; ,
\eel{eqn:netreact}
satisfying the constraints of conservation of charge and lepton number.
Here the positrons are immediately annihilated, while the 
electron neutrinos escape the Sun essentially without
reacting with matter and therefore represent an immediate energy loss.
The actual path by which this net reaction takes place involves
different sequences of reactions, depending on the temperature
(for details, see for example Bahcall, 1989).
These reactions differ substantially in the neutrino energy loss and hence
in the energy actually available to the star.

The change in composition resulting from \Eq{eqn:netreact}
largely drives solar evolution.
Until fairly recently, `standard' solar model calculations did not
include any other effects that changed the composition.
However, Noerdlinger (1977) pointed out the potential
importance of diffusion of helium in the Sun.
Strong evidence for the importance of diffusion and settling has
since come from helioseismology (see Section~\ref{sec:infsound})
and these processes are now generally included in the calculations.%
\footnote{{\eg},
Wambsganss (1988),
Cox {\etal}\ (1989),
Proffitt and Michaud (1991),
Proffitt (1994),
Guenther {\etal}\ (1996),
Richard {\etal}\ (1996),
Gabriel (1997),
Morel {\etal}\ (1997),
and Turcotte {\etal}\ (1998).}
Specifically, the rate of change of the hydrogen abundance
is written
\be
{\partial X \over \partial t}
= {\cal R}_{\rm H} + {1 \over r^2 \rho} {\partial \over \partial r}
\left [ r^2 \rho \left(D_{\rm H} {\partial X \over \partial r} 
+ V_{\rm H} X \right) \right] \; ;
\eel{eqn:diffus}
here ${\cal R}_{\rm H}$ is the rate of change 
in the hydrogen abundance from nuclear reactions,
$D_{\rm H}$ is the diffusion coefficient and
$V_{\rm H}$ is the settling speed.
Similar equations are of course satisfied for the abundances of
other elements.
In \Eq{eqn:diffus}, the term in $D_{\rm H} \partial X / \partial r$
tends to smooth out composition gradients, whereas the term
in the settling velocity leads to separation, hydrogen rising
towards the surface and heavier elements including helium sinking
towards the interior.

The basic equations of stellar structure and evolution,
Eqs (\ref{eqn:solstruc}) and (\ref{eqn:diffus}), are relatively simple;
also, the numerical techniques for solving them are well established
and well tested in the case of solar models.
However, the apparent simplicity hides a great deal of complexity,
often combined under the heading of `microphysics'.
To complete the equations, their right-hand sides must be
expressed in terms of the basic variables $\{p, m, T, L, X_i\}$,
where $X_i$ denotes the abundances of the relevant elements.
This requires expressions for the density $\rho$ and other
thermodynamic variables, for the opacity $\kappa$, for
the energy generation rate $\epsilon$ and the rates of change
of composition ${\cal R}_i$, as well as for the diffusion and
settling coefficients.
At the level of precision required for solar modeling,
each of these components involves substantial physical subtleties.
The thermodynamic quantities are obtained from an equation of state,
which as a minimum requirement (although not always met) must
satisfy thermodynamic consistency.
Two conceptually very different formulations are in common use:
one is the so-called `chemical picture' where the equation of state
is based on an expression for the free energy
of a system consisting of atoms, ions, {\etc}, containing the
relevant physical effects; 
the second is the `physical picture',
which assumes as building blocks only fundamental particles
(nuclei and electrons), and treats density effects by means of a
systematic expansion
(for reviews, see for example D\"appen, 1998; D\"appen and Guzik, 2000).
A representative and commonly used example of the chemical picture
is the so-called MHD equation of state (Mihalas {\etal}, 1988).
The physical picture has been implemented by the OPAL group
(Rogers {\etal}, 1996).
In the opacity calculation the detailed distribution of the atoms
on ionization and excitation states must be taken into account,
obviously requiring a sufficiently accurate equation of state
(see D\"appen and Guzik, 2000).
The most commonly used opacity tables are those of the OPAL group
(Iglesias and Rogers, 1996).
Computation of the energy generation and composition changes
obviously requires nuclear cross sections, the determination
of which is greatly complicated by the low typical reaction
energies relevant to stellar interiors;
recently, two major compilations of nuclear parameters have
been published by Adelberger {\etal}\ (1998) and Angulo {\etal}\ (1999).
Additional complications result from the partial screening of
the Coulomb potential of the reacting nuclei by the stellar plasma;
the so-called weak-screening approximation (Salpeter, 1954)
is still in common use.%
\footnote{A careful analysis of Salpeter's result was provided by
Br\"uggen and Gough (1997).
For different treatments, see, for example, Gruzinov and Bahcall (1998) and
Shaviv and Shaviv (2001).
Bahcall {\etal}\ (2002) gave a critical discussion of these issues.}
Expressions for the diffusion and settling coefficients
have been provided by, for example, Michaud and Proffitt (1993)
and Thoul {\etal}\ (1994).

In the Sun, convection occurs in the outer about 29\% of the solar radius; 
this is visible on the solar surface in the form of motion and other
fluctuations in the so-called granulation and supergranulation.
In the convectively unstable regions, modeling requires a relation
to determine the convective energy transport from
the local structure; particularly important is the superadiabatic
gradient, {\ie}, the difference between the actual temperature gradient
$\nabla$ and the adiabatic value $\nablaad$, which controls both
the dynamics of the convective motion and the net energy transport.
In model calculations this relation is typically
obtained from simple recipies, 
and characterized by one or more parameters that determine convective efficacy.
A characteristic example is the mixing-length treatment (B\"ohm-Vitense, 1958),
parametrized by the mixing-length parameter $\alphac$ which measures
the mean free path of convective eddies in units of the local
pressure scale height.
Also, it is common to neglect the dynamical effects of convection,
generally described as a turbulent pressure.
In most of the solar convection zone, convection is so efficient
that the actual temperature gradient is very close to the adiabatic value.
Near the surface, however, where the density is low, a fairly
substantial superadiabatic gradient is required to transport the energy.
The effect of the parametrization of the convection treatment through,
{\eg}, $\alphac$ is to control the degree of superadiabaticity and hence,
effectively, the adiabat of the nearly adiabatic part of the convection zone
({\cf} Gough \& Weiss, 1976).

A more realistic description of the uppermost part of the convection zone
is possible through detailed three-dimensional and
time-dependent hydrodynamical simulations, taking into account
radiative transfer in the atmosphere ({\eg}, Stein and Nordlund, 1998a).
Such simulations successfully reproduce the observed surface
structure of solar granulation ({\eg}, Nordlund and Stein, 1997),
as well as detailed profiles of lines in the solar radiative
spectrum, without the use of parametrized models of turbulence
(Asplund {\etal}, 2000a).
The simulations only cover a very small fraction of the solar radius, 
and are evidently far too time-consuming to be included in 
general solar modeling.
%However, it has been demonstrated that at least some aspects of
%convection, as obtained from mixing-length theory, agree
%with the detailed calculations (Abbett {\etal}, 1997).
Rosenthal {\etal}\ (1999) extrapolated an averaged simulation
through the adiabatic part of the convection zone by means of
a model based on the mixing-length description, demonstrating that
the adiabat predicted by the simulation was essentially consistent
with the depth of the solar convection zone as determined from
helioseismology (see Section~\ref{sec:solstruc}).
Also, Li {\etal}\ (2002) developed an extension of mixing-length theory,
including effects of turbulent pressure and kinetic energy,
based on numerical simulations of near-surface convection.

\begin{figure}[htbp]
\begin{center}
\inclfig{7.2cm}{\fig/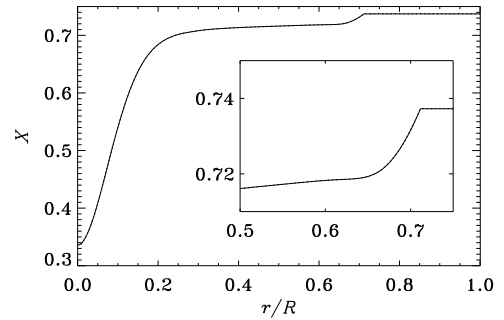}{Hydrogen abundance}
\end{center}
%\vskip -1cm
\caption{
\figlab{fig:X-mod}
Hydrogen mass fraction $X$ as a function of fractional radius in a model
of the present Sun (Model S of Christensen-Dalsgaard {\etal}, 1996).
The insert shows details of the behavior near the base of the convection
zone.
}
\end{figure}

The computation of a model of the present Sun typically starts
from the so-called zero-age main sequence, where the model
can be assumed to be of uniform composition, with nuclear reactions
providing the energy output; however, models have also been computed
starting during the earlier phase of gravitational contraction
({\eg}, Morel {\etal}, 2000).
The model is characterized by the mass (generally assumed to be constant
during the evolution) and the initial composition, specified by
the abundances $X_0$, $Y_0$ and $Z_0$.
In addition, parameters characterizing convective energy transport,
such as the mixing-length parameter $\alphac$, must be specified.
The model at the age of the present Sun must match the present solar
radius and luminosity, as well as the observed ratio $\Zs/\Xs$
of the abundance of heavy elements to hydrogen at the surface.
This is achieved by adjusting $\alphac$ and $Y_0$, which largely
control the radius and luminosity, and the initial heavy-element
abundance $Z_0$.

To illustrate some properties of models of the present Sun,
Fig.~\ref{fig:X-mod} shows the hydrogen-abundance profile $X$.%
\footnote{Extensive sets of variables for Model S of 
Christensen-Dalsgaard {\etal}\ (1996) are available at\hfill\break
{\tt http://astro.ifa.au.dk/$\sim$jcd/solar\_models/}.}
The abundance is uniform in the outer convection zone, 
extending from the surface to $r = 0.711 R$, which is
fully mixed; as a result of helium settling, $X$ has increased
by about 0.03 relative to its initial value of 0.709.
Just below the convection zone, helium settling has caused a sharp
gradient in the hydrogen abundance.
In the inner parts of the model the hydrogen abundance has been
reduced due to nuclear fusion.
Detailed tables of model quantities, from a slightly different
calculation, were provided by Bahcall and Pinsonneault (1995).

\subsection{Solar neutrinos} \plabel{sec:solneutrino}

As indicated in \Eq{eqn:netreact}, hydrogen fusion in the Sun
unavoidably produces electron neutrinos.
It is easy to estimate, from the solar energy flux,
that the total flux of solar neutrinos at the Earth is around
$7 \times 10^{10} \cm^{-2} \s^{-1}$.
This depends little on the details of the nuclear reactions in the
solar core, as long as the solar energy output derives solely from
nuclear reactions.
However, the energy spectrum of the neutrinos depends sensitively on
the branching between the various reactions.
This is particularly true of the highest-energy neutrinos, which
are produced by a relatively rare and very temperature-sensitive reaction.
This is of crucial importance to attempts to detect neutrinos from the Sun.

A detailed description of the issues related to solar neutrinos,
including their detection, was given by Bahcall (1989).
More recent reviews have been provided by, for example, 
Haxton (1995), Castellani {\etal}\ (1997),
Kirsten (1999), and Turck-Chi\`eze (1999).
Until recently, three classes of experiments had been carried out to detect 
solar neutrinos.
The first experiment, where the $\nu_{\rm e}$ reacted with chlorine,
was established by R. Davis in the Homestake Gold Mine, South Dakota,
and yielded its initial results in 1968 (Davis {\etal}, 1968),
providing an upper limit on the capture rate of 3 SNU  (Solar Neutrino Units;
1 SNU corresponds to $10^{-36}$ reactions per target atom per second).
This was substantially below the expected flux
({\eg}, Bahcall, Bahcall \& Shaviv, 1968).
The latest average measured value is 
${\rm 2.56 \pm 0.16(statistical) \pm 0.16(systematic)}$ SNU
(Cleveland {\etal}, 1998);
this is to be compared to typical model predictions of around 8 SNU
({\eg}, Bahcall {\etal}, 2001; Turck-Chi\`eze {\etal}, 2001a).

This experiment is most sensitive to high-energy neutrinos, and hence
the predictions depend on the solar central temperature to a high power.
Thus attempts to explain the discrepancy, known as the `solar neutrino problem',
generally aimed at lowering the core temperature of the model,
for example by postulating a rapidly rotating core such that the central
pressure, and therefore the central temperature, would be reduced
by centrifugal effects ({\eg}, Bartenwerfer, 1973;
Demarque {\etal}, 1973).
Another suggestion was an inhomogeneous composition, the interior being
lower in heavy elements than the convection zone; 
this would reduce the opacity and hence the core temperature
({\eg}, Joss, 1974).
A similar effect would result if energy transport in the Sun 
were to take place in part by non-radiative means, 
such as through motion of postulated weakly interacting massive 
particles ({\eg}, Faulkner and Gilliland, 1985; Spergel and Press, 1985;
Gilliland {\etal}, 1986).
Substantial mixing of the core was also proposed; by increasing the amount
of hydrogen in the core, this would reduce the temperature required to
generate the solar luminosity and hence reduce the neutrino flux
({\eg}, Ezer and Cameron, 1968; Bahcall, Bahcall, and Ulrich, 1968; 
Schatzman {\etal}, 1981).
An interesting variant on this idea, appropriately called `the solar spoon',
was proposed by Dilke and Gough (1972): 
according to this the solar core was mixed about a million years ago
due to the onset of instability to oscillations,
and the present luminosity derives in part from the
readjustment following this mixing, reducing the rate of nuclear energy
generation and hence the neutrino flux.
Detailed calculations have confirmed the required instability
({\eg}, Christensen-Dalsgaard {\etal}, 1974; Boury {\etal}, 1975);
however, it has not been definitely determined whether or not the
subsequent nonlinear development of the oscillations may lead to mixing.

It should be emphasized that such non-standard models are constructed
to satisfy the constraint of the observed solar radius and luminosity;
thus, although they may account for the observed neutrino flux, there
is no independent way of testing them or choosing between them on the
basis of `classical' observations.
This is clearly a rather unsatisfactory situation.
As discussed in Section~\ref{sec:solstruc}, helioseismology
has provided tests of these non-standard models.

Other experiments have confirmed the discrepancy between the observed
neutrino flux and the predictions of standard solar models.
Measurements at the Kamiokande and Super-Kamiokande facilities
of neutrino scattering on electrons in water,
which detect only the rare high-energy neutrinos,
yield a flux smaller by about a factor two than
the standard models ({\eg}, Fukuda {\etal}, 2001); these measurements
are sensitive to the direction of arrival of the neutrinos and 
in this way confirm their solar origin.
Detection also of the lower-energy neutrinos has been made 
in the GALLEX and SAGE experiments 
through neutrino capture in gallium.
For GALLEX the resulting measured detection rate is
${\rm 77.5 \pm 6.2 (statistical) \pm 4.5 (systematic)}$ SNU
(Hampel {\etal}, 1999) while the result for SAGE is
${\rm 75.4 \pm 6.9 (statistical) \pm 3.2 (systematic)}$ SNU 
(Gavrin, 2001; see also Abdurashitov {\etal}, 1999);
these are again substantially lower than the model predictions
of around 130 SNU.

Although these discrepancies clearly raise doubts about solar modeling,
their origin may instead be in the properties of the neutrinos.
In addition to the electron neutrino, two other types of neutrinos,
the muon neutrino $\nu_\mu$ and the tau neutrino $\nu_\tau$, are known.
If neutrinos have finite mass these three types may couple, and
hence the electron neutrinos generated in the solar core may
be converted into neutrinos of the other types, to which current
experiments are less sensitive.
A mechanism of this nature, the so-called MSW effect, was proposed by
Wolfenstein (1978) and Mikheyev and Smirnov (1985).
Here the neutrinos oscillate between the different states through
interaction with matter in the Sun;
by choosing appropriately the relevant parameters, it is possible
to bring the measured and computed neutrino capture rates into agreement.
A confirmation that such a mechanism may operate has been obtained
through measurements of oscillations of muon neutrinos generated
in the Earth's atmosphere ({\eg}, Fukuda {\etal}, 1998).
For a recent overview of neutrino oscillations, 
see Bahcall {\etal}\ (1998).

Very recently new measurements have been announced from the
Sudbury Neutrino Observatory, which strongly support the presence
of neutrino oscillations and are consistent with the standard 
solar model (Ahmad {\etal}, 2001).
Here measurements of high-energy neutrinos are made through the
interaction with deuterium, in the form of heavy water.
This reaction is only sensitive to $\nu_{\rm e}$.
The measured flux is significantly lower than the flux obtained 
at Super-Kamiokande through
electron scattering, which has some sensitivity to $\nu_\mu$ and $\nu_\tau$.
Thus the difference between the two measurements provides an
indirect measure of the conversion of $\nu_{\rm e}$ into 
$\nu_\mu$ and $\nu_\tau$, and hence of the flux of neutrinos
originating from the Sun.
The result agrees, within errors, with standard solar models.

Given this striking confirmation of the existence of neutrino oscillations,
the emphasis of solar neutrino research is shifting towards using
the measurements to constrain the properties of the neutrinos.
This evidently requires secure constraints on the rate of neutrino
production in the Sun.
In Section \ref{sec:helneutrino} I return to the possible importance 
of helioseismology in this regard.

\subsection{The rotation of the Sun} \plabel{sec:solrot}

As mentioned in Section~\ref{sec:sunprop}, the 
solar surface displays differential rotation, 
the rotation period varying from around 25~d at the equator
to more than 30~d near the poles.
Different measures of the rotation give somewhat different results.
For example, the rotation rates of magnetic features are generally
a few per cent higher than the photospheric
rate as determined from Doppler-velocity
measurements (for a recent review, see Beck, 2000).
As the magnetic field is likely anchored at some depth beneath the
solar surface, this suggests the presence of an increase in rotation
rate with depth.

There is as yet no firm theoretical understanding of the rotation
of the Sun and its evolution with time.
It is normally assumed that stars rotate rapidly when they are
formed and subsequently slow down;
indeed, one observes a strong correlation between age and rotation
rate amongst solar-type stars ({\eg}, Skumanich, 1972).
The loss of angular momentum probably takes place through a stellar
wind, magnetically coupled to the outer convection zone
({\eg}, Mestel, 1968).
However, it is not clear how the convection zone is coupled rotationally
to the radiative interior or how angular momentum may be transported from
the deep interior towards the surface.
Thus while the convection zone is braked, the star might still retain
a rapidly rotating core.
In fact, evolution calculations taking rotation into account,
and assuming angular-momentum transport in the interior as a
result of hydrodynamical instabilities, have found the rotation
of the deep interior of the model of the present Sun to be
several times higher than the surface rotation rate
({\eg}, Pinsonneault {\etal}, 1989;
Chaboyer {\etal}, 1995).
A sufficiently rapidly rotating core could affect solar structure;
also, the resulting distortion of the Sun's external gravitational field
might compromise tests of Einstein's theory of general relativity
based on observations of planetary motion ({\eg}, Dicke, 1964;
Nobili and Will, 1986).
Finally, the instabilities invoked to transport angular momentum could also
lead to partial mixing of the solar interior, hence affecting its evolution.
Thus it is evidently important to obtain secure information about
the solar internal rotation and the evolution of stellar rotation.

The rotation within the convection zone, and hence the surface differential
rotation, is likely controlled by angular-momentum transport
by the convective motions.
Early hydrodynamical models ({\eg}, Glatzmaier, 1985;
Gilman and Miller, 1986) indicated that rotation depends predominantly
on the distance to the rotation axis, as suggested by the
Taylor-Proudman theorem ({\eg}, Pedlosky, 1987; see also Miesch, 2000).
Thus the observed surface variation with latitude would translate into
a decrease in rotation rate with depth, at the solar equator, 
in apparent conflict with the inferences from different measures
of surface rotation.
However, these and other models are certainly far from resolving
all the relevant scales of convection, and hence the results
must still be regarded as somewhat uncertain.
I return to these problems in Section~\ref{sec:helsun},
in the light of the helioseismic inferences of solar internal rotation.

\subsection{Solar magnetic activity} \plabel{sec:activity}

Because of proximity of the Sun, phenomena on its surface and in its
atmosphere can be studied in great, and often bewildering, detail
(for a recent detailed overview, see Schrijver and Zwaan, 2000).
These phenomena are closely related to magnetic fields and occasionally
give rise to explosions and ejections into the solar wind
of matter and magnetic fields which may harm satellites in orbit
near the Earth and interfere with radio communication and power grids.
Thus there is substantial practical interest in a better understanding
of the solar magnetic activity and, if possible, predictions of eruptions.

At the photospheric level
the most visible manifestation of the activity are the sunspots, which
have been observed fairly systematically over the last four centuries.
Sunspots are areas of somewhat lower temperature, and hence lower
luminosity, than the rest of the photosphere.
Here convective energy transport is partly suppressed by a strong magnetic
field emerging through the solar surface;
typical field strengths are up to 0.4 Tesla.
Sunspots often occur in pairs with opposite polarity, which may
correspond to a loop of magnetic flux anchored in the solar interior.

The most striking aspect of the sunspots and other related phenomena
is the variation with time: 
the number of sunspots vary with a period of roughly 11 years.
Observations of the solar magnetic field show that it reverses between
sunspot minima; hence the full, magnetic solar cycle has a period
of 22 years. 
However, there are considerable variations in the length of the cycle
and the number of spots at solar maximum activity.
Interestingly, there were virtually no sunspots during the period
1640 -- 1710 (the so-called Maunder minimum),
where the Sun was already observed regularly
({\eg}, Ribes and Nemes-Ribes, 1993; Hoyt and Schatten, 1996).

The origin of the solar magnetic activity and its variation with time
is likely to involve interactions, often described as dynamo processes,
between rotation and motion of
the solar plasma within or just beneath the solar convection zone
({\eg}, Gilman, 1986;
Choudhuri, 1990;
Parker, 1993; 
Cattaneo, 1997;
Charbonneau and MacGregor, 1997).
Thus an understanding of the cause of the solar cyclic variation
depends on knowledge about the solar internal rotation.

\section{Stellar oscillations}

In order to understand the diagnostic potential of solar oscillations,
some basic insight into the properties of stellar oscillations is required.%
\footnote{A much more detailed description of general stellar oscillations
was provided by Unno {\etal}\ (1989), while Gough (1993) discussed
aspects more directly relevant to helio- and asteroseismology.
The classical review by Ledoux and Walraven (1958) still repays careful study.}
The observed
oscillations have extremely small amplitudes and hence can be described
as linear perturbations, around the solar models
resulting from evolution calculations.
As a result, the frequencies provide a direct diagnostic of the
properties of the solar interior: given a solar model, the relevant
aspects of the frequencies can be computed very precisely, and
the differences between the observed and the computed frequencies
can be related to the errors in the model.

\subsection{Equations and boundary conditions}

\subsubsection{Some basic hydrodynamics}

A hydrodynamical system is characterized by specifying
the physical quantities  as functions of position $\boldr$ and time $t$.
These properties include, {\eg},
the local density $\rho(\boldr,t)$, the local pressure
$\pres(\boldr,t)$, %and any other thermodynamic quantity that may be needed,
as well as the local instantaneous velocity $\boldv(\boldr,t)$.
%Here $\boldr$ denotes the position vector to a given point in space.
%and the description
%therefore corresponds to what is seen by a stationary observer.
%This is known as the so-called \textsl{Eulerian} description.
%In addition, we shall also use the 
%so-called {\it Lagrangian} description, following the motion of a
%given parcel of fluid.
%In addition to the time derivative
%$\partial / \partial t$ at fixed $\boldr$, 
%{\ie}, as seen by a stationary observer, 
%it is often convenient to consider the derivative 
%$\dd / \dd t$ observed when following the motion;
%the latter is also known as the material (or Stokes) time derivative.
%The local velocity is obviously determined by the rate of
%change of the position $\boldr$ of a fluid parcel:
%\be
   %\boldv(\boldr,t)={\dd {\boldr}\over \dd t} \; .
%\eel{E3.0}
%This may furthermore be used to relate the 
%two time derivatives of some quantity $\phi$:
%\be
%  {\dd \phi\over \dd t}=\pard{\phi}{t}_{\boldr}+\vectordel\phi\scalarprod
%    {\dd \boldr\over \dd t}=\wpard{\phi}{t}+\boldv\scalarprod\vectordel\phi\,.
%\eel{E3.1}
%
For helioseismology, the most important aspects of the system concern
its mechanical properties.
Conservation of mass is expressed by {\it the equation of continuity}:
\be
 \wpard{\rho}{t} + \diverge(\rho\boldv) = 0 \; .
\eel{E3.4}
In stellar interiors the viscosity in the gas can generally be neglected,
and the relevant forces are in most cases just pressure and gravity.
Then {\it the equations of motion} (also known as Euler's equations)
can be written as
%\be
%    \rho\,{\dd \boldv \over \dd t} = -\vectordel \pres + \rho \,\boldg \; ,
%\eel{E3.7}
%or, using \Eq{E3.1},
\be
   \rho\,\left( \wpard{\boldv}{t}+\rho\boldv\scalarprod\vectordel\boldv 
\right) =
      -\vectordel \pres+\rho \, \boldg \,,
\eel{E3.8}
where, on the left-hand side, the quantity in brackets is the
time derivative of velocity in a fluid parcel following the motion.
The first term on the right-hand side is the
surface force, given by the pressure $\pres$,
while the second term is given by
the gravitational acceleration $\boldg$,
obtained from the gradient of the gravitational potential $\Phi$,
$\boldg = - \vectordel\Phi$,
where $\Phi$ satisfies Poisson's equation,
$\nabla^2 \Phi = 4\pi G\rho$.

To complete the description, we need to relate $p$ and $\rho$.
In general, this requires consideration of
the energetics of the system,
as described by the first law of thermodynamics.
%By applying it to a volume of unit mass (and hence of volume $1/\rho$), 
%moving with the fluid,
%we obtain {\it the energy equation}
%\be
%  {\dd Q\over \dd t}
%  ={\dd E\over \dd t}+p{\dd \over \dd t} \left({1 \over \rho}\right) =
%  {\dd E\over \dd t}-{p\over\rho^2}\,{\dd \rho\over \dd t} \; ,
%\eel{E3.13}
%Here $\dd Q/\dd t$ is the rate of heat loss or gain per unit mass of material,
%and $E$ the internal energy per unit mass.
%The equation may be cast in a more convenient form by
%using thermodynamic identities ({\eg}, Cox \& Giuli, 1968):
%\be
%  {\dd Q\over \dd t} =
%  {1\over\rho(\Gamma_3-1)}\,\left({\dd \pres\over \dd t}-
%          {\Gamma_1 \pres\over\rho}\,{\dd \rho\over \dd t}\right)  \; ,
%\eel{E3.14a}
%where $\Gamma_1 = (\partial\ln p / \partial\ln \rho)_s$ and
%$\Gamma_3-1 = (\partial\ln T / \partial\ln \rho)_s$.
%The heating $Q$ comes from nuclear reactions as well as from
%the divergence of the energy flux through the star.
However, in most of the star the time scale for energy exchange
%(essentially the thermal time scale of the star) 
is much longer than the relevant pulsation periods. 
Then the motion is essentially adiabatic, satisfying 
{\it the adiabatic approximation}
\be
{\dd \pres\over \dd t} = {\Gamma_1 \pres\over\rho} {\dd \rho\over \dd t} \; ,
\eel{E3.19q}
where $\Gamma_1 = (\partial \ln p / \partial \ln \rho)_s$,
and $\dd / \dd t$ denotes the time derivative following the motion.
We shall use this approximation in most of the analysis of solar oscillations.
It breaks down near the stellar surface, where the local thermal time
scale becomes very short.
However, as discussed in Section~\ref{sec:oscilprop} this is only one amongst
a number of problems in the treatment of this region, which must
be taken into account in the analysis of the observed solar oscillation
frequencies.

\subsubsection{The linear approximation} \plabel{sec:linosc}

We now regard the oscillations as small perturbations
around a stationary equilibrium model,
assumed to be a normal spherically symmetric stellar evolution model.
Thus it satisfies
Eqs (\ref{eqn:hydrost}) and (\ref{eqn:mass}) of stellar structure,
with
\be
\boldg_0 = - {G m_0 \over r^2} \bolda_r \; ,
\eel{eqn:eqgrav}
where equilibrium quantities are characterized by subscript `0',
and $\bolda_r$ is a unit vector in the radial direction.

To describe the oscillations we write, for example, pressure as
\be
    \pres(\boldr,t) = \pres_0(\boldr)+\pres'(\boldr,t) \; ,
\eel{E3.34}
where $\pres'$ is a small perturbation.
Here $\pres'$ is the
\textsl{Eulerian} perturbation, that is, the perturbation at a given
spatial point.
In addition to the velocity $\boldv$, we introduce
the displacement $\bfdelr$ of fluid elements resulting from the 
perturbation, such that $\boldv = \partial \bfdelr / \partial t$.
It is also convenient to consider
\textsl{Lagrangian} perturbations, in a reference frame following the motion. 
The Lagrangian perturbation to pressure, for example, may be calculated as
\be
     \delta \pres (\boldr) = \pres(\boldr+\bfdelr)-\pres_0(\boldr)
       = \pres'(\boldr)+\bfdelr\scalarprod\vectordel \pres_0\,.
\eel{E3.35}
%Equation (\ref{E3.35}) is equivalent to the relation (\ref{E3.1}) between the
%local and the material time derivative. 

To obtain the lowest-order
(linear) equations for the perturbations, we insert
expressions such as \Eq{E3.34} into the full equations,
subtract equilibrium
equations, and neglect quantities of order higher than one in
$\pres'$, $\rho'$, $\boldv$, etc.
For the continuity equation the result is,
%\be
%   \wpard{\rho'}{t}+\diverge(\rho_0\boldv) = 0 \; ,
%\eel{E3.37}
%or, by integrating with respect to time,
after integration with respect to time,
\be
    \rho' + \diverge(\rho_0\,\bfdelr) = 0 \,.
\eel{E3.38}
%Note that this equation may also be written as 
%[using the analogue to \Eq{E3.35}]
%\be
%   \delta\rho + \rho_0\,\diverge(\bfdelr) = 0 \; ,
%\eel{E3.39}
%which corresponds to \Eq{E3.5}.
The equations of motion become
\be
   \rho_0\,{\partial^2  \bfdelr\over\partial t^2} =\rho_0\,\wpard{\boldv}{t}
      = -\vectordel \pres' + \rho_0\boldg' + \rho'\boldg_0 \; ,
\eel{E3.40}
where, obviously, $\boldg' = - \vectordel\Phi'$.
The perturbation $\Phi'$ to
the gravitational potential satisfies the perturbed Poisson equation
\be
   \del^2\Phi' = 4\pi G\rho' \; .
\eel{E3.41}

We finally assume the adiabatic approximation, \Eq{E3.19q}, to obtain
\be
   \wpard{\,\delta \pres}{t}
   -{\Gamma_{1,0} \pres_0\over\rho_0}\,\wpard{\,\delta\rho}{t} =0 \; ,
\eel{E3.45q}
or, by integrating over time and expressing it
%\be
   %\delta \pres = {\Gamma_{1,0} \pres_0\over\rho_0}\,\delta \rho\,.
%\eel{E3.46}
on Eulerian form,
\be
   \pres'+ \bfdelr \scalarprod \vectordel \pres_0
   = {\Gamma_{1,0} \pres_0\over\rho_0}
           (\rho'+\bfdelr\scalarprod \vectordel\rho_0) \,.
\eel{E3.47}

\subsubsection{Equations of linear adiabatic stellar oscillations}

Assuming a spherically symmetric and time-inde\-pen\-dent equilibrium,
the solution is separable in time, and in the
angular coordinates $(\theta, \phi)$ of the spherical polar coordinates
$(r, \theta, \phi)$
(where $\theta$ is co-latitude, {\ie}, the angle from the polar axis,
and $\phi$ is longitude).
Then, time dependence is naturally expressed as a harmonic function,
characterized by a frequency $\omega$;
%since, furthermore, it simplifies the analysis to work in terms
%of complex variables, 
for instance, the pressure perturbation is written on complex form as
\be
\pres'(r, \theta, \phi, t) =
\Re [\tilde \pres ' (r) f ( \theta , \phi ) \exp (- i \omega t) ] \; .
\eel{pres-per0}
Here $f(\theta, \phi)$, which remains to be specified,
describes the angular variation of the solution and,
as indicated, the amplitude function $\tilde \pres '$ 
is a function of $r$ alone.
For simplicity, I also drop the subscript `0' on equilibrium quantities.

Given a time dependence of this form, Eqs (\ref{E3.40})
can be written as
\be
    \omega^2  \bfdelr
      = {1 \over \rho} \vectordel \pres' - \boldg' 
      - {\rho'\over \rho} \boldg \; ,
\eel{E3.40x}
which has the form of a linear eigenvalue problem, $\omega^2$ being the
eigenvalue.
Indeed, the right-hand side can be regarded as a linear operator
$\force(\bfdelr)$ on $\bfdelr$:
in the adiabatic approximation
$\pres'$ is related to $\rho'$ by \Eq{E3.47},
and $\rho'$, in turn, can be obtained from $\bfdelr$ by using
\Eq{E3.38};
also, given $\rho'$, $\Phi'$ and hence $\boldg'$ can be obtained
by integrating \Eq{E3.41}.
I return to this formulation of the problem in Section~\ref{sec:varprin}, below.

To obtain the proper form of $f (\theta, \phi)$
in \Eq{pres-per0},
we first express the displacement vector as
$$
\bfdelr = \xi_r \bolda_r + \boldxih \; ,
$$
where $\boldxih$ is the tangential component of the displacement.
We now take the tangential divergence $\divh$ of the equations of motion,
and use the tangential part of the continuity equation to
eliminate $\divh \boldxih$.
In the resulting equation, derivatives with respect to $\theta$ and
$\phi$ only occur in the combination $\nablah^2$, where
$$
\nablah^2  = {1 \over r^2 \sin \theta} {\partial \over \partial \theta}
\sin \theta {\partial \over \partial \theta}  +
{1 \over r^2 \sin^2 \theta} {\partial^2 \over \partial \phi^2}
$$
is the tangential part of the Laplace operator.
The same is obviously true of Poisson's equation. %
%\footnote{It may be shown that the full energy equation, in the case where the
%adiabatic approximation is not made, results in the same behavior.}
This shows that separation in the angular variables can be
achieved in terms of a function $f(\theta, \phi)$ which
is an eigenfunction of $\nablah^2$,
\be
\nablah^2 f =- {1 \over r^2} \Lambda f \; ,
\eel{hor-lap}
where $\Lambda$ is a constant.
A complete set of solutions to this eigenvalue problem
are the spherical harmonics,
\be
f ( \theta , \phi ) = (-1)^m c_{l m} P_l^m ( \cos \theta )
\exp (i m \phi ) \equiv Y_l^m ( \theta , \phi ) \; ,
\eel{spher-harm}
where $P_l^m$ is a Legendre function and
$c_{l m}$ is a normalization constant,
such that the integral of $| Y_l^m |^2$ over the unit sphere is unity.
Here $l$ and $m$ are integers, such that $-l \le m \le l$ and
$\Lambda = l(l+1)$.

With this separation of variables the pressure perturbation,
for example, can be expressed as
\be
\pres'(r, \theta, \phi, t) =
\sqrt {4 \pi} 
\Re [\tilde \pres ' (r) Y_l^m ( \theta , \phi ) \exp (- i \omega t) ] \; .
\eel{pres-per}
Also, it follows from the equations of motion that the displacement
vector can be written as
\bea
\label{displ}
\bfdelr &=&
\sqrt {4 \pi} \Re \left\{
\left[ {\tilde \xi}_r (r) Y_l^m (\theta ,\phi ) \bolda_r
\right. \right. \\
& & \left. \left.
+ {\tilde \xih (r) \over L} \left( {\partial Y_l^m \over \partial \theta}
\bolda_{\theta}
+ {1 \over \sin \theta} {\partial Y_l^m \over \partial \phi}
\bolda_{\phi} \right)\right] \exp (- i \omega t ) \right\} \; , \nonumber
\eeanl{displ}
where
\be
\tilde \xih (r) = {L \over r \omega^2} \left( {1 \over \rho} \tilde \pres '
+ \tilde \Phi ' \right) \; ,
\eel{hor-disp}
and $L = \sqrt{l(l+1)}$;
in \Eq{displ} $\bolda_\theta$ and $\bolda_\phi$ are unit vectors
in the $\theta$ and $\phi$ directions.
With this definition $\tilde\xi_r$ and $\tilde\xih$ are
essentially the root-mean-square radial and horizontal displacements.

In investigations of the properties of the oscillations it is often
convenient to approximate locally their spatial behavior by a
plane wave, $\exp( i \, \boldk \cdot \boldr)$, where the local wavenumber
$\boldk$ can be separated into radial and tangential components
as $\boldk = k_r \bolda_r + \boldkh$.
From \Eq{hor-lap} it then follows that
\be
\kh^2 \simeq {l(l+1) \over r^2} \; ,
\eel{hor-wave}
where $\kh = |\boldkh|$.
Thus, for example, the horizontal surface wavelength of the mode
is given by
\be
\lambda_{\rm h} = {2 \pi \over \kh} \simeq {2 \pi R \over \sqrt{l(l+1)}} \; ;
\eel{hor-wavelength}
in other words, $l$ is approximately the number of wavelengths
around the stellar circumference.
This identification is very useful in the asymptotic analysis
of the oscillations.
Also, it follows from, {\eg}, \Eq{pres-per} that $m$
measures the number of nodes around the equator.

\begin{figure}[htbp]
\begin{center}
\inclfig{8.6cm}{\fig/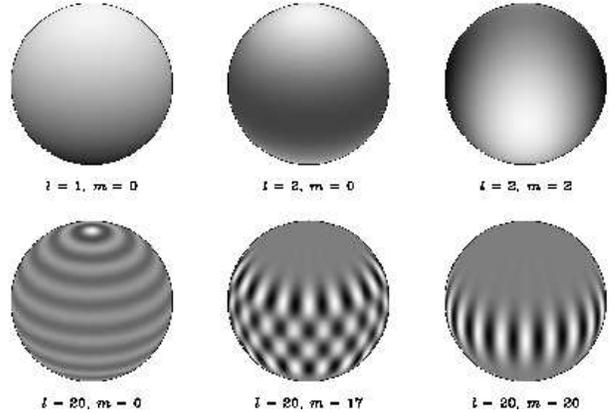}{Spherical harmonics}
\end{center}
\caption{
\figlab{fig:ylm}
Examples of spherical harmonics, labelled by the degree $l$ and
azimuthal order $m$.
For clarity the polar axis has been inclined $30^\circ$ relative
to the plane of the page.
}
\end{figure}

A few examples of spherical harmonics are shown in Fig.~\ref{fig:ylm}.
It should be noticed that with increasing degree the sectoral modes,
with $m = \pm l$, become increasingly confined near the equator.

Given the separation of variables, the equations of adiabatic stellar pulsation
are reduced to ordinary differential equations for the
amplitude functions;
writing the equations in terms of the variables
$\{\xi_r, \; \pres', \; \Phi', \; \dd \Phi' / \dd r\}$
(where I have dropped the tildes)
it is straightforward to obtain
\bea
{\dd \xi_r \over \dd r }
= & - & \left( {2 \over r }
+ {1 \over \Gamma_1 p} {\dd \pres \over \dd r }\right) \xi_r
+ {1 \over \rho c^2} \left({S_l^2 \over \omega^2} - 1 \right) \pres' 
\nonumber \\ 
& + & {l ( l +1) \over \omega^2 r^2} \Phi ' \; ,
\eeal{E4.54}
\be
{\dd \pres' \over \dd r }= \rho ( \omega^2 - N^2 ) \xi_r +
{1 \over \Gamma_1 p} {\dd \pres \over \dd r} \pres'
- \rho {\dd \Phi ' \over \dd r } \; ,
\eel{E4.55}
and
\be
{1 \over r^2} {\dd \over \dd r }\left( r^2 {\dd \Phi ' \over \dd r }\right) =
 4 \pi G \left( {\pres' \over c^2} + {\rho \xi_r \over g }N^2 \right)
+ {l ( l + 1) \over r^2} \Phi ' \; .
\eel{E4.57}
Here 
\be
c^2 = {\Gamma_1 \pres \over \rho}
\eel{E4.57q}
is the squared adiabatic sound speed,
and I have introduced the characteristic frequencies
$S_l$ and $N$ (the so-called Lamb and buoyancy frequencies),
defined by
\be
S_l^2 = {l ( l +1) c^2 \over r^2} \simeq \kh^2 c^2 \; ,
\eel{E4.53}
and
\be
N^2 = g \left( {1 \over \Gamma_1 p} {\dd \pres \over \dd r }
- {1 \over \rho }{\dd \rho \over \dd r }\right) \; .
\eel{E4.56}
%The physical meaning of these frequencies will become clearer
%in Section~\ref{sec:asymp}, below.

The equations must be combined with boundary conditions: two of these ensure
regularity at the center, $r = 0$, which is a regular singular point
of the equations.
One condition enforces continuity of $\Phi'$ and its gradient at the
surface, $r = R$.
Finally, the surface pressure perturbation must satisfy
a dynamical condition.
In its most simple form it imposes zero pressure perturbation
on the perturbed surface, {\ie},
\be
\delta \pres = 0 \qquad {\rm at} \qquad r = R \; .
\eel{surf-p}

The fourth-order system of differential equations, Eqs (\ref{E4.54}) --
(\ref{E4.57}), and the boundary conditions
define an eigenvalue problem which has solutions only for
selected discrete values of $\omega$. 
Thus for each $(l, m)$ we obtain a set of eigenfrequencies
$\omega_{nlm}$, distinguished by their radial order $n$.

It should be noticed that in the present case of a spherically 
symmetric star the frequencies are degenerate in azimuthal order:
the definition of $m$ is tied to the orientation of the coordinate
system which, for a spherically symmetric star, can have no
physical significance.
Indeed, the equations and boundary conditions do not depend on $m$.
Thus, in analyzing the effects of the spherically symmetric 
structure of the Sun, the frequencies are characterized solely by $l$ and $n$;
the relation between structure and these {\it multiplet frequencies}
$\omega_{nl}$ is discussed in Sections~\ref{sec:oscilprop} -- \ref{sec:varprin}.
As discussed in Section~\ref{sec:rotsplit}, 
the degeneracy in $m$ is lifted by rotation.

\subsection{Properties of oscillations} \plabel{sec:oscilprop}

From the point of view of helio- and asteroseismic investigations,
it is important to realize which aspects of stellar structure
are accessible to study, in the sense of having a direct effect
on the oscillation frequencies.
Within the adiabatic approximation it follows
from Eqs (\ref{E4.54}) -- (\ref{E4.57}) that the frequencies
are completely determined by specifying $p$, $\rho$, $g$ and $\Gamma_1$
as functions of the distance $r$ to the center.
However, assuming that the equations of stellar structure
are satisfied, $p$, $g$ and $\rho$ are related by 
Eqs (\ref{eqn:hydrost}), (\ref{eqn:mass}) and (\ref{eqn:eqgrav}).
Thus specifying just $\rho(r)$ and $\Gamma_1(r)$, say,
completely determines the adiabatic oscillation frequencies.
Conversely, the observed frequencies only provide direct
information about these `mechanical' quantities.
To constrain other properties of the stellar interior,
additional information has to be included, 
such as the equation of state or
Eqs (\ref{eqn:tgrad}) and (\ref{eqn:energy})
determining the temperature gradient and luminosity
({\eg}, Gough and Kosovichev, 1990).
It is evident that the inferences obtained in such investigations
may suffer from uncertainties in, for example, the assumed physics.

The observed solar oscillations are in most cases predominantly of
acoustic nature, and hence there frequencies are most sensitive to
sound speed. 
To interpret helioseismic inferences of sound speed in terms of 
quantities more directly related to the properties of solar models,
it is instructive to note that 
equation of state of stellar interiors is reasonably well approximated by
that of a perfect, fully ionized gas,
according to which $\pres = \kB \rho T / (\mu m_{\rm u})$;
here $\kB$ is Boltzman's constant,
$m_{\rm u}$ is the atomic mass unit,
and $\mu$ is the so-called mean molecular weight, related to the
abundances $X$ and $Z$ of hydrogen and heavy elements by
$\mu \simeq 4 /( 3 + 5 X - Z)$.
In this approximation, also, $\Gamma_1 = 5/3$.
Thus
\be
c^2 \simeq {\Gamma_1 \kB T \over \mu m_{\rm u}} \; ,
\eel{approx-sound}
{\ie}, the sound speed is essentially determined by $T/\mu$.
To obtain separate estimates of $T$ and $\mu$, additional constraints
on the model are required.

The near-surface layers of the Sun present special problems which have
so far not been resolved.
Modelling of the structure of these layers is complicated by the presence
of convective motions with Mach numbers approaching 0.5,
in the uppermost few hundred km of the convection zone.
Results of detailed three-dimensional and time-dependent
hydrodynamical simulations
have been incorporated in a solar model 
used to compute oscillation frequencies, resulting in
some improvement in the agreement with the observed frequencies
(Rosenthal {\etal}, 1999);%
\footnote{Similar results were obtained by Li {\etal}\ (2002) using
an extension of mixing-length theory calibrated against numerical simulations.}
however, in general simple prescriptions,
which are certainly inadequate,
are used for the treatment of convection in this region.
The adiabatic approximation used in most computations of solar oscillation
frequencies is not valid near the surface.
Even in the cases where nonadiabatic calculations have been carried out
({\eg}, Guzik and Cox, 1991; Guenther, 1994),
these suffer from neglect, or inadequate treatment, of the perturbations
to the convective flux; 
furthermore, the perturbation to the turbulent pressure is usually ignored.
These potential problems with the models must be kept in mind
when the observed and computed frequencies are compared.
However, it is important to note that they are in all cases confined
to a very thin region near the solar surface.

\begin{figure}[htbp]
\begin{center}
\inclfig{8.0cm}{\fig/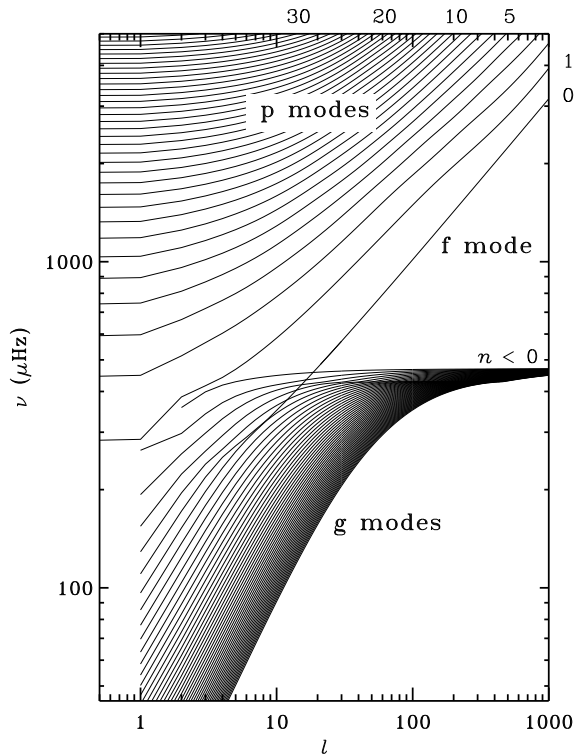}{Numerical frequencies for solar model}
\end{center}
%\vskip 0.5cm
\caption{
\figlab{fig:freqsun}
Cyclic frequencies $\nu = \omega / 2 \pi$, as
functions of degree $l$, computed for a normal solar model.
Selected values of the radial order $n$ have been indicated.
}
\end{figure}

Figure~\ref{fig:freqsun}
illustrates adiabatic oscillation frequencies computed for a solar model.
For clarity modes of given radial order $n$ have been connected.
With a few unconfirmed exceptions (see Section~\ref{sec:obsgmodes})
the observed solar oscillations have frequencies in excess of 500 $\muHz$
({\eg}, Schou, 1998a; Bertello {\etal}, 2000; 
Finsterle and Fr\"ohlich, 2001; Garc\'{\i}a {\etal}, 2001),
and hence correspond to the modes labelled `p modes' and,
at relatively high degree `f modes'.
As discussed in more detail in the following section, the former
are standing acoustic waves, whereas the latter behave essentially as
surface gravity waves.
The modes labelled `g modes' are internal gravity waves.
As indicated, it is conventional to assign positive and negative
radial orders $n$ to p and g modes, respectively, with $n = 0$ for f modes.
With this definition, frequency is an increasing function of $n$ for given $l$;
also, in most cases $|n|$ corresponds to the number of radial nodes in
the radial component of the displacement,
excluding a possible node at the center.

In Figure~\ref{fig:freqsun} it appears that the f-mode curve crosses
the g-mode curves;
in fact, if $l$ is regarded as a continuous variable,%
\footnote{This is clearly mathematically permissible, although only
the integral values of $l$ have a physical meaning.}
it is found that the interaction takes place through {\it avoided crossings}
where the frequencies approach very closely without actually crossing
({\eg} Christensen-Dalsgaard 1980).
This type of behavior is commonly seen for stellar oscillation frequencies,
as a parameter characterizing the solution is varied ({\eg} Osaki 1975). 
It is also well-known in, for example, atomic physics;
an early and very clear discussion of the behavior of eigenvalues
in the vicinity of an avoided crossing was given by
von Neuman \& Wigner (1929).

\subsection{Asymptotic behavior of stellar oscillations} \plabel{sec:asymp}

Although it is relatively straightforward to solve the equations of
adiabatic stellar oscillation,
approximate techniques play a major role in the interpretation of
observations of solar and stellar oscillations.
They provide insight into the relation between the observations and
the properties of the stellar interiors, which can inspire more 
precise analyses.
Also, since the observed solar modes are in many cases of high order,
asymptotic expressions are sufficiently precise to provide useful
quantitative results.

\subsubsection{Properties of acoustic modes}

Most of the modes observed in the Sun are essentially acoustic modes,
often of relatively high radial order.
In this case an asymptotic description can be obtained
very simply, by approximating the modes locally by plane sound waves,
satisfying the dispersion relation
$$
\omega^2 = c^2 |\boldk|^2 \; ,
$$
where $\boldk = k_r \bolda_r + \boldkh$ is the wave vector.
Thus the properties of the modes are entirely controlled by the
variation of the adiabatic sound speed $c(r)$.
To describe the radial variation of the mode,
we 
%separate $\boldk$ into radial and horizontal components
%$k_r \bolda_r$ and $\boldkh$ and 
use \Eq{hor-wave} to obtain
\be
k_r^2 = {\omega^2 \over c^2} - {L^2 \over r^2} 
= {\omega^2 \over c^2} \left(1 - {S_l^2 \over \omega^2} \right) \; .
\eel{vert-wave}
%where again $L = \sqrt{l(l+1)}$
%and $S_l$ is defined in \Eq{E4.53};
%thus this relation provides the physical meaning of $S_l$.
This equation can be interpreted very simply in geometrical
terms through the behavior of rays of sound,
as illustrated in Fig.~\ref{fig:rays}.
With increasing depth beneath the surface of a star temperature,
and hence sound speed, increases.
As a result, waves that are not propagating vertically
are refracted, as indicated in \Eq{vert-wave} by the
decrease in $k_r$ with increasing $c$; 
the horizontal component $|\boldkh|$ of the wave vector, in contrast,
increases with decreasing $r$.
Thus the rays bend, as shown in Fig.~\ref{fig:rays}.
The waves travel horizontally at the lower turning point, 
$r = \rt$, where $\omega = S_l$ and hence $k_r = 0$, {\ie},
\be
{c(\rt) \over \rt} = {\omega \over L} \; .
\eel{turn-point}
For $r < \rt$, $k_r$ is imaginary and the wave decays exponentially.

\begin{figure}[htbp]
\begin{center}
\inclfig{4.0cm}{\fig/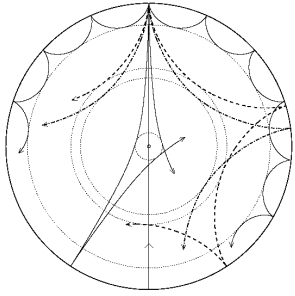}{Wave propagation}
\end{center}
\caption{
\figlab{fig:rays}
Propagation of rays of sound in a cross-section of the
solar interior. The ray paths are bent by the increase in sound speed
with depth until they reach {\em the inner turning point}
(indicated by the dotted circles) where they undergo total internal
refraction.
At the surface the waves are reflected by the rapid decrease in density.
}
\end{figure}

%This ray description illustrates the behavior of acoustic waves
%propagating through the star.
The normal modes observed as global oscillations on the stellar surface
arise through interference between waves propagating in this manner.
In particular, they share with the waves the total internal
reflection at $r = \rt$.
It follows from \Eq{turn-point} that the lower turning point is
located the closer to the center, the lower is the degree 
or the higher is the frequency.
Radial modes, with $l = 0$, penetrate the center,
whereas the modes of highest degree observed in the Sun, with $l \gwig 1000$,
are trapped in the outer small fraction of a per cent of the solar radius.
Thus the oscillation frequencies of different modes
reflect very different parts of the Sun;
it is largely this variation in sensitivity which allows the
detailed inversion for the properties of the solar interior
as a function of position (see also Sections~\ref{sec:solstruc}
and \ref{sec:infrot}).

Equation (\ref{vert-wave}) can be used to justify an approximate,
but extremely useful, expression for the frequencies of acoustic oscillation.
The requirement of a standing wave in the radial direction implies
that the integral of $k_r$ over the region of propagation, between
$r = \rt$ and $R$, must be an integral multiple of $\pi$, apart
from possible effects of phase changes at the end-points of the interval:
\be
(n + \alpha) \pi \simeq \int_{\rt}^R k_r \dd r
\simeq \int_{\rt}^R {\omega \over c} 
\left(1 - {S_l^2 \over \omega^2} \right)^{1/2} \dd r \; ,
\eenl{eqn:ppropag}
where $\alpha$ contains the phase changes at the points of reflection.
This may also be written as
\be
{\pi ( n +\alpha ) \over \omega } \simeq
F \left( {\omega \over L} \right) \; ,
\eel{E7.80}
where
\be
F(w) = \int_{\rt}^R \left( 1 - {c^2 \over w^2 r^2 } \right)^{1/2}
{\dd r \over c } \; .
\eel{E7.81}
That the observed frequencies of solar oscillation satisfy 
the simple functional relation given by \Eq{E7.80}
was first found by Duvall (1982);
this relation is therefore commonly known as the {\it Duvall law}.

\subsubsection{A proper asymptotic treatment}

Although instructive, this derivation is hardly satisfactory,
in either a mathematical or physical sense.
It ignores the fact that the oscillations are not purely acoustic of
nature, and neglects effects of variations of stellar structure
with position.
Also, effects near the stellar surface leading to reflection of the
waves are simply postulated.

A more satisfactory description can be based on asymptotic analyses of
the oscillation equations, Eqs (\ref{E4.54}) -- (\ref{E4.57}).
The modes observed in the Sun
are either of high radial order or high degree.
In such cases it is often possible, in approximate analyses, to make 
the so-called {\it Cowling approximation},
where the perturbation $\Phi'$ to the gravitational potential
is neglected (Cowling, 1941).
This can be justified, at least partly,%
\footnote{The validity of this argument under all circumstances
is not entirely obvious, however; see Christensen-Dalsgaard and Gough (2001).}
by noting that for modes
of high order or high degree, and hence varying rapidly as
a function of position, the contributions from regions where
$\rho'$ have opposite sign largely cancel in $\Phi'$.
%in the solution to Poisson's equation, \Eq{E3.41}.
In this approximation, the order of the equations is reduced to two, 
making them amenable to standard asymptotic techniques
({\eg}, Ledoux, 1962; Vandakurov, 1967; Smeyers, 1968).
A convenient formulation has been derived by Gough 
(see Deubner and Gough, 1984; Gough, 1993):
in terms of the quantity
\be
\Psi = c^2 \rho^{1/2} \diverge \bfdelr \; ,
\eel{E7.60}
the oscillation equations can be approximated by
\be
{\dd^2 \Psi  \over \dd r^2} = - K(r) \Psi \; ,
\eel{E7.75}
where
\be
K(r) = {\omega^2  \over c^2}
\left[ 1 - {\omegac^2  \over \omega^2}
- {S_l^2  \over \omega^2} \left( 1 - {N^2  \over \omega^2} \right) \right]
\; .
\eel{E7.76}
Here $N^2$ and $S_l^2$ were defined in Eqs (\ref{E4.53}) and (\ref{E4.56}),
and {\it the acoustical cut-off frequency} $\omegac$ is given by
\be
\omegac^2
= {c^2  \over 4 H^2} \left( 1 - 2 {\dd H \over \dd r }\right) \; ,
\eel{E7.73}
where $H = -( \dd \ln \rho / \dd r)^{-1}$ is the density scale height.

In addition to the modes determined by \Eq{E7.75},
there are modes for which
$\diverge \bfdelr \simeq 0$; these modes clearly cannot
be analyzed in terms of $\Psi$.
They approximately correspond to surface gravity waves,
with frequencies satisfying
\be
\omega^2 \simeq g \kh \; ,
\eel{E7.67}
and are usually known as {\it f modes}.
I return to them in Section~\ref{sec:fgmodes}.

\begin{figure}[htbp]
\begin{center}
\inclfig{8.0cm}{\fig/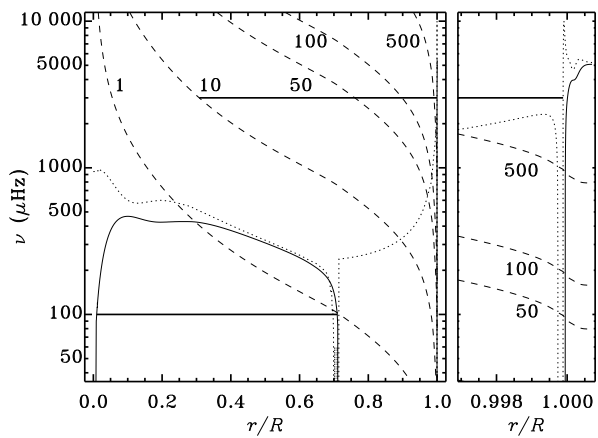}{Characteristic frequencies}
\end{center}
\vskip 0.5cm
\caption{
\figlab{fig:charsun}
Characteristic frequencies $N/2 \pi$ (solid line),
$\omegac/2 \pi$ (dotted line) and $S_l/2 \pi$ (dashed lines,
labelled by $l$)
for $l = 1$, 10, 50, 100 and 500.
The frequencies have been computed for Model S 
of Christensen-Dalsgaard {\etal}\ (1996).
The heavy horizontal lines mark the trapping regions of
a g mode of frequency $100 \muHz$ and a p mode of frequency $3000 \muHz$
and degree $l = 10$.
}
\end{figure}

The physical meaning of \Eq{E7.75} becomes clear if we
make the identification $K = k_r^2$ where, as before, $k_r$
is the radial component of the local wave number.
Accordingly, a mode 
oscillates as a function of $r$ in regions where $K > 0$;
such regions are referred to as regions of propagation.
The mode is evanescent, decreasing or increasing exponentially,
where $K < 0$.
The detailed behavior of the mode is thus controlled by 
the value of the frequency, relative to the characteristic
frequencies $S_l$, $N$ and $\omegac$.
%Note in particular that if $\omega^2 \gg \omegac^2, N^2$
%we approximately recover \Eq{vert-wave} for $k_r^2$;
%this is the limit in which the approximation of the mode
%by plane sound waves is valid.
%More generally, the points where $K = 0$, marking the transition between
%the oscillatory and evanescent behavior, are called turning points.

Figure~\ref{fig:charsun}
illustrates the characteristic frequencies in a model of the present Sun.
It is evident that $\omegac$ is large only near the stellar surface,
where the density scale height is small.
In the range of observed solar oscillations the frequencies are
higher than the buoyancy frequency; 
thus, roughly speaking, modes have
an oscillatory behavior where $\omega > S_l$ and $\omega > \omegac$.
Another type of propagation occurs at low frequency, in a region where
$\omega < N$.
Examples of propagation regions corresponding to these two
cases are marked in Fig.~\ref{fig:charsun}.
Modes corresponding to the former case are called {\it p modes};
it follows from the analysis given above that they are 
essentially standing sound waves, where the dominant restoring
force is pressure.
Modes corresponding to the latter cases are called {\it g modes};
here the dominant restoring force is buoyancy, and the modes have
the character of standing internal gravity waves.

Equations (\ref{E7.75}) and (\ref{E7.76}) are in a form well suited
for JWKB analysis.%
\footnote{For Jeffreys, Wentzel, Kramers and Brillouin, who were amongst
the first to use such techniques.
Applications to quantum mechanics, were discussed, for example,
by Schiff (1949).}
The result is that the modes satisfy
\be
\omega \int_{r_1}^{r_2} \left[ 1 -
{\omega_{\rm c}^2  \over \omega^2}
- {S_l^2  \over \omega^2}
\left( 1 - {N^2  \over \omega^2} \right) \right]^{1/2} {\dd r \over c}
\simeq \pi (n - 1/2) \; ,
\eel{E7.79}
%$$
where $r_1$ and $r_2$ are adjacent zeros of $K$ such that
$K > 0$ between them.
%This expression is valid in general, for both p modes and g modes.
%However, substantial simplifications can be achieved in each of
%the two cases, as discussed below.

\subsubsection{Asymptotic properties of p modes} \label{sec:pasymp}

For the p modes, we may approximately neglect the term in $N$
and, except near the surface, the term in $\omegac$.
Thus we recover \Eq{vert-wave};
in particular, the location of the lower turning point 
is approximately given by \Eq{turn-point}.
Near the surface, on the other hand, $S_l \ll \omega$ 
for small or moderate $l$ and may be neglected
({\cf}\ Fig.~\ref{fig:charsun});
thus the location $r = R_{\rm t}$ of the
upper turning point is determined by $\omega \simeq \omegac$.
Physically, this corresponds to the reflection of the waves 
where the wavelength becomes comparable to the local density scale height.
It should also be noticed from Fig.~\ref{fig:charsun}
that $\omegac$ approximately tends to a constant in the stellar atmosphere.
%Although $\omegac$ displays a sharp peak
%(associated with the region of strong superadiabaticity
%near the top of the convection zone)
%rising substantially higher than this limiting atmospheric value,
%the trapping of the acoustic modes is dominated by the atmosphere.
%It follows that 
Modes with frequencies exceeding the atmospheric value of $\omegac$
are only partially trapped,
losing energy in the form of running waves in the solar atmosphere;
hence they may be expected to be rather strongly damped.

If we assume that $|N^2/\omega^2| \ll 1$, \Eq{E7.79} simplifies to
\be
\omega \int_{r_1}^{r_2} \left( 1 -
{\omega_{\rm c}^2  \over \omega^2}
- {S_l^2  \over \omega^2} \right)^{1/2} {\dd r \over c}
\simeq \pi (n - 1/2) \; ,
\eel{E7.28}
where, as discussed above, $r_1 \simeq \rt$ and $r_2 \simeq R_{\rm t}$.
Further simplification results by noting that since $\omegac/\omega \ll 1$,
except near the upper turning point,
the integral may be expanded, yielding
\be
\omega \int_{\rt}^R \left( 1 -
{S_l^2  \over \omega^2} \right)^{1/2} {\dd r \over c}
\simeq \pi [n + \alpha (\omega) ] \; 
\eel{E7.28q}
({\eg}, Christensen-Dalsgaard and P\'erez Hern\'andez, 1992).
Here we again assumed that $S_l \ll \omega$ near the upper turning point;
consequently $\alpha$ depends only on frequency and
results from the expansion of the near-surface behavior of $\omegac$.
Thus we recover Eqs (\ref{E7.80}) and (\ref{E7.81}),
previously obtained from a simple analysis of sound waves.
From a physical point of view, the assumption on $S_l$
ensures that the waves travel nearly vertically near the surface;
thus their behavior is independent of their horizontal structure,
leading to a phase shift depending solely on frequency.

For low-degree modes these relations may be simplified even further,
by noting that in the integrand in \Eq{E7.81} 
$( \ldots )^{1/2}$ differs
from unity only close to the lower turning point which, for
these modes, is situated very close to the center.
As a result it is possible to expand the integral to obtain,
to lowest order,
that $F(w) \simeq \int_0^R \dd r / c - w^{-1} \pi / 2 $.
%thus \Eq{E7.80} may be approximated by 
%\be
%\omega  = {(n  + L/2 + \alpha ) \pi  \over
%\int_0^R {\dd r / c} } \; .
%\eel{E7.40}
Furthermore, a more careful analysis shows that for low-degree modes
$L$ should be replaced by%
\footnote{Note that, in any case,
except at the lowest degrees this is an excellent approximation
to the original definition of $L$;
thus in the asymptotic discussions
I shall use the two definitions interchangeably.}
$l + 1/2$ ({\eg}, Vandakurov, 1967; Tassoul, 1980).
Thus from \Eq{E7.80} we obtain
\be
\nu_{nl} \equiv {\omega_{nl}  \over 2 \pi}
 \simeq \left( n + {l \over 2 }+ {1 \over 4 }+ \alpha \right) \Delta \nu \; ,
\eel{E7.41}
where
$\Delta \nu = [ 2 \int_0^R \dd r / c ]^{-1}$
is the inverse of twice the sound travel time between the
center and the surface.
This equation predicts a uniform spacing $\Delta \nu$
in $n$ of the frequencies of low-degree modes.
Also, modes with the
same value of $n + l /2$ should be almost degenerate,
$\nu_{nl}  \simeq \nu_{n-1 \,l +2}$.
This frequency pattern was first observed for the solar
five-minute modes of low degree by Claverie {\etal}\ (1979)
and may be used in the search for stellar oscillations of solar type.

The {\it deviations} from the simple relation (\ref{E7.41}) have
considerable diagnostic potential.
By extending the expansion of \Eq{E7.81}, leading to \Eq{E7.41},
to take into account the variation of $c$ in the core one finds 
%a departure from the approximate
%frequency coincidence obtained in \Eq{E7.43}
(Gough, 1986; see also Tassoul, 1980)
\be
d_{nl} \equiv \nu_{n l} - \nu_{n-1\,l + 2}
 \simeq - ( 4 l + 6 )
 {\Delta \nu  \over 4 \pi^2 \nu_{nl}}
 \int_0^R {\dd c  \over \dd r }{\dd r \over r } \; ;
\eel{E7.45}
here the integral is predominantly weighted towards the center of
the star, as a result of the factor $r^{-1}$ in the integrand.
This behavior
provides an important diagnostic of the structure of stellar cores.
In particular, we note that, according to \Eq{approx-sound},
the core sound speed is reduced as $\mu$ increases with the conversion
of hydrogen to helium as the star ages.
As a result, $d_{nl}$ is reduced, thus providing a measure of the 
evolutionary state of the star
({\eg}, Christensen-Dalsgaard, 1984, 1988; Ulrich, 1986;
Gough and Novotny, 1990; see also Gough, 2001a).

It is interesting to investigate the effects on 
the frequencies of small changes to the model.
Such frequency changes may be estimated quite simply by
linearizing the Duvall law in differences
$\delta\omega_{nl}$ in $\omega_{nl}$,
$\delta_r c(r)$ in $c(r)$ and
$\delta\alpha (\omega )$ in $\alpha(\omega )$.
The result can be written
(Christensen-Dalsgaard {\etal}, 1988)
\be
S_{nl}{{\delta\omega_{nl}}\over{\omega_{nl}}}
\simeq \Hone\left({{\omega_{nl} } \over L}\right) +\Htwo(\omega_{nl} ) \; ,
\eel{E7.103}
where
\be
S_{nl}=\int_{\rt}^R
\left( 1-{{L^2c^2}\over{r^2\omega_{nl}^2}}\right)^{-1/2}
{{\dd r}\over c} -\pi {{\dd\alpha}\over{\dd\omega}} \; ,
\eel{E7.102}
\be
\Hone({w})=\int_{\rt}^R \left( 1-{{c^2}\over{r^2{w}^2}}
\right)^{-1/2}{{\delta_r c}\over c}{{\dd r}\over c} \; ,
\eel{E7.104}
%$$
and
\be
\Htwo(\omega )={{\pi}\over{\omega}}\delta\alpha (\omega) \; .
\eel{E7.105}
Christensen-Dalsgaard, Gough, and Thompson (1989)
noted that $\Hone(\omega/ L)$
and $\Htwo(\omega )$ can be obtained separately, to within a constant,
by means of a double-spline fit of the expression (\ref{E7.103})
to p-mode frequency differences.
The dependence of $\Hone$ on $\omega/ L$ is determined by
the sound-speed difference throughout the star;
in fact, it is straightforward to verify that the contribution
from $\Hone$ is essentially just an average of $\delta_r c/c$,
weighted by the sound-travel time along the rays characterizing the mode.
The contribution from
$\Htwo(\omega )$ depends on differences in the upper layers of the models.
Thus, in particular, it contains the effects of the near-surface
errors discussed in Section~\ref{sec:oscilprop}.

The preceding, relatively simple, asymptotic analysis has been
improved in several investigations.
For modes of high degree the expansion leading to a frequency-dependent
phase function $\alpha(\omega)$ in \Eq{E7.28q}
is no longer valid;
Brodsky and Vorontsov (1993) showed how the analysis could be
generalized to obtain the $l$-dependence of $\alpha$.
For modes of low degree or relatively low frequency the
perturbation to the gravitational potential can no longer
be ignored, and it may furthermore be necessary to include
the effect of the buoyancy frequency in the asymptotic 
dispersion relation
({\eg}, Vorontsov, 1989, 1991; Gough, 1993).
Finally, the usual asymptotic expansion, as used for example to
obtain \Eq{E7.45}, is somewhat questionable in the core of the
star where conditions vary on a scale comparable with the
wavelengths of the modes;
here other formulations may be more appropriate
({\eg}, Roxburgh and Vorontsov, 1994a, 2000ab, 2001).
However, for the present review the simpler expressions are 
generally adequate.

\subsubsection{f and g modes} \plabel{sec:fgmodes}

In addition to p modes, the observations of solar oscillations
also show f modes of moderate and high degree.
As discussed above, these modes are approximately divergence-free,
with frequencies given by ({\cf}\ Eq.\ \ref{E7.67})
\be
\omega^2 \simeq g_{\rm s} \kh = {G M \over R^3} L \; ,
\eel{eqn:fmode}
where $g_{\rm s}$ is the surface gravity.
It may be shown that the displacement eigenfunction is approximately
exponential,
$\xi_r \propto \exp(\kh r)$,
as is the case for surface gravity waves in deep water.
According to \Eq{eqn:fmode} the frequencies of these modes
are independent of the internal structure of the star;
this allows the modes to be uniquely identified in the observed
spectra, regardless of possible model uncertainties.
A more careful analysis must take into account the fact that
gravity varies through the region over which the mode has
substantial amplitude;
this results in a weak dependence of the
frequencies on the density structure (Gough, 1993).
%; Chitre, Christensen-Dalsgaard \& Thompson, 1998).

I finally briefly consider the properties of g modes.
It follows from Fig.~\ref{fig:charsun} that these are trapped
in the radiative interior and behave exponentially in the convection zone.
In fact, they have their largest amplitude close to the solar center and
hence are potentially very interesting as probes of conditions in the
deep solar interior.
High-degree g modes are very effectively trapped by the exponential decay
in the convection zone and are therefore unlikely to be visible at the surface.
However, for low-degree modes the trapping is relatively inefficient,
and hence the modes might be expected to be observable, if they were
excited to reasonable amplitudes.
The behavior of the oscillation frequencies can be obtained from \Eq{E7.79}.
In the limit where $\omega \ll N$ in much of the radiative interior
this shows that the modes are uniformly spaced in oscillation
{\it period}, with a period spacing that depends on degree.

\subsection{Variational principle} \plabel{sec:varprin}

The formulation of the oscillation equations given in \Eq{E3.40x}
is the starting point for powerful analyses of general properties of
stellar pulsations.
For convenience, we write the equation as
\be
\omega^2 \bfdelr = \force(\bfdelr) \; ,
\eel{E5.01}
where the right-hand side is the linearized force per unit mass, which,
as discussed in Section~\ref{sec:linosc}, can be regarded
as a linear operator on $\bfdelr$.

The central result is that \Eq{E5.01}, applied to
adiabatic oscillations, defines a {\it variational principle}.
Specifically, by multiplying the equation by $\rho \bfdelr^*$
(`$*$' denoting the complex conjugate)
and integrating over the volume $V$ of the star,
we obtain
\be
\omega^2 = {\int_V \bfdelr^* \scalarprod \force(\bfdelr) \rho \dd V
\over
\int_V |\bfdelr|^2 \rho \dd V} \; .
\eel{E5.02}
We now consider adiabatic oscillations which
satisfy the surface boundary condition given by \Eq{surf-p}.
In this case it may be shown that the right-hand side
of \Eq{E5.02} is stationary with respect to small 
perturbations to the eigenfunction $\bfdelr$
({\eg}, Chandrasekhar, 1964).
%From a physical point of view, the assumptions ensure that
%the pulsating star is a conservative mechanical system;
%in particular, when $\delta \pres = 0$ there are no forces applied
%to the star from the outside.
%The stationarity then just reflects Hamilton's principle
%for a conservative system.

A very important application of this principle concerns
the effect on the frequencies of perturbations to the
equilibrium model or other aspects of the physics of the oscillations.
Such perturbations can in general be expressed as a perturbation
$\delta \force$ to the force in \Eq{E5.01}.
It follows from the variational principle that their
effect on the frequencies can be determined as
\be
\delta \omega^2 
= {\int_V \bfdelr^* \scalarprod \delta \force(\bfdelr) \rho \dd V
\over
\int_V |\bfdelr|^2 \rho \dd V} \; ,
\eel{E5.03}
evaluated using the eigenfunction $\bfdelr$ of the unperturbed force operator.
Applications of this expression to rather general situations
were considered by Lynden-Bell and Ostriker (1967).

Equation (\ref{E5.03}) provides the basis for determining
the relation between differences in structure and
differences in frequencies between the Sun and solar models.
%generalizing the asymptotic expressions in 
%Eqs (\ref{E7.102}) -- (\ref{E7.105}).
As discussed in Section~\ref{sec:oscilprop},
the oscillation frequencies are determined by a suitable
pair of model variables, {\eg},
the pair $(c^2, \rho)$, which reflects the acoustic nature
of the observed modes.
The differences between the structure of the Sun and a model can then
be characterized by the differences
$\deltar c^2/c^2 = [c_\odot^2 (r) - c_{\rm mod}^2(r)]/c^2(r)$
and $\deltar \rho/\rho = [\rho_\odot (r) - \rho_{\rm mod}(r)]/\rho(r)$.
In particular, the perturbation $\delta \force$ can be expressed
in terms of $\delta_r c^2 /c^2$ and $\delta_r \rho/\rho$,
through appropriate use of the linearized versions of 
Eqs (\ref{eqn:hydrost}) and (\ref{eqn:mass}) 
({\eg}, Gough and Thompson, 1991), resulting in a linear
relation for the frequency change in terms of the structure differences.

The analysis in terms of $\delta_r c^2/c^2$ and $\delta_r \rho/\rho$
only captures the differences between the Sun and the model to the
extent that they relate to the hydrostatic structure of the Sun.
As discussed in Section~\ref{sec:oscilprop}, inadequacies in the treatment
of the physics of the modes, such as non\-adiabatic effects,
contribute in the near-surface layers of the Sun.
These can also be represented as perturbations $\delta \force_{\rm surf}$,
such that $\delta\force_{\rm surf}(\bfdelr)$ is significant only in the
superficial layers.
For modes of low or moderate degree the eigenfunctions depend little
on degree in this region, as discussed in Section~\ref{sec:pasymp}.
Assuming that $\delta \force_{\rm surf}$ does not depend explicitly
on $l$, it follows that for these modes 
$\int_V \bfdelr^* \scalarprod \delta \force_{\rm surf}(\bfdelr) \rho \dd V$
depends little on $l$;
hence, according to \Eq{E5.03}, 
the effects of the near-surface problems may in general be
expected to be of the form
\be
\left( {\delta \omega_{nl} \over \omega_{nl}} \right)_{\rm surf}
\simeq I_{nl}^{-1} \Fsurf(\omega_{nl}) \; ,
\eel{near-surface}
where $I_{nl} = \int_V |\bfdelr|^2 \rho \dd V$,
the denominator in \Eq{E5.03}, is known as the mode inertia
and, as indicated, $\Fsurf$ depends only on frequency.
We note also that at relatively low frequency
the relevant superficial layers are outside the upper turning point
determined by $\omega = \omegac$ ({\cf}\ Fig.~\ref{fig:charsun})
and hence the modes are evanescent in this region.
Thus we expect the effects of the near-surface problems
to be small for low-frequency modes 
({\eg}, Christensen-Dalsgaard and Thompson, 1997).

%\begin{figure}[htbp]
%\begin{center}
%\leavevmode\epsfxsize=9cm\epsfbox{\fig/fig2-11.eps}
%\end{center}
%\caption{
%\figlab{fig:qnl}
%The inertia ratio $Q_{nl}$, defined in \Eq{def-q},
%against frequency $\nu$, for f and p modes in a normal solar model.
%Each curve corresponds to a given degree $l$,
%selected values of which are indicated.
%}
%\end{figure}

The mode inertia still depends on both degree and frequency:
in particular, modes of high degree and/or low frequency
are trapped closer to the the solar surface
({\cf}\ Eq.\ \ref{turn-point}), involve a smaller fraction of
the Sun's mass and hence have a smaller $I_{nl}$.
Thus high-degree modes are affected more strongly by the
near-surface errors than are low-degree modes at the same frequency.
To eliminate this essentially trivial effect, it is instructive
to consider frequency differences scaled by $I_{nl}$.
This may be done conveniently by scaling the frequencies by
\be
Q_{nl} \equiv {I_{nl} \over \bar I_0 (\omega_{nl})} \; ,
\eel{def-q}
where $\bar I_0 (\omega_{nl})$ is the inertia of a hypothetical radial mode
(with $l = 0$)
with frequency $\omega_{nl}$, obtained by interpolation to
that frequency in the inertias for the actual radial modes.
This effectively reduces the frequency shift to the effect on a
radial mode of the same frequency.
%The behavior of $Q_{nl}$ is illustrated in Fig.~\ref{fig:qnl}:
%evidently the variation in mode inertia with degree is quite
%substantial for high degrees.
Examples of scaled frequency differences will be shown later.

From the preceding analysis it finally follows that the
frequency differences between the Sun and the model,
assuming that the differences are so small that a linear representation
is adequate, can be written as
\bea
{\delta \omega_{nl} \over \omega_{nl}}  & = &
\int_0^R \left[ K_{c^2, \rho}^{nl}(r) {\delta_r c^2 \over c^2}(r)
+ K_{\rho,c^2}^{nl}(r) {\delta_r \rho \over \rho}(r) \right] \dd  r
\nonumber \\
& & + I_{nl}^{-1} \Fsurf(\omega_{nl})  \; ,
\eeal{E5.81}
where the kernels $K_{c^2, \rho}^{nl}$ and $K_{\rho,c^2}^{nl}$,
which result from manipulating $\delta\force$,
are computed from the eigenfunctions of the reference model
({\eg}, Dziembowski {\etal}, 1990;
D\"appen {\etal}, 1991; Gough \& Thompson, 1991).
This relation forms the basis for inversions of the oscillation frequencies
to determine solar structure (see Section~\ref{sec:solstruc}).

The similarity of \Eq{E5.81} to the asymptotic expressions
in Eqs (\ref{E7.102}) -- (\ref{E7.105}) should be noted.
In both cases the frequency differences are separated into contributions
from the bulk of the Sun (in the asymptotic case characterized solely by
the sound-speed difference) and from the near-surface layers, the latter
depending essentially only on frequency after appropriate scaling;
indeed, it may be shown that $S_{nl}$ and $I_{nl}$ are closely related.

\subsection{Effects of rotation} \plabel{sec:rotsplit}

So far, we have considered only oscillations of a spherically symmetric star;
in this case, the frequencies are independent of the azimuthal order $m$.
Departures from spherical symmetry lift this degeneracy,
causing a frequency splitting according to $m$.

The most obvious, and most important, such departure is rotation;
early studies of the effect of rotation were presented by
Cowling and Newing (1949) and Ledoux (1949, 1951).
A simple description can be obtained by first noting that,
according to Eqs (\ref{spher-harm}) and 
(\ref{displ}), the oscillations depend on longitude $\phi$
and time $t$ as $\cos(m \phi - \omega t)$, {\ie},
as a wave running around the equator.
We now consider a star rotating with angular velocity $\Omega$ and a
mode of oscillation with frequency $\omega_0$ in a frame rotating
with the star; 
the coordinate system is chosen with polar axis along the
axis of rotation.
Letting $\phi'$ denote longitude in this frame, the oscillation
therefore behaves as $\cos(m \phi' - \omega_0 t)$.
The longitude $\phi$ in an inertial frame is related to $\phi'$
by $\phi' = \phi - \Omega t$;
consequently, the oscillation as observed from the inertial frame
depends on $\phi$ and $t$ as
$$
\cos(m \phi - m \Omega t - \omega_0 t) \equiv \cos ( m \phi -\omega_m t) \; ,
$$
where $\omega_m = \omega_0 + m \Omega$.
Thus the frequencies are split according to $m$,
the separation between adjacent values of $m$ being simply
the angular velocity;
this is obviously just the result of the advection of the
wave pattern with rotation.

This simple description contains the dominant physical effect,
{\ie}, advection,
of rotation on the observed modes of oscillation, but it suffers from
two problems:
it assumes solid-body rotation, whereas the Sun rotates differentially;
and it neglects the effects, such as the Coriolis force,
in the rotating frame.
In a complete description in an inertial frame,
including terms linear in the angular velocity,%
\footnote{In the solar case the centrifugal force and other
effects of second or higher order in $\Omega$,
including the distortion of the equilibrium structure,
can be neglected to a good approximation.}
\Eq{E3.40x} must be replaced by
\be
    \omega^2  \bfdelr
      = {1 \over \rho} \vectordel \pres' - \boldg' - {\rho'\over \rho} \boldg 
      +2 m \omega \Omega \bfdelr - 2 i \omega \boldOmega \times \bfdelr \; ,
\eel{E5.11}
where $\boldOmega$ is the rotation vector, of magnitude 
$\Omega$ and aligned with the rotation axis.
The first term resulting from rotation is the 
contribution from advection, as discussed above,
whereas the last term is the Coriolis force.

The terms arising from rotation obviously correspond to a
perturbation to the force operator $\force$ in \Eq{E5.01};
from \Eq{E5.03} the effect on the oscillation frequencies can be
obtained on the form
\be
\omega_{nlm} = \omega_{nl0} + 
m \int_0^R \int_0^\pi K_{nlm}(r,\theta) \Omega(r, \theta) 
r \dd  r \dd \theta \; ,
\eel{rot-split}
where the kernels $K_{nlm}$ can be calculated from the eigenfunctions
for the non-rotating model.
The kernels depend only on $m^2$, so that the rotational splitting 
$\omega_{nlm} - \omega_{nl0}$ is an odd function of $m$.
Also, the kernels are symmetrical around the equator;
%(this follows immediately in the approximation where the kernels
%are determined by $\rho |\bfdelr|^2$);
as a result, the rotational splitting is only sensitive to the
component of $\Omega$ which is similarly symmetrical. 

\begin{figure}[htbp]
\begin{center}
\inclfig{8.6cm}{\fig/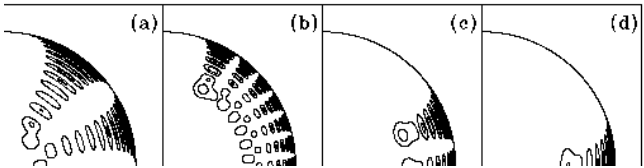}{Rotational kernels}
\end{center}
%\vskip -1cm
\caption{
\figlab{fig:rotker}
Contour plots of rotational kernels $K_{nlm}$ in a solar quadrant.
The modes all have frequencies near 2~mHz; the following pairs of
$(l, m)$ are included: a) $(5,2)$; b) $(20,8)$; c) $(20,17)$; and
d) $(20,20)$.
}
\end{figure}

The general expression for the rotational kernels is 
quite complicated and will not be given here
(see, for example, Hansen {\etal}, 1977; Cuypers, 1980; Gough, 1981).
Examples of kernels are shown in Fig.~\ref{fig:rotker}.
The extent in the radial direction is essentially determined by
the location of the lower turning point, $r  = \rt$ 
({\cf}\ Eq.\ \ref{turn-point}).
The latitudinal extent is determined by the properties of the
Legendre functions $P_l^m$;
it follows from their asymptotic behavior that the kernels
are confined between latitudes $\pm \cos^{-1}(|m|/L)$.
Thus, as also reflected in the behavior of the spherical harmonics
(Fig.~\ref{fig:ylm}) modes with low $|m|$ extend over essentially
all latitudes, whereas modes with $m \simeq \pm l$ are confined close
to the equator.

%\begin{figure}[htbp]
%\begin{center}
%\leavevmode\epsfxsize=11cm\epsfbox{\fig/fig1-10.eps}
%\end{center}
%\caption{
%\figlab{fig:rot-ker}
%Kernels $K_{n l }$ for the frequency splitting
%caused by spherically symmetric rotation ({\cf}\ eq. \ref{E8.38}),
%for a model of the present Sun.
%In a) is plotted $R K_{n l} (r)$ for a mode with
%$l = 1$, $n = 22$ and $\nu = 3239 \muHz$.
%The maximum value of $R K_{n l} (r)$ is 62.
%In b) is shown the same mode, on an expanded vertical scale,
%(continuous line)
%together with the modes
%$l = 20$, $n = 17$, $\nu = 3375 \muHz$ (short-dashed line), and
%$l = 60$, $n = 10$, $\nu = 3234 \muHz$ (long-dashed line).
%Notice that the kernels almost vanish inside
%the turning-point radius $\rt$,
%and that there is an accumulation just outside the turning point.
%}
%\end{figure}

If $\Omega = \Omega(r)$
is assumed to be a function of $r$ alone,
the corresponding kernels do not depend on $m$,
so that \Eq{rot-split} predicts a uniform frequency splitting in $m$.
This is often written on the form
\be
\delta \omega_{nlm}  \equiv 
\omega_{nlm} - \omega_{nl0} =
m \beta_{nl} \int_0^R K_{nl} (r) 
\Omega (r) \dd r \; ,
\eel{E8.37}
%$$
where
$K_{nl}$ is unimodular, {\ie}, $\int K_{nl} (r) \dd r = 1$.

For stars rotating substantially more rapidly than the Sun terms
of higher order in $\Omega$ must be taken into account.
Terms quadratic in $\Omega$, such as the centrifugal distortion,
give rise to frequency perturbations that are even functions of $m$,
also changing the mean frequency of the multiplet
({\eg}, Gough and Thompson, 1990),
while cubic terms may be important in cases of modes closely
spaced in frequency, such as result from the asymptotic behavior
of low-degree p modes ({\cf} Section~\ref{sec:pasymp}).
A detailed discussion of these effects was given by
Soufi, Goupil \& Dziembowski (1998);
they can give rise to complex oscillation spectra, considerably
complicating mode identification for rapidly rotating stars.

\subsection{The causes of solar oscillations} \plabel{sec:solcause}

Given the assumption of adiabatic oscillations, no information is
obtained about the possible damping or driving of the modes:
the equations are conservative and do not involve any energy
exchange between the oscillations and the flow of energy 
in the equilibrium model.
Calculations taking into account nonadiabatic effects
%starting with Baker and Kippenhahn (1962), 
have investigated the linear stability
of stellar oscillations; this is determined by the imaginary part
$\omegai$ of the complex frequency $\omega$, modes with positive
$\omegai$ being unstable.
It is found that many types of stars, for example the classical Cepheids,
have unstable modes; 
the instability results from favourable phase relations between the compression
and the perturbation to the heat flux in the oscillations,
often caused by suitable variations in the opacity.

Early nonadiabatic calculations of solar oscillations 
({\eg}, Ando and Osaki, 1975)
found that modes in the observed range of frequencies were in fact unstable.
These calculations, however, used a simplified treatment of
radiative transfer in the outer layers of the Sun and, more importantly,
neglected effects of convection.
Balmforth (1992a) carried out nonadiabatic calculations of solar
oscillations, including convective effects;
these were described by expressions, based on mixing-length theory,
for the perturbations induced
by stellar pulsation to the convective flux and turbulent stresses,
developed from an original formulation of Gough (1977a).
He found that all the modes were damped,
an important contribution to the damping coming from the perturbation
to the turbulent pressure.%
\footnote{It should be noted, however, that Xiong {\etal}\ (2000)
found some solar modes to be unstable, using a different formulation for
the convective effects.}
%Independent evidence against instability of the solar oscillations was
%obtained by Kumar and Goldreich (1989) who showed that 
%amplitude limitation through nonlinear interactions
%amongst the solar modes, had they been unstable,
%would be unlikely to reproduce the observed amplitude distribution.

This motivates a search for driving mechanisms external to the
oscillations, the most natural source being the vigorous
convection near the solar surface, where motion at near-sonic speed
may be expected to be a strong source of acoustic waves (Lighthill, 1952).
Stein (1967)
applied this to the interpretation of the solar five-minute oscillations.
An early estimate of the expected amplitude of global modes excited
by this mechanism was made by Goldreich and Keeley (1977).

Since each mode feels the effect of a very large number of 
turbulent eddies, acting at random, the combined effect is
that of a stochastic forcing of the mode.
To illustrate the properties of the resulting oscillations
I consider a very simple model of this process
(Batchelor, 1956; see also Christensen-Dalsgaard, Gough, and Libbrecht, 1989),
consisting of a simple damped oscillator 
of amplitude $A(t)$, forced by a random function $f(t)$,
and hence satisfying the equation %
%\footnote{An equation of essentially this form may in fact be obtained
%from the full oscillation equations, including the convective forcing,
%by projecting onto the eigenmodes.}
\be
{\dd^2 A \over \dd t^2} + 2 \eta {\dd A \over \dd t} + \omega_0^2 A = f(t) \; ;
\eel{EC.73}
here $\eta$ is the linear damping rate, $\eta = - \omegai$.
This equation is most easily dealt with in terms of its
Fourier transform.
Introducing the Fourier transforms $\tilde A(\omega)$
and $\tilde f(\omega)$ by
$ \tilde A(\omega) = \int A(t) e^{i \omega t} \dd t$,
$\tilde f(\omega) = \int f(t) e^{i \omega t} \dd t$, we obtain from
\Eq{EC.73} 
\be
-\omega^2 \tilde A - 2 i \eta \omega \tilde A
+ \omega_0^2 \tilde A = \tilde f \; .
\eel{EC.75}
This yields the power spectrum of the oscillator as
\be
P(\omega) = |\tilde A(\omega)|^2
= { |\tilde f(\omega)|^2 \over
(\omega_0^2 - \omega^2)^2 + 4 \eta^2 \omega^2} \; .
\eel{EC.76}
Near the peak in the spectrum,
where $|\omega - \omega_0| \ll \omega_0$
the average power of the oscillation is therefore given by
\be
\langle P(\omega) \rangle \simeq
{1 \over 4 \omega_0^2} { P_f(\omega) \over
(\omega - \omega_0)^2 + \eta^2} \; ,
\eel{EC.77}
where $ P_f(\omega) = \langle |\tilde f(\omega) |^2 \rangle $
is the average power of the forcing function.

Since $ P_f(\omega)$ is often a slowly
varying function of frequency, the frequency dependence
of $\langle P(\omega) \rangle$ is dominated by
the denominator in \Eq{EC.77}.
The resulting profile is therefore approximately {\it Lorentzian},
with a width determined by the linear damping rate $\eta$.
Consequently, under the assumption of stochastic
excitation one can make a meaningful comparison
between computed damping rates and observed line widths.

It should be noted that this model makes definite predictions about
the oscillation amplitudes, from the power available in the forcing function.
This depends on the details of the interaction between
convection and the oscillations, with contributions both from
Reynolds stresses and entropy fluctuations generated by
convection ({\eg}, Goldreich and Kumar, 1990;
Balmforth, 1992b;
Goldreich {\etal}, 1994; Samadi {\etal}, 2001;
Stein and Nordlund, 2001).
The excitation varies strongly with frequency as a result both of the
structure of the eigenfunction and the temporal spectrum of convection,
hence accounting for the frequency dependence of the mode amplitudes.
However, since the horizontal scale of convection near the solar surface
is much smaller than the horizontal wavelength of the oscillations,
the interaction is likely to depend little on the degree $l$ of the modes;
thus, as is indeed observed, we expect excitation of modes at all degrees
within the relevant frequency range,
with amplitudes that depend relatively little on degree
except at high degree ({\eg}, Christensen-Dalsgaard and Gough, 1982; 
Woodard {\etal}, 2001).

The observed line profiles show significant departures from 
the Lorentzian shape, in the form of asymmetries.
These can be understood from more complete models of the
excitation, taking into account that the dominant contributions
to the forcing are spatially localized to relatively thin
regions beneath the solar surface
({\eg}, Duvall {\etal}, 1993a;
Gabriel, 1993, 2000;
Roxburgh and Vorontsov, 1995;
Abrams and Kumar, 1996;
Nigam and Kosovichev, 1998;
Rast and Bogdan, 1998; Rosenthal, 1998).
The observed asymmetry can be used to constrain the depth and other
properties of the excitation
(Chaplin and Appourchaux, 1999; Kumar and Basu, 1999;
Nigam and Kosovichev, 1999)
and hence obtain information about subsurface convection.

\section{Observation of solar oscillation}

Solar oscillations manifest themselves in the solar atmosphere
in different ways:
the displacement causes the atmosphere to move,
changes in the energy transport in the outer layers of the Sun cause
oscillations in the solar energy output,
while oscillations in the atmospheric temperature are reflected
in the properties of the solar spectral lines.
Each of these effects may be used to observe the oscillations;
since they all reflect the same underlying modes they should evidently
yield the same oscillation frequencies.
The choice of observing technique is then determined by a combination
of technical considerations and noise properties, including the
effects of the Earth's atmosphere for ground-based observations,
and effects of other variations in the solar atmosphere.
A detailed review of techniques for helioseismic observations and data
analysis was given by Brown (1996).

The combined oscillation velocity amplitude in the five-minute range at
any given point on the solar surface, as detected by
Leighton {\etal}\ (1962), is around $500 \m \s^{-1}$.
However, this results from the random combination of signals
from of order $10^7$ individual modes.
The velocity amplitude for each mode is at most around $10 \cm \s^{-1}$.
The corresponding amplitude in relative intensity perturbations is
a few parts per million. 
Thus extreme sensitivity is required to carry out detailed observations
of the oscillations.
Furthermore, the observations have to deal with other fluctuations
in the solar atmosphere, such as resulting from near-surface convection
and solar activity, of far higher magnitude.
That it is even possible to extract the small oscillation signal is in 
large measure due to the high spatial and temporal 
coherence of the oscillations,
with lifetimes extending over several weeks to months;
in contrast, other phenomena in the solar atmosphere typically
have low coherence in space and time.
Thus, by integrating over the solar disk and analyzing data over
extended periods in time, the solar `noise' is suppressed and
the oscillations can be isolated;
even so, in current observations of solar oscillations the effects
of random solar fluctuations are probably the dominant source 
of background noise.
To achieve the noise suppression and the required frequency resolution
the observations are typically analyzed coherently over several months;
furthermore, temporal gaps in the data introduce frequency sidebands
in the power spectrum which complicate the determination of the frequencies,
and hence data with minimal interruptions are highly desirable.
This immediately points to the need for the combination of data
from several sites around the Earth, to compensate for the day/night
cycle, or for observations from space.

\subsection{Observing techniques}

The most detailed observations of solar oscillations
have been carried out in line-of-sight
velocity, measured from the Doppler shift of lines in the solar spectrum.
As illustrated in Fig.~\ref{fig:dop-obs}, this may be done by measuring
the intensity in two bands on either side of a suitable spectral line.
If the intensities are recorded by means of an imaging detector,
the result is a velocity image, measuring simultaneously
the motion of the solar surface with potentially high spatial resolution.
Alternatively, by passing integrated light from the Sun through
the filter to the detector, one obtains a disk-averaged velocity,
corresponding to observing the Sun as a star.

The main challenge in the observations is to provide a stable
determination of the wavelength intervals defining the two intensities.
In an ingenious technique for disk-averaged observations,
the filter is replaced by a scattering cell, where light is scattered
from the Zeeman-split components of a line in sodium or potassium
vapour placed in the field from a permanent magnet
({\eg}, Fossat and Ricort, 1975; Brookes {\etal}, 1976).
Here the wavelength bands are determined mainly by the strength of
the field, which is very stable, with little sensitivity 
to other properties of the instrument.
A variant of this technique ({\eg}, Cacciani and Fofi, 1978)
can be used as a magneto-optical transmission filter for
spatially resolved observations ({\eg}, Rhodes {\etal}, 1986;
Tomczyk {\etal}, 1995).

The perhaps most extensively developed technique
for spatially resolved observation is derived from
the so-called Fourier Tachometer ({\eg}, Brown, 1984).
Here the line shift is obtained from four measurements
in narrow bands across a given spectral line. 
This allows the definition of a measure that is essentially
linear in the line-of-sight velocity, over the considerable
range of velocities encountered over the solar surface.
In the actual implementations the spectral bands are
defined by Michelson interferometers.
Examples of Doppler images obtained using
this technique are shown in Fig.~\ref{fig:dop-imag}.

\begin{figure}[htbp]
\begin{center}
\inclfig{6.0cm}{\fig/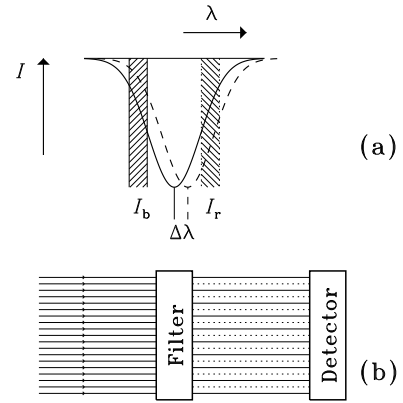}{Schematic Doppler observations}
\end{center}
%\vskip -1cm
\caption{
\figlab{fig:dop-obs}
Schematic illustration of Doppler-velocity observations.
In a), the line-of-sight velocity shifts the line in
wavelength by $\Delta \lambda$, from the continuous to the
dashed position.
This changes the intensities $\Ib$ and $\Ir$ measured
in the narrow wavelength intervals shown as hatched, as well as
the ratio $(\Ib - \Ir)/(\Ib + \Ir)$ which provides
a measure of the shift and hence of the velocity.
Panel b) illustrates the experimental setup.
The filter alternates between letting light in the $\Ib$
and $\Ir$ bands through.
If an imaging detector is used, the resulting images in
$\Ib$ and $\Ir$ can be combined into a Doppler image.
}
\end{figure}

A conceptually very simple way to study the oscillations
is to observe them in broad-band intensity or irradiance.
In practice, fluctuations in the Earth's atmosphere
render such observations very difficult from the ground;
however, the technique has been highly successful from
space ({\eg}, Woodard and Hudson, 1983;
Toutain and Fr\"ohlich, 1992).

A very substantial number of helioseismic observing facilities have
been established (see also the review by Duvall, 1995).
To limit effects of gaps in the data, networks of observing stations
are used.
The BiSON ({\bf Bi}rmingham {\bf S}olar {\bf O}scillation
{\bf N}etwork; Chaplin {\etal}, 1996) network was established in 1981
and now consists of six stations; it carries out disk-averaged
velocity observations by means of potassium-vapour resonant-scattering cells.
Spatially resolved velocity observations are obtained with the
GONG ({\bf G}lobal {\bf O}scillation {\bf N}etwork {\bf G}roup;
Harvey {\etal}, 1996) six-station network, based on the
Fourier-tachometer technique, which has been operational since 1995;
this is funded by the US National Science Foundation,
%and has headquarters in Tucson, AZ,
but involves a large international collaboration.
Valuable data are also being provided by the LOWL instrument of
the High Altitude Observatory (Tomczyk {\etal}, 1995) on Mauna Loa,
Hawaii, using a magneto-optical filter; this has recently been 
extended to a two-station network, with the addition of an instrument
on Tenerife, in the Canary Islands.
Other ground-based networks include the IRIS (Fossat, 1991)
and TON (Chou {\etal}, 1995) networks.

\begin{figure}[htbp]
\begin{center}
\incltwofig{4cm}{\fig/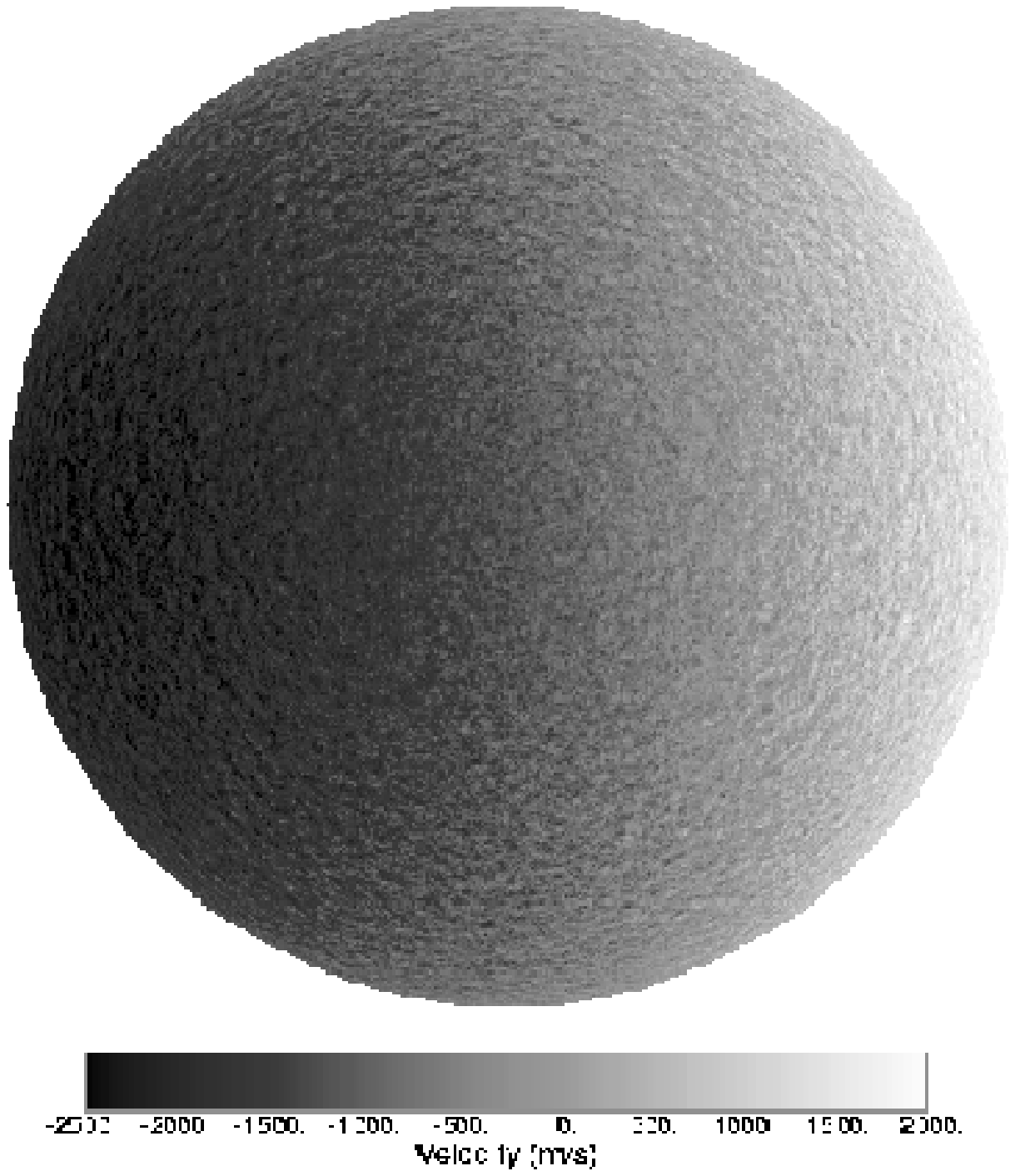}{\fig/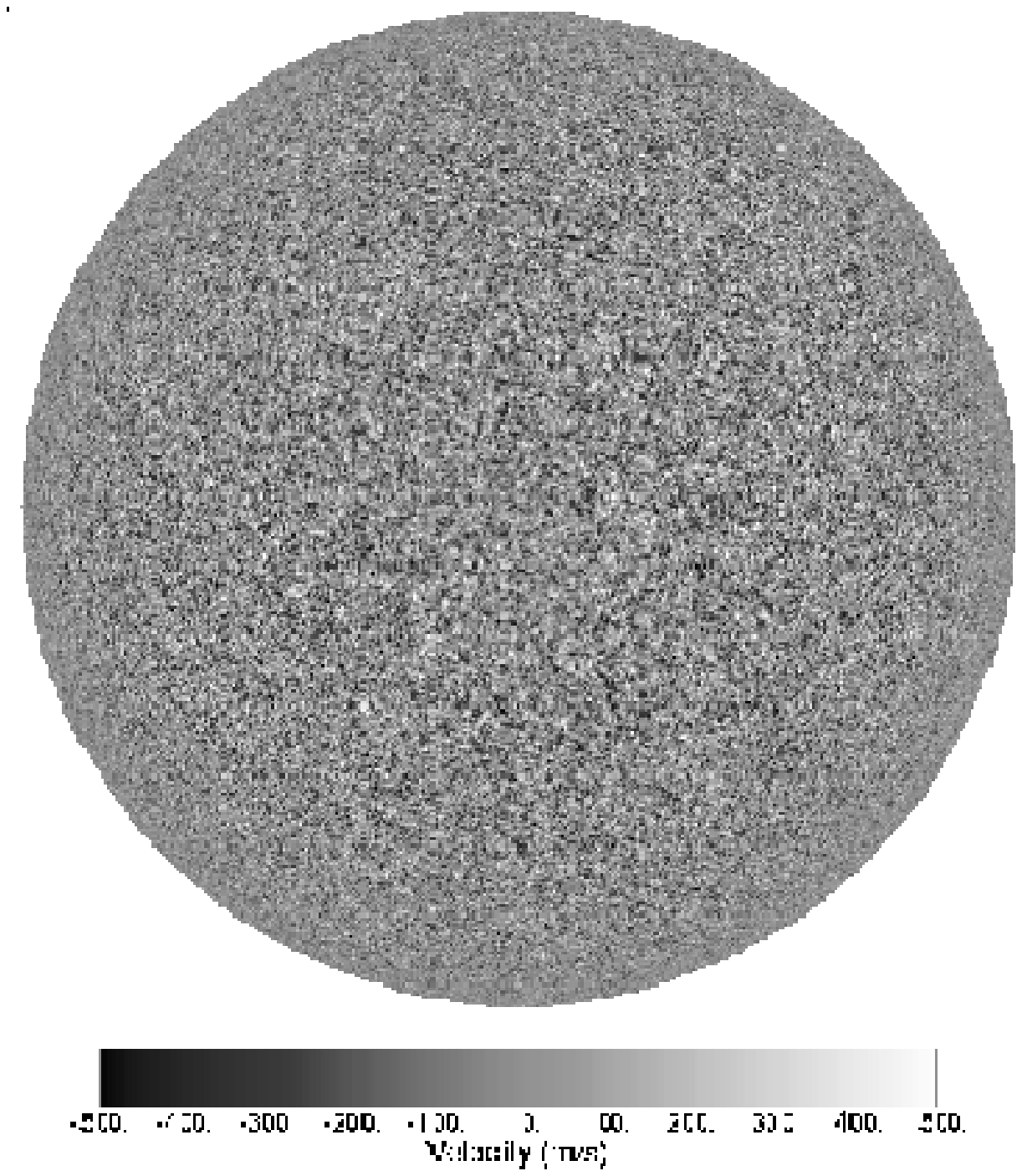}{Doppler images}
\end{center}
\caption{
\figlab{fig:dop-imag}
Doppler images obtained with the MDI instrument (Scherrer {\etal}, 1995).
To the left is the original image, with a greyscale ranging
from $-2000 \m \s^{-1}$ (dark) to $2000 \m \s^{-1}$ (light).
This is dominated by solar rotation.
After removing rotation by averaging (right-hand image) the
mottling associated primarily with solar oscillations becomes apparent;
here the greyscale ranges from
$-500 \m \s^{-1}$ (dark) to $500 \m \s^{-1}$ (light).
}
\end{figure}

Space observations from a suitable orbit completely avoid the problem
of periodic interruptions of the data.
Major contributions have been made from the SOHO spacecraft
(Domingo {\etal}, 1995),
a joint project between ESA and NASA; it was launched in 1995 and
started scientific observations in 1996 from an orbit close to
the first Lagrange point between the Earth and the Sun.
SOHO carries three helioseismic instruments.
GOLF ({\bf G}lobal {\bf O}scillations at {\bf L}ow {\bf F}requency;
Gabriel {\etal}, 1995, 1997) aims in particular at detecting 
low-frequency modes, possibly including g modes, in disk-averaged observations.
It was designed as a resonant-scattering Doppler-velocity instrument,
using sodium vapour; however technical problems have led to the
observations now being carried out in intensity variations in the blue
wing of the sodium spectral line.
VIRGO ({\bf V}ariability of solar {\bf IR}radiance and
{\bf G}ravity {\bf O}scillations; Fr\"ohlich {\etal}, 1995, 1997)
measures solar irradiance, disk-integrated intensities in three
different wavelength regions, and intensity with limited spatial resolution.
An important goal of the instrument is again the search for g modes,
with the hope that these might be more easily detectable in intensity data
than in velocity data.
Finally, the SOI/MDI ({\bf S}olar {\bf O}scillations {\bf I}nvestigation --
{\bf M}ichelson {\bf D}oppler {\bf I}ma\-ger; Scherrer {\etal}, 1995;
Rhodes {\etal}, 1997)
uses a technique based on the Fourier Tachometer, the spectral
bands being defined by a pair of tunable Michelson interferometers.
This provides observations of Doppler velocity over the entire solar disk
with a spatial resolution of 2~arc~sec, corresponding to 
independent velocity measurements
over a total number of about $800\,000$ locations, %
%\footnote{A roughly comparable resolution in terrestrial seismology 
%would require 
%a similar number of seismographs uniformly distributed over a hemisphere
%of the Earth.}
allowing detailed study of oscillations of degrees up to about 1000.
%In addition, the instrument can make observations over a part of the
%solar disk with a resolution of 1.25~arc~sec.

\subsection{Analysis of oscillation data} \plabel{sec:anal}

Regardless of the observing technique, the signal contains contributions
from the broad range of modes that are excited in the Sun
({\cf}\ Section~\ref{sec:solcause}).
The goal of the analysis is to extract from this signal, as a function
of position on the solar disk and time,
information about the properties of the solar interior, such as the
structure and internal motions, and about the properties of the
excitation of the oscillations.
In principle, this may be thought of as fitting to the observations
an overall model encompassing all the relevant features.
In practice, the analysis must be carried out in several steps,
at each step taking into account the properties of the intermediate data
resulting from the preceding steps.

Here I concentrate on the determination of the properties of global
modes of solar oscillation, most importantly their oscillation frequencies
$\omega_{nlm}$, and the subsequent analysis of the frequencies.
Alternative analysis techniques, aimed at investigating local properties
of the solar interior, are discussed in Section~\ref{sec:lochel}.

\subsubsection{Spatial analysis}

The first substantial step in the analysis is to separate as far
as possible the contributions from the individual spherical harmonics $Y_l^m$.
Oscillations in broad-band or line intensity behave essentially 
as spherical harmonics as functions of $\theta$ and $\phi$ on the solar disk.
%although the variations of the background intensity,
%known as {\it limb darkening}, across the disk must be taken into account.
For observations in Doppler velocity, the signal is the projection
of the velocity field on the line of sight.
The surface velocity field for a single mode is determined by 
Eqs (\ref{displ}) and (\ref{hor-disp}) and is characterized
by the ratio $\xih(R)/\xi_r(R)$.
%A useful approximation to this ratio can be obtained in the Cowling
%approximation, where the perturbation $\Phi'$ to the gravitational
%potential is neglected;
%using also the surface boundary condition in \Eq{surf-p},
%as well as \Eq{eqn:hydrost}, we obtain
%\be
%{\xih(R) \over \xi_r(R)} \simeq {G M \over R^3 } {L \over \omega_{nl}^2} 
%\simeq {\omega_{0l}^2 \over \omega_{nl}^2 } \; ,
%\eel{amp-ratio}
%where $\omega_{0l}$ is the frequency of the f mode of the given degree
%({\cf}\ Eq.\ \ref{eqn:fmode}).
%From this it follows, for example,
It may be shown, however,
that at the observed solar frequencies and low or moderate degree
the oscillations are predominantly in the radial direction.
Thus it is common in the analyses to ignore the horizontal component
of velocity.

Here I consider Doppler observations in more detail, assuming the
velocity to be purely in the radial direction.
For simplicity I furthermore take the axis of the spherical harmonics
to be in the plane of the sky, orthogonal to the line of sight.
Then the observed Doppler signal $\VD$ can be written as
\bea
\label{dopsig}
 \VD(\theta , \phi, t) & = &
\sin\theta\cos\phi
\sum_{n, l, m} A_{nlm} c_{lm} P_l^m ( \cos\theta) \times \\
& & \times \cos[m \phi - \omega_{nlm} t - \beta_{nlm}] \; .  \nonumber
\eeanl{dopsig}
Here the factor $\sin\theta \cos\phi$ arises from
the projection of the velocity onto the line of sight.
%The intensity signal has a similar form, although without the
%projection factor.
To isolate modes corresponding to a given
spherical harmonic, with $(l, m) = (l_0, m_0)$, say,
the signal is integrated over the area $A$ of the
solar disk, with a suitable weight $W_{l_0 m_0}(\theta, \phi)$,
yielding
\bea
\label{filtsig}
V_{l_0 m_0}(t) &  = &
\int_A \VD(\theta, \phi, t) W_{l_0 m_0} (\theta, \phi) \dd A \\
 & = & \sum_{n, l, m} S_{l_0 m_0 l m} A_{nlm}
 \cos[ \omega_{nlm} t + \beta_{nlm,l_0 m_0}] \; . \nonumber
\eeanl{filtsig}
The response function $S_{l_0 m_0 l m}$ and
the combined phase $\beta_{nlm,l_0 m_0}$ are obtained from
integrals of the projected spherical harmonics weighted by $W_{l_0 m_0}$.

The goal of this spatial analysis is obviously to isolate
a single spherical harmonic in the time string $V_{l_0 m_0}(t)$,
{\ie}, to have, as far as possible,
that $S_{l_0 m_0 l m} \propto \delta_{l_0 l} \delta_{m_0 m}$,
where $\delta_{ij}$ is the Kronecker delta.
From the orthogonality of the $Y_l^m$ over the unit sphere, 
it may be expected that 
$W_{l_0 m_0} \simeq Y_{l_0}^{m_0}$ is suitable.
Indeed, had data been available over the entire solar surface,
and apart from the velocity projection factor, complete isolation
of a single spherical harmonic would have been possible.
In practice, however, $V_{l_0 m_0}(t)$ contains contributions
also from neighboring $(l, m)$.
This so-called {\it leakage} substantially complicates the
subsequent determination of the oscillation frequencies.
Examples of the leakage matrix 
$S_{l_0 m_0 l m}$ are illustrated in Fig.~\ref{fig:leaks}.

\begin{figure}[htbp]
\begin{center}
\inclfig{5.5cm}{\fig/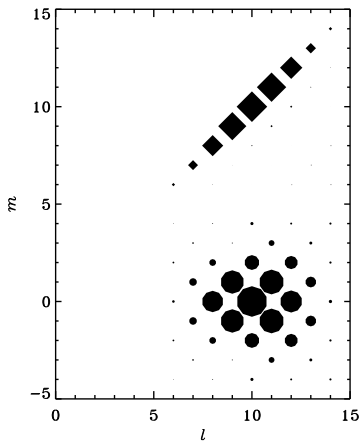}{Leakage matrix}
\end{center}
\caption{
\figlab{fig:leaks}
Leakage matrices $S_{l_0 m_0 l m}$ for $(l_0, m_0) = (10,0)$
(circles) and $(l_0, m_0) = (10,10)$ (diamonds),
as functions of $(l, m)$.
The size of the symbols is proportional to $S_{l_0 m_0 l m}$.
}
\end{figure}

A special case of weighting is obtained in disk-averaged observations;
in this case the signal is dominated by modes of low degree,
$l \lwig 4$, with no explicit separation between the azimuthal orders
({\eg}, Dziembowski, 1977; Christensen-Dalsgaard and Gough, 1982).
However, since the solar rotation axis is always close to the
plane of the sky, it follows from the symmetry of the spherical
harmonics that such observations are essentially insensitive
to modes where $l - m$ is odd.

In practice, the analysis involves a number of steps.
The observed solar Dopplergram is transferred to a co-latitude --
longitude grid aligned with the solar rotation axis, taking into
account the variation with time of the orientation of the rotation
axis relative to the observer.
Also, to speed up the calculation of the required very large number
of integrals in \Eq{filtsig} the integration in longitude $\phi$
is carried out by means of a Fast Fourier Transform.
Some details of these procedures were described by Brown (1985, 1988).

\begin{figure}[htbp]
\begin{center}
\inclfig{8.2cm}{\fig/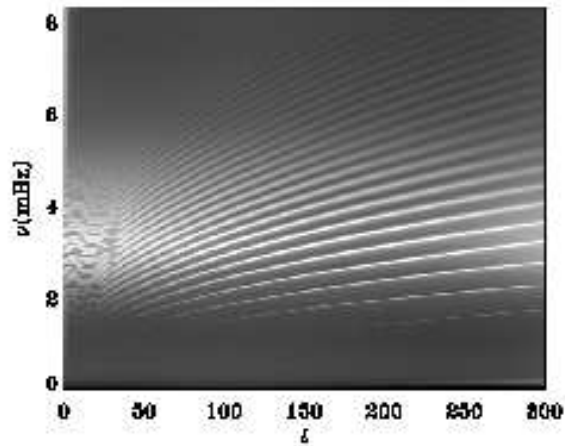}{SOI/MDI spectrum}
\end{center}
\caption{
\figlab{fig:mdi-lnu}
Power spectrum of velocity observations from the SOI/MDI experiment
on the SOHO spacecraft.
The ridges of power concentration correspond to separate radial orders,
starting at the lowest frequency with the f mode, with $n = 0$.
}
\end{figure}

\subsubsection{Temporal analysis} \plabel{sec:tempanal}

The next step in the analysis is to isolate the individual modes,
characterized by radial orders $n$, in the time string $V_{l_0 m_0}(t)$.
This is done through Fourier analysis of $V_{l_0 m_0}(t)$.
The result can be illustrated in a so-called $l - \nu$ diagram,
such as is shown in Fig.~\ref{fig:mdi-lnu},
where the power is plotted against target degree $l_0$ and frequency%
\footnote{It is conventional to analyze observed frequencies 
in terms of {\it cyclic frequencies} $\nu = \omega/2 \pi$.}
$\nu$.
This clearly shows the concentration of power in ridges,
each corresponding to a given value of $n$
({\cf}\ Fig.~\ref{fig:freqsun}).
A clearer impression of the power distribution is obtained
by plotting the power as a function of frequency, for a given
target degree.
As a special example, Fig.~\ref{fig:bis-power} shows a power
spectrum obtained from disk-averaged observations from the BiSON network.
It is evident that the power is indeed concentrated in very narrow 
peaks, hardly resolved at low frequencies;
this reflects the intrinsic damping times of the modes
which at the lowest frequencies exceed several months.
At the maximum power, the amplitude per mode is around $15 \cm \s^{-1}$.
It should be noticed also that the spectrum reflects the 
asymptotic frequency behavior for low-degree p modes
[{\cf}\ Eqs~(\ref{E7.41}) and (\ref{E7.45})]:
thus several cases of pairs of modes with $l = 0, 2$
or $l = 1, 3$ can be identified.

From such spectra, the frequencies and other parameters of the individual modes
can be obtained by fitting.
This must take into account the statistical nature of the 
power distribution, resulting from the stochastic excitation
({\cf}\ Section~\ref{sec:solcause}), and assuming a parametrized
form of the average line profile;
although in principle asymmetrical profiles should be considered,
most analyses to date have been based on Lorentzian profiles
characterized by their widths and amplitudes
(but see Toutain {\etal}, 1998; Chaplin {\etal}, 1999a; Thiery {\etal}, 2000).
The fits are further complicated by the leakage of power from other
$(l, m)$ into the spectrum being analyzed.%
\footnote{For descriptions of the analysis techniques and
the complications encountered, see for example 
Anderson {\etal}\ (1990), Schou (1992),
Hill {\etal}\ (1996),
Appourchaux, Gizon \& Rabello-Soares (1998), and
Appourchaux, Rabello-Soares \& Gizon (1998);
an overview was provided by Schou (1998b).}

\begin{figure}[htbp]
\begin{center}
\inclfig{8.6cm}{\fig/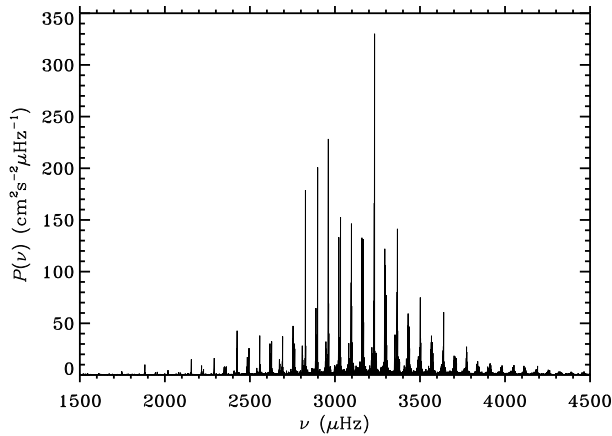}{BiSON spectrum}
\end{center}
\caption{
\figlab{fig:bis-power}
Power spectrum of solar oscillations,
obtained from Doppler observations
in light integrated over the disk of the Sun.
The ordinate is normalized to show velocity power per frequency bin.
The data were obtained from six observing stations
and span approximately four months.
(See Elsworth {\etal}, 1995a.)
}
\end{figure}

To illustrate the quality of present data on solar oscillations,
Fig.~\ref{fig:mdi-freq} shows observed mean multiplet frequencies
$\nu_{nl}$, obtained from the MDI instrument
(Kosovichev {\etal}, 1997).
Over a large part of the diagram the errors, even when multiplied
by 1000, are barely visible; the relative error $\sigma(\nu)/\nu$
is below $5 \times 10^{-6}$ for more than 1000 multiplets.
It is this extreme accuracy, in measured quantities related directly
to the properties of the solar interior, which allows detailed investigations
of solar internal structure.

The ridges in Fig.~\ref{fig:mdi-freq} extend to a limit
where the natural line width of the modes is comparable to the
separation between modes of adjacent degree;
beyond this limit neighboring modes partially merge as a result
of the spatial leakage, and a strict separation of modes
in frequency becomes difficult or impossible (Howe and Thompson, 1998).
At higher degree the mode frequencies must be inferred from
the location of ridges containing overlapping contributions from
several modes, the relative importance of which depend on the leakage matrix.
Thus the frequency determination requires accurate calculation of the
leakage matrix, taking also the horizontal component of velocity into account
({\eg}, Rabello-Soares {\etal}, 2001).
Although progress has been made in this area ({\eg}, Rhodes {\etal}, 2001),
more work is required for the determination of 
fully reliable high-degree frequencies.

\begin{figure}[htbp]
\begin{center}
\inclfig{8.6cm}{\fig/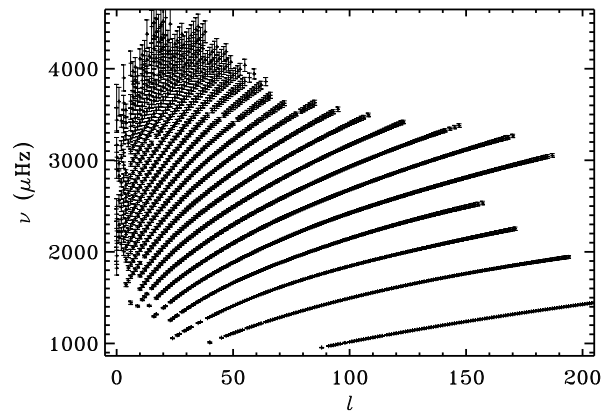}{Multiplet frequencies}
\end{center}
\caption{
\figlab{fig:mdi-freq}
Observed mean multiplet frequencies of solar oscillations,
from 144 days of MDI observations.
The error bars correspond to 1000 standard deviations.
The smallest relative errors $\sigma(\nu)/\nu$ are below $3 \times 10^{-6}$.
}
\end{figure}

The frequency splittings $\Delta \nu_{nlm} = \nu_{nlm} - \nu_{nl}$
contain information about solar internal rotation and other
possible departures from spherical symmetry
({\cf}\ Section~\ref{sec:rotsplit}).
Although full utilization of the information contained in the
oscillation data requires use of the individual frequencies
$\nu_{nlm}$, the determination of these frequencies is
often difficult or impossible.
Thus it is customary to represent the frequency splittings by
polynomial expansions
\be
\nu_{nlm}
= \nu_{nl0}
+ \sum_{j=1}^{j_{\rm max}} a_j (n,l) \, \PP_j^{(l)} {(m)} \; ,
\eel{acoeff}
in terms of the so-called $a$ coefficients $a_j(n,l)$;
here the $\PP_j^{(l)}$ are polynomials of degree $j$ which satisfy
the orthogonality relation 
$\sum_m \PP_i^{(l)}(m) \PP_j^{(l)}(m) = 0$ for $i \neq j$
({\eg}, Ritzwoller and Lavely, 1991;
Schou {\etal}, 1994).
Explicit expressions for these polynomials were given by Pijpers (1997).
It follows from Section~\ref{sec:rotsplit}
that to lowest order rotation gives rise to odd $a$ coefficients.
The even $a$ coefficients correspond to departures from spherical
symmetry in solar structure, as well as
to quadratic effects of rotation. %on the oscillations.
%Examples of odd $a$ coefficients are shown in Fig.~\ref{fig:mdi-split}.

\subsubsection{Solar g modes?} \plabel{sec:obsgmodes}

As discussed in Section~\ref{sec:fgmodes},
observation of g modes would provide very important information
about the properties of the solar core.
Indeed, the search for solar g modes has been an important theme
in the development of helioseismology.
The early indications of a 160-min signal in solar data
({\cf} Section~\ref{sec:history}) hinted that such modes might
be present and led to continued efforts to detect them.
An important aspect in these searches was the uniform period
spacing that is predicted by asymptotic theory
({\eg}, Delache and Scherrer, 1983; Fr\"ohlich and Delache, 1984).
Unfortunately, although further indications of g modes were presented
by Gabriel {\etal}\ (1998), the reality of these detections, and
the precise nature of the modes, has not yet been definitely
established.
In particular, Appourchaux {\etal}\ (2000a), analyzing several different
data sets, obtained stringent upper limits to the amplitudes of solar
g modes, substantially lower than the early claims and barely consistent
with the results of Gabriel {\etal}\ (1998).

\subsection{Helioseismic inversion} \plabel{sec:helinv}

Given the observed frequencies, an important goal is to infer
localized properties about the solar interior from them
through {\it inversion}.
Several inversion techniques have been developed for this purpose.%
\footnote{For reviews, see, for example, Gough and Thompson (1991),
and Gough (1996a).}
Here I first illustrate general principles by considering 
the somewhat idealized case
of inference of a spherically symmetric angular velocity $\Omega (r)$
from observed rotational splittings ({\cf}\ Eq.\ \ref{E8.37}),
and then discuss the techniques that are applied in
more realistic cases.

\subsubsection{Principles of inversion} \plabel{sec:prininv}

In the simple rotational inversion problem the data are of the form
\be
\Delta_i = \int_0^R K_i (r) \Omega (r) \dd r + \epsilon_i \; ,
\qquad i = 1, \ldots, M \; ,
\eel{E9.1}
%$$
where, for notational simplicity, I represent the pair $(n, l )$ by
the single index $i$;
$M$ is the number of modes in the data set considered,
$\Delta_i$ is the scaled rotational splitting 
$m^{-1} \beta_{nl}^{-1} \delta \omega_{nlm}$,
%such that the kernels $K_i$ are unimodular,
and $\epsilon_i$ is the observational error in $\Delta_i$.
The goal of the inversion is to determine an approximation
$\Ombar (r_0)$ to the true angular velocity, as a function 
of position $r_0$ in the Sun.
Inversion is often carried out through linear operations on the data.
Hence for each $r_0$ there exists a set of
{\it inversion coefficients} $c_i(r_0)$ such that
\be
\Ombar ( r_0 ) = \sum_i c_i ( r_0 ) \Delta_i 
= \int_0^R \CK(r_0, r) \Omega (r) \dd r \; ,
\eel{E9.2}
%$$
using \Eq{E9.1} and ignoring the error;
here the {\it averaging kernel} $\CK(r_0, r)$ is given by
\be
\CK(r_0, r) = \sum_i c_i ( r_0 ) K_i (r) \; .
\eel{E9.4}
%$$

The inversion coefficients and averaging kernels clearly
depend on the choice of inversion method, and of possible parameters
that enter into the method;
indeed, the inversion may be thought of as a way to determine
coefficients and averaging kernels such as to obtain as
much information about the angular velocity as possible.
%On the other hand, once the method and parameters have been
%chosen, the coefficients and averaging kernels are independent
%of the data.
%Hence they can be used to make a data-independent comparison
%of different inversion methods;
%this was the approach taken by Christensen-Dalsgaard {\etal}\ (1990).

The averaging kernels provide an indication
of the resolution of the inversion; it is clearly desirable
to achieve averaging kernels that are sharply peaked around 
$r = r_0$, and with small amplitude far away from that point.
%As a quantitative measure of resolution it is common to
%use a width of $\CK(r_0, r)$ obtained as the distance
%between the quartile points, defined such that one quarter of the
%area of $\CK(r_0, r)$, regarded as function of $r$, lies
%below the lower, and one quarter of the area above the upper,
%quartile point.
The inversion coefficients give information about the
propagation of errors from the data to the solution $\Ombar(r_0)$.
In particular, if the errors $\epsilon_i$
are assumed to be uncorrelated, with standard errors $\sigma ( \Delta_i )$,
the standard error in the result of the inversion satisfies
\be
\sigma [ \Ombar ( r_0 ) ]^2 = \sum_i c_i (r_0 )^2 \sigma ( \Delta_i )^2
 \; .
\eel{E9.6}
%$$
%This may also conveniently be characterized by the
%{\it error magnification}, defined as
%\be
%\Lambda(r_0) = {\sigma [ \Ombar ( r_0 ) ] \over \sigma_{\rm rms}} \; ,
%\eel{E9.6q}
%where $\sigma_{\rm rms}$ is the root-mean-square error in the data.
The optimization of the inversion techniques requires
a trade-off between width of the averaging kernels
and the error.

In the techniques of {\it optimally localized averages},
developed by Backus and Gilbert (1970),
the coefficients $c_i ( r_0 )$ are chosen such as to make
$\CK(r_0, r)$ approximate as far as possible a delta function
$\delta (r - r_0 )$ centered on $r_0$;
then $\Ombar ( r_0 )$ provides an approximation to $\Omega ( r_0 )$.
In one version this is achieved by determining
the coefficients $c_i ( r_0 )$ such as to minimize
\be
\int_0^R \CJ(r_0, r) \CK(r_0, r)^2 \dd r
 + \mu \sum_{i} c_i(r_0)^2 \sigma(\Delta_i)^2 \; ,
\eel{E9.7}
%$$
subject to the constraint
\be
\int_0^R \CK(r_0, r) \dd r = 1 \; .
\eel{E9.8}
%$$
Here $\CJ (r_0, r)$ is a weight function which is small close
to $r = r_0$ and large elsewhere;
a common choice is $\CJ(r_0, r) = (r - r_0)^2$.
Furthermore, $\mu$ is a parameter which, as discussed
below, must be adjusted to optimize the result.

Minimizing the first term in the expression (\ref{E9.7}) subject to \Eq{E9.8}
ensures that $\CK(r_0, r)$ is large close to $r_0$,
where the weight function $\CJ(r_0, r)$ is small, and
small elsewhere.
This is precisely the required ``delta-ness'' of the combined kernel.
%However, with no further constraints,
%the optimization of the combined kernel 
%may result in numerically large coefficients of opposite sign.
%Hence, the variance in $\Ombar$, given by \Eq{E9.6},
%would be large.
The effect of the second term in \Eq{E9.7}
is to restrict $\sigma^2 ( \Ombar )$.
The size of $\mu$ determines the relative importance of
the localization and the size of the variance in the result.
Hence, $\mu$ must be determined to ensure a trade-off
between the localization and the error,
measured by the width of $\CK(r_0, r)$ 
and $\sigma[\Ombar(r_0)]$, respectively.

Pijpers and Thompson (1992, 1994) developed a
computationally more efficient method, 
%where only one matrix inversion is required for all target radii.
where the inversion coefficients are determined
by matching $\CK(r_0, r)$ to a prescribed
{\it target function} $\CT(r_0, r)$.
They dubbed this the {\it SOLA} technique
(for Subtractive Optimally Localized Averaging),
to distinguish it from the {\it MOLA} technique
(for Multiplicative Optimally Localized Averaging) discussed above.
Specifically, the coefficients $c_i(r_0)$ are
determined by minimizing
\be
\int_0^R \left[ \CK(r_0, r) - \CT(r_0, r)\right]^2 \dd r
+ \mu \sum_{i} c_i(r_0)^2 \sigma(\Delta_i)^2 \; ,
\eel{E9.14}
%$$
where again $\mu$ is a trade-off parameter.
In addition, the width of $\CT(r_0, r)$ functions as a parameter,
in most cases depending on $r_0$, of the method.%
\footnote{It was argued by Thompson (1993) that for inversion
of acoustic data the resolution width is proportional to 
the local sound speed $c$; thus the target width is often
chosen to be proportional to $c(r_0)$.}
As before, the inclusion of the last term in \Eq{E9.14}
serves to limit the error in the solution.
%In the resulting set of linear equations, the coefficient matrix
%is independent of $r_0$.
%Thus it can be inverted or, more efficiently,
%suitably factored, once and for all;
%after this the determination of the coefficients at
%each target point $r_0$ is virtually free.
%Compared with the MOLA technique the computational
%effort is therefore reduced by roughly a factor
%given by the number of target locations.
An important advantage of the technique is the ability
to choose the target function such as to tailor the
averaging kernels to have specific properties.

A second commonly used class of techniques are the
regularized least-squares, or Tikhonov,
methods (see, for example, Craig and Brown, 1986).
Here the solution $\Ombar ( r )$ is parameterized,
for example as a piecewise constant
function on a grid $r_0 < r_1 < \ldots < r_N$, 
with $\Ombar(r) = \Omega_j$ on the interval $[r_{j-1}, r_j]$;
the parameters $\Omega_j$ are determined through a least-squares
fit to the data.
In general, this procedure is regularized to obtain a smooth solution,
by including in the minimization a term which
restricts the square of $\Ombar$, or
the square of its first or second derivative.
Thus, for example one may minimize
\bea
& & \sum_i \sigma(\Delta_i)^{-2} 
\left[\int_0^R K_i(r) \Ombar(r) \dd r - \Delta_i \right]^2 \nonumber \\
& & + \mu^2 \int_0^R \left({ \dd^2 \Ombar \over \dd r^2} \right)^2 \dd r \; ,
\eeal{E9.18}
%$$
where in the last term a suitable discretized approximation
to $\dd^2 \Ombar / \dd r^2$,
in terms of the $\Omega_j$, is used.
The minimization of \Eq{E9.18} clearly leads to a
set of linear equations for $\Ombar_j$, defining the solution;
however, it is still the case that the solution is linearly
related to the data and hence is characterized by
inversion coefficients and averaging kernels ({\cf}\ Eq.\ \ref{E9.2}).
By restricting the second derivative the last term in \Eq{E9.18}
suppresses rapid oscillations in the solution, and hence
ensures that it is smooth;
the weight $\mu^2$ given to this term serves as a trade-off parameter,
determining the balance between resolution and error for this method.

Christensen-Dalsgaard {\etal}\ (1990)
made a comparison of different inversion techniques as applied to this
problem, in terms of their error and resolution properties.
Useful insight into the properties of inversion techniques can
be obtained from analyzing the inverse problem by means of
(Generalized) Singular Value Decomposition
({\eg}, Hansen, 1990, 1994;
Christensen-Dalsgaard {\etal}, 1993).
This can also be used to develop efficient algorithms for the inversion,
through pre-processing of the problem
(Christensen-Dalsgaard and Thompson, 1993;
%Basu, Christensen-Dalsgaard \& Thompson, 1997a).
Basu {\etal}, 1997a).

It is evidently important to consider the statistical properties
of the inferences obtained through helioseismic inversion.
This requires reliable information about the statistics of
the data (oscillation frequencies or frequency splittings),
which may not always be available.
An important example is correlation between data errors;
although the correlation matrix has been estimated in a few cases
({\eg}, Schou {\etal}, 1995),
off-diagonal elements are generally not taken into account in the inversion.
Yet it was demonstrated by Gough (1996a) and Gough and Sekii (2002)
that this might have serious effects on the inferences.
Howe and Thompson (1996) noted the importance of taking into account also the
error correlation between different points in the inference.
Careful evaluations of statistical aspects of helioseismic
inversion were provided by Genovese {\etal}\ (1995)
and Gough {\etal}\ (1996a).

\subsubsection{Inversion for solar rotation} \plabel{sec:rotinv}

In reality, we wish to infer the solar internal angular velocity
$\Omega(r, \theta)$ as a function of both distance $r$ to the center
and co-latitude $\theta$.
Inversions to do so can be based directly
on \Eq{rot-split}, although quite often the expansion of the
rotational splittings in $a$ coefficients ({\cf}\ Eq.\ \ref{acoeff})
is used; it is straightforward to show that the odd $a$ coefficients
are related to $\Omega(r, \theta)$ by relations similar to
\Eq{rot-split}, with kernels that may be determined from
the kernels $K_{nlm}(r, \theta)$.
The inversion methods discussed above
can be immediately generalized to the two-dimensional case of
inferring functions of $(r, \theta)$, including the definitions
of inversion coefficients and averaging kernels ({\eg}, Schou {\etal}, 1994).
The main difficulty, compared to the one-dimensional case, is the
amount of data that must be dealt with;
while inversion for solar structure, based on average multiplet
frequencies, requires the analysis of typically at most a few thousand
frequencies, 
the splittings or $a$ coefficients used for rotational inversion 
number tens of thousands.
For this reason early investigations were typically carried out with the
so-called 1.5-dimensional methods ({\eg}, Brown {\etal}, 1989)
where $\Omega(r, \theta)$ was expanded suitably in $\theta$,
reducing the problem to a series of one-dimensional inversions for the
expansion coefficients as functions of $r$.
However, with the development of computer power, and even more with
the development of efficient algorithms taking advantage of
the detailed structure of the problem
({\eg}, Larsen, 1997; Larsen and Hansen, 1997),
the fully two-dimensional inversions are entirely feasible and commonly used.
An overview of inversion methods and further references were given
by Schou {\etal}\ (1998), who also carried out tests of the
inversion procedures based on artificial data.

\subsubsection{Inversion for solar structure} \plabel{sec:strucinv}

Inversion for solar structure is conceptually more complicated than
the rotational inversion.
In the case of rotation, the basic relation between the unknown angular
velocity and the data is linear to a high approximation.
In the structure case, on the other hand, the corresponding relation
between structure and multiplet frequencies is highly nonlinear.
This is dealt with through linearization, on the assumption that
a solar model is available which is sufficiently close to the actual
solar structure;
then the inversion can be based on \Eq{E5.81}.
This is of a form similar to the simple inverse problem in \Eq{E9.1}, 
although with additional terms,
and can be analyzed using extensions of the methods discussed in
Section~\ref{sec:prininv}.

Least-squares inversion can be carried out by
parametrizing 
the unknown functions $\deltar c^2/c^2$, $\deltar \rho/\rho$ and 
$\Fsurf$, the parameters being determined
through regularized least-squares fitting similar to \Eq{E9.18}
({\eg}, Dziembowski {\etal}, 1990; Antia and Basu, 1994a);
as shown by Basu and Thompson (1996) this allows tests
for possible systematic errors in the data through investigation of
the residuals.
However, most inversions for solar structure differences have applied
generalizations of the optimally-localized average techniques,
by constructing linear combinations of the relations (\ref{E5.81})
with coefficient $c_i(r_0)$ chosen to isolate a specific feature of
the structure.
To infer $\deltar c^2/c^2$, for example, this is achieved with the SOLA 
method by
replacing the expression (\ref{E9.14}) to be minimized by
\bea
&& \int_0^R \left[ \CK_{c^2,\rho}(r_0, r) - \CT(r_0, r)\right]^2 \dd r
+ \beta \int_0^R \CC_{\rho, c^2}(r_0, r)^2 \dd r \nonumber \\
&& + \mu \sum_i \sigma_i c_i(r_0) c_j(r_0) \; ,
\eeal{E9.32}
%$$
where again $i$ numbers the multiplets $(n,l)$, and $\sigma_i$
is the standard error of $\delta \omega_i/\omega_i$.
Here the averaging kernel is now
\be
\CK_{c^2, \rho}(r_0, r) = \sum_i c_i(r_0) K_{c^2, \rho}^i(r)
\eel{E9.33}
%$$
and I have introduced the {\it cross-term kernel}
\be
\CC_{\rho, c^2}(r_0, r) = \sum_i c_i(r_0) K_{\rho, c^2}^i(r) \; ,
\eel{E9.34}   
which controls
the (undesired) contribution of $\deltar \rho/\rho$ to the solution.
As in the rotation case, the minimization of the expression (\ref{E9.32})
ensures that $\CK_{c^2, \rho}(r_0, r)$ approximates the target
$\CT(r_0, r)$ while suppressing the contributions from the cross term and
the data errors;
the effect of the term in $\Fsurf$ is reduced by chosing the coefficients
to satisfy in addition the constraints
\be
\sum_i c_i(r_0) I_i^{-1} \psi_\lambda(\omega_i) = 0 \; ,
\lambda = 0, \ldots, \Lambda \; ,
\eel{E9.35}
%$$
for a suitably chosen set of functions $\psi_\lambda$,
typically taken to be polynomials of order $\lambda$
({\eg}, D\"appen {\etal}, 1991; Kosovichev {\etal}, 1992).
A detailed discussion of implementation details, including the choice
of the trade-off parameters $\beta$ and $\mu$ and of the properties of the
target function, was provided by Rabello-Soares {\etal}\ (1999).

For high-degree modes the surface effects are no longer functions
of frequency alone, as demonstrated by Brodsky \& Vorontsov (1993).
Di Mauro {\etal}\ (2002) have developed a generalization of the
constraints (\ref{E9.35}), based on the asymptotic expressions
of Brodsky \& Vorontsov, allowing suppression
of the surface term for modes of degree as high as 1000.
The resulting inversion enabled resolution of the upper few per cent
of the solar radius, including the helium and parts of the hydrogen
ionization zones, of great interest in connection with investigation
of the equation of state of the solar plasma 
(see also Section~\ref{sec:helphys}).

\section{Helioseismic investigation of solar structure} \plabel{sec:solstruc}

The average multiplet frequencies $\nu_{nl}$ carry information 
about the spherically symmetric component of solar structure.
This can be used to test solar models and obtain information about
the properties of matter in the solar interior.
As noted in Section~\ref{sec:oscilprop} only quantities
such as density $\rho$, adiabatic exponent $\Gamma_1$ or sound speed
$c$ are immediately constrained by the frequencies;
constraints on other aspects of the solar interior structure require
further assumptions about the models.

Already the early observations of high-degree modes
(Deubner, 1975; Rhodes {\etal}, 1977) provided significant
constraints on the solar interior.
Although these modes are trapped in the outer part of the convection zone,
they are sensitive to its general adiabatic structure,
and comparison between the observed and computed frequencies
indicated that the convection zone was deeper than previously
assumed (Gough, 1977b; Ulrich and Rhodes, 1977).
Furthermore, it was pointed out that the frequencies were sensitive to
details of the equation of state 
({\eg}, Berthomieu {\etal}, 1980; Lubow {\etal}, 1980).
The detection of low-degree modes, penetrating to the solar
core, allowed tests of more profound aspects of the models, including
effects of changes aimed at reducing the neutrino flux
({\cf}\ Section~\ref{sec:solneutrino}).
An early result was the likely exclusion of solar models
with abundances of helium and heavier elements substantially
below the standard values (Christensen-Dalsgaard and Gough, 1980b).
Elsworth {\etal}\ (1990) obtained strong evidence against non-standard
models involving either mixing or energy transport by weakly
interacting massive particles.
More generally, it is now clear that all models that have been
proposed to reduce the solar neutrino flux to the observed values
through modifications to solar structure are inconsistent with the
helioseismic data.

The determination of frequencies for a broad range degrees
by Duvall and Harvey (1983) opened up the possibility for
inversions to determine the structure of substantial parts of the
solar interior.
Gough (1984a) noted that \Eq{E7.81} for the function $F(w)$,
determined from observed quantities by \Eq{E7.80}, could be inverted,
%to yield
%\be
%r = R  \exp \left[ -{ 2 \over \pi} \int_{a_{\rm s}}^a
%\left( w^{-2} - a^{-2} \right)^{- 1/2} { \dd F \over \dd w }
%\dd w \right] \; ,
%\eel{E7.108}
%where $a = c/r$ and $a_{\rm s}$ is its surface value.
%This implicitly 
without any reference to a solar model,
to determine the sound speed $c$ as a function of $r$.%
\footnote{A very similar technique for geophysical inversion
was presented by Brodski\v{\i} and Levshin (1977).}
This was applied to solar data by Christensen-Dalsgaard {\etal}\ (1985)
to infer the sound speed in much of the solar interior, testing
the method by applying it to frequencies of solar models.
The results showed clear indications of the base of the convection zone, 
as a change in curvature in $c(r)$;
the discrepancies in the radiative interior between the Sun and the model
could be interpreted as a deficit in the opacity in the model,
as was subsequently confirmed by the opacity calculations
by, for example, Iglesias {\etal}\ (1992).

Equations (\ref{E7.80}) and (\ref{E7.81}) were derived from a very simple
form of the asymptotic analysis, and hence the resulting inversion
suffers from systematic errors.
These can be substantially reduced by basing the inverse analysis
on higher-order or otherwise improved asymptotic descriptions
({\eg}, Vorontsov and Shibahashi, 1991; Marchenkov {\etal}, 2000),
maintaining the advantage of being independent of a solar model.
Alternatively, the systematic errors can to some extent be eliminated
by carrying out a {\it differential} asymptotic inversion,
based on a fit of \Eq{E7.103} to frequency differences between
the Sun and a model (Christensen-Dalsgaard, Gough, and Thompson, 1989);
given the resulting $\Hone(w)$, \Eq{E7.104} may be inverted
analytically to infer the sound-speed difference between the Sun and the model.
%\be
%{\deltar c \over c} = - { 2 a \over \pi }{ \dd \over \dd \ln r }
%\int_{ a_{\rm s} }^a ( a^2  -  w^2 )^{ - 1/2 } \Hone ( w )  \dd w \; .
%\eel{E7.109}
This technique was used by Christensen-Dalsgaard {\etal}\ (1991)
to determine the depth of the solar convection zone as
$d_{\rm cz} = (0.287 \pm 0.003) R$, a result also obtained independently
by Kosovichev and Fedorova (1991);
the inference has later been confirmed and substantially tightened
by Basu and Antia (1997) and Basu (1998) from fits of $\Hone(w)$
to sequences of models.
Using the differential asymptotic technique, 
Christensen-Dalsgaard, Proffitt, and Thompson (1993) demonstrated 
that the inclusion of helium settling very substantially reduced
the sound-speed differences between solar models and the Sun.

\subsection{Inferences of sound speed and density} \plabel{sec:infsound}

I now consider in more details the results of inferring solar internal
structure from the oscillation frequencies.
In much of the discussion I use as reference Model S of
Christensen-Dalsgaard {\etal}\ (1996); this falls within the category of
`standard solar models' (see Section~\ref{sec:standardsol})
and has been used quite extensively in helioseismic investigations.

The simplest way to test a solar model is to consider
differences between observed frequencies and those of the model.
In Fig.~\ref{fig:freqdif}, panel (a) shows relative differences between
observed frequencies presented by Basu {\etal}\ (1997b)
and those of Model S.
Although there is some scatter, the differences depend predominantly
on frequency, and furthermore they are quite small at low frequency.
According to Section~\ref{sec:varprin} this suggests
that the dominant contributions to the differences are located in
the near-surface layers of the model.
This is confirmed by considering differences scaled by $Q_{nl}$
(panel b), where most of the scatter has been suppressed.
Indeed, giving the simplifications involved in the modeling
of the near-surface structure, and the use of adiabatic frequencies,
it is hardly surprising that differences of this magnitude are obtained.

\begin{figure}[htbp]
\begin{center}
\inclfig{8.6cm}{\fig/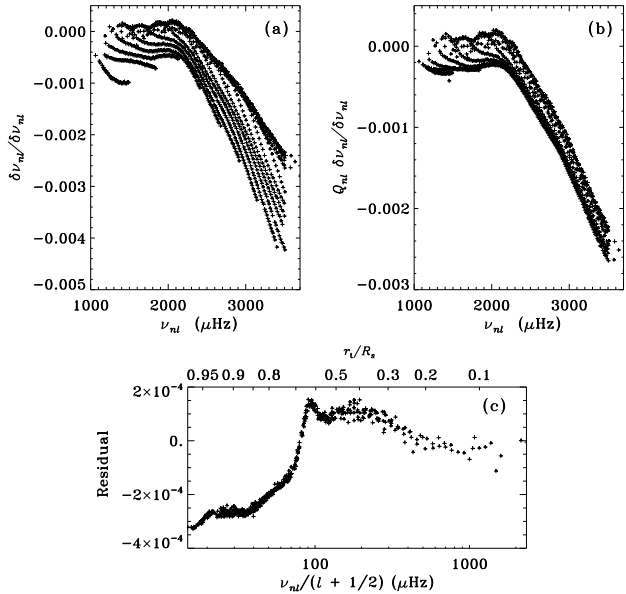}{Frequency differences}
\end{center}
\caption{
\figlab{fig:freqdif}
(a) Relative frequency differences, in the sense (observation) -- (model).
The observations are a combination of BiSON whole-disk measurements
({\eg}, Elsworth {\etal}, 1994) and LOWL observations (Tomczyk {\etal}, 1995),
as described by Basu {\etal}\ (1997b), while the computed frequencies
are for Model S.
(b) The same, but scaled by the inertia ratio $Q_{nl}$
(see Section~\ref{sec:varprin}).
(c) Scaled differences after subtraction of the fitted $\Htwo(\nu)$,
plotted against $\nu_{nl}/(l+ 1/2)$ which determines the lower
turning point $\rt$, shown as the upper abscissa.
}
\end{figure}

Even after scaling, there remains some scatter in the differences,
suggesting a dependence on the depth of penetration of the mode
and hence the presence of differences between the structure of
the Sun and the model that are not confined to the near-surface layers.
These effects can be isolated by subtracting a function of
frequency fitted to the points in Fig.~\ref{fig:freqdif}b.
The residual (see Fig.~\ref{fig:freqdif}c)
is clearly highly systematic; the small intrinsic
scatter reflects both the extremely small observational error
and the extent to which frequency differences can be represented
by \Eq{E7.103}.
It is evident that the behavior changes drastically for modes
penetrating just beneath the base of the convection zone,
with $\rt/R \lwig 0.7$;
this suggests that there may be substantial differences between the
Sun and the model in this region.

Inversion for the differences in structure, without making asymptotic
approximations, was discussed in Section~\ref{sec:strucinv}.
Typical results of such inversions, using the SOLA method, are shown in 
Fig.~\ref{fig:csq-inv}.
To illustrate the resolution properties of the inversion, 
panel (c) shows selected averaging kernels.
It is evident that the inversion has indeed succeeded in resolving
the sound-speed difference between the Sun and the model in considerable detail.
Also, the 1-$\sigma$ formal errors in the results are extremely small, 
below $2 \times 10^{-4}$ in the bulk of the model, owing to the precision of
the observed multiplet frequencies.
Other, similar results were obtained by, for example,
Gough {\etal}\ (1996b) and Kosovichev {\etal}\ (1997).

\begin{figure}[htbp]
\begin{center}
\inclfig{7.5cm}{\fig/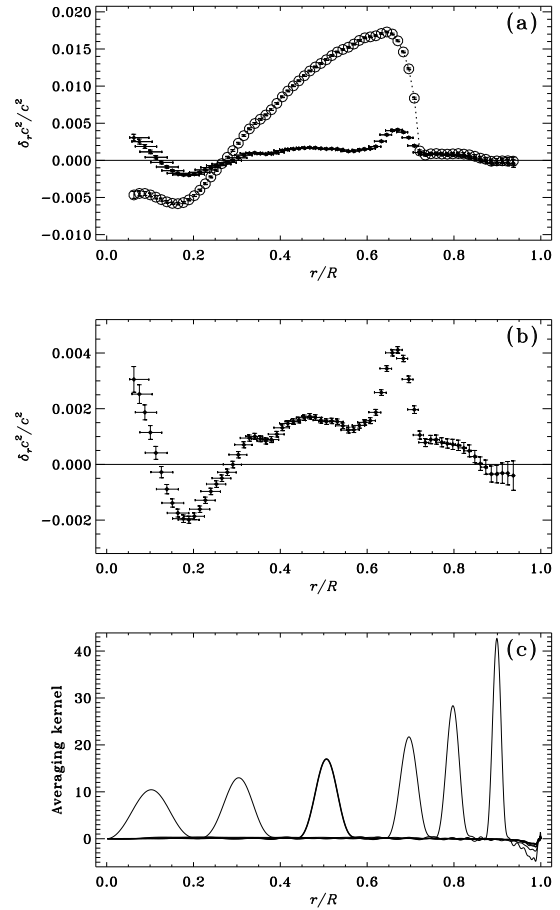}{Sound-speed inferences}
\end{center}
\caption{
\figlab{fig:csq-inv}
Results of sound-speed inversion.
(a) Difference in squared sound speed, in the sense (Sun) -- (model),
inferred from inversion
of the differences between the observed BiSON and LOWL frequencies
and the frequencies of two solar models: closed circles are for Model S,
and open circles for a similar model, but ignoring element diffusion and
settling.
(b) Results for Model S, on an expanded scale.
The vertical error bars are 1-$\sigma$ errors on the inferred
differences, while the horizontal bars provide a measure of
the resolution of the inversion.
(c) Selected averaging kernels $\CK(r_0,r)$, for fractional
target radii $r_0/R = 0.1, 0.3, 0.5, 0.7, 0.8$ and $0.9$.
(Adapted from Basu {\etal}, 1997b.)
}
\end{figure}

The inferred difference between the solar and the Model S sound speed
({\cf}\ Fig.~\ref{fig:csq-inv}b)
is striking.
First of all, the overall magnitude should be noted:
the difference is everywhere below about $5 \times 10^{-3}$,
indicating that $c^2$ of the model agrees with that of the Sun
to within 0.5 per cent from below $0.1 R$ to very near the surface.
It is important to recall that the model calculation contains no
free parameters which have been adjusted to achieve this level of agreement.
It is true that the computation of solar models has been improved 
as a result of the constraints imposed by the steadily improving
helioseismic data, through the inclusion of settling
as well as through improved equation of state and opacities;
as an example, Fig.~\ref{fig:csq-inv}a compares inversion
relative to Model S with the use of a corresponding model which
does not include diffusion and settling.
In this sense the current models have been developed
as a result of the helioseismic data.
However, the improvements in physics have not been tailored towards
fitting the data; it is remarkable that they have nonetheless
resulted in a fit as good as the one shown in Fig.~\ref{fig:csq-inv}b.
%This is a remarkable example of the power of physical modeling to
%reproduce the properties of such a complex object as an evolved star.

It should noticed, however, that the differences, although small,
are highly significant.
Particularly prominent is the peak in $\deltar c^2/c^2$ just below
the convection zone. 
This is a feature shared by all recent investigations, based on a
variety of data and `standard' solar model calculations;
interestingly, recent updates to
the opacities and the solar initial composition have tended to
increase the discrepancies between the Sun and standard solar models.
Similarly, the negative $\deltar c^2/c^2$ around $r = 0.2 R$ is
a common feature to most inferences.
On the other hand, the results in the inner core, for $r \lwig 0.1 R$,
show some variation between different data sets, although the inferred
differences are in all cases of a magnitude similar to that shown in 
Fig.~\ref{fig:csq-inv}b.
The inferences certainly show that standard calculations are inadequate.
I return to possible causes for the discrepancies in Section~\ref{sec:helsun}.

Extensive comparisons have been carried out between
solar models and the results of helioseismic inversions,
to investigate effects of changes in the physics of the solar interior.%
\footnote{See, for example, 
Dziembowski {\etal}\ (1994),
Richard {\etal}\ (1996),
Turck-Chi\`eze {\etal}\ (1997),
Brun {\etal}\ (1998, 1991),
Fiorentini {\etal}\ (1999),
Morel {\etal}\ (1999),
Bahcall {\etal}\ (2001),
Guzik {\etal}\ (2001),
Neuforge-Verheecke {\etal}\ (2001).}
Basu {\etal}\ (2000) showed that the 
inferred solar structure depends little on the assumed reference model,
thus confirming that the linearization in \Eq{E5.81} is justified.
A detailed analysis of the sensitivity of the helioseismic results
to the composition profile and aspects of the nuclear energy generation
was presented by Turck-Chi\`eze {\etal}\ (2001b).

Although the most general information about the solar interior is
obtained from inverse analyses, as discussed above,
other techniques may be more sensitive to specific features of the
solar interior.
In particular, localized features in the Sun cause oscillatory
perturbations in the frequencies, as a function of mode order,
resulting from the change in the phase of the eigenfunctions at
the location of the feature as the order is varied ({\eg}, Gough, 1990).
An interesting example is the rapid change in
the temperature gradient at the base of the convection zone,
which has a distinct signature in the oscillation frequencies.
Analyses of the observed frequencies
have been used to show that convective penetration into the radiative
region below the convection zone,
at least assuming a relatively simple model of the resulting
structure, has at most a very limited extent
({\eg}, Basu {\etal}, 1994;
Monteiro {\etal}, 1994;
Roxburgh and Vorontsov, 1994b).

\subsection{Physics and composition of the solar interior} \plabel{sec:helphys}

The precision of the observed frequencies allows us to go beyond
the determination
of the sound speed, to investigate finer details of the physics of the
solar interior. 
An important aspect is the equation of state, particularly in the regions
of partial ionization which to a large extent are found in the convection zone.
This part of the Sun has substantial advantages
for helioseismic investigations:
since the stratification is very nearly
adiabatic, apart from a thin region near the top,
the structure of the convection zone depends essentially only on the
equation of state and composition, while it is not directly
affected by the opacity. 
The potential for helioseismic determination of the convection-zone
composition and tests of the equation of state was recognized
by Gough (1984b) (see also D\"appen and Gough, 1986).
An important and potentially detectable effect
of the thermodynamic state and composition arises from $\Gamma_1$
which is suppressed relative to the value of $5/3$ for a fully ionized ideal gas
in the zones of partial ionization of abundant elements ({\eg}, D\"appen, 1998).
In particular,
determination of the helium abundance is in principle possible
because the reduction in $\Gamma_1$ in the second
ionization zone of helium obviously depends on the abundance of helium.

Investigations of these ionization zones can be carried out
in terms of the asymptotic description of the oscillations
in Eqs~(\ref{E7.28q}) or (\ref{E7.103}), where the effects of
the near-surface regions are contained in the phase functions
$\alpha(\omega)$ or $\Htwo(\omega)$.%
\footnote{{\eg},
Brodski\v{\i} and Vorontsov (1989),
Baturin and Mironova (1990),
Marchenkov and Vorontsov (1990),
Pamyatnykh {\etal}\ (1991),
Christensen-Dalsgaard and P\'erez Hern\'andez (1992), 
Gough and Vorontsov (1995).}
As discussed above, the relatively sharp variation of $\Gamma_1$
in the second helium ionization zone causes an oscillation in
the frequencies, reflected in the phase functions,
of a magnitude that depends on the helium abundance.
Determinations of the solar envelope helium abundance
by means of such asymptotic methods 
were carried out by Vorontsov {\etal}\ (1991),
Antia and Basu (1994b), 
and P\'erez Hern\'andez and Christensen-Dalsgaard (1994).
Furthermore, the phase functions may provide powerful diagnostics
of the equation of state in the near-surface region
({\eg}, Vorontsov {\etal}, 1992; Baturin {\etal}, 2000).

To discuss the potential of helioseismology for
testing composition and thermodynamic properties,
beyond the asymptotic approximation,
we note that the sound speed is determined by $p$, $\rho$ and $\Gamma_1$
({\cf}\ Eq. \ref{E4.57q}), where, in turn,
$\Gamma_1 = \Gamma_1(p, \rho, Y, Z)$ may be obtained
from the thermodynamical properties of the gas and the composition;
allowance should be made, however, 
for a possible error $(\delta \Gamma_1)_{\rm int}$
in the equation of state used in the calculation of the reference model,
where $(\delta \Gamma_1)_{\rm int}$ is the difference in $\Gamma_1$
between the values obtained with the solar and the model equations of state,
at fixed $p, \rho, Y, Z$.%
\footnote{For simplicity I neglect the effect of $Z$ in the following;
in any case it is constrained (at least in the convection zone)
by the spectroscopic measurements.}
Then \Eq{E5.81} can be rewritten,
expressing $\deltar c^2$ in terms of $\deltar p$, $\deltar \rho$,
$\deltar Y$ and $(\delta \Gamma_1)_{\rm int}$;
it is convenient to express the result in terms of $u = p/\rho$,
using also Eqs (\ref{eqn:hydrost}) and (\ref{eqn:mass}), to obtain
\bea
{\delta \nu_{nl} \over \nu_{nl}}
& = & \int  K^{nl}_{u,Y}{ \deltar u \over u}\dd r +
\int K^{nl}_{Y,u} \deltar Y \dd r \nonumber \\
& & + \int  K^{nl}_{c^2, \rho}
\left({\delta \Gamma_1 \over \Gamma_1} \right)_{\rm int} \dd r +
{\Fsurf(\nu_{nl})\over I_{nl}}
\eeal{eosdif}
(see also Basu and Christensen-Dalsgaard, 1997).
If it is assumed that the model equation of state is
adequate, such that $(\delta \Gamma_1 / \Gamma_1)_{\rm int}$
is negligible, 
\Eq{eosdif} may be inverted to determine $\deltar Y$
in the helium ionization zones ({\eg}, Kosovichev {\etal}, 1992);
since the convection zone is fully mixed, this provides a measure
of the convection-zone value $\Ye$ of the helium abundance.
In a regularized least-squares inversion, for example, 
$\deltar Y$ may be assumed to be constant and hence taken
outside the integral in equation \Eq{eosdif} as a single parameter
({\eg}, Dziembowski {\etal}, 1990, 1991).

In general, potential errors in the equation of state must be
taken into account.
Basu and Christensen-Dalsgaard (1997) showed how the differences
in equation of state might be taken explicitly into account
in the inversion, albeit at the expense of an increase in
the error in the solution;
they also pointed out that the inversion might be
carried out to determine the intrinsic difference in $\Gamma_1$ between
the solar and model equations of state.

To illustrate the sensitivity of such investigations,
Fig.~\ref{fig:gaminv1} shows the results of inversions
for $\Gamma_1$ in the entire solar interior
(Elliot and Kosovichev, 1998).
The most striking aspect are the differences in the solar core
which are clearly resolved.
These demonstrate that the inference is sensitive to
the relativistic effects in the treatment of the electrons,
which were neglected in the original MHD equation of state
used in the top panel, but included in the corrected version
used in the bottom panel.%
\footnote{Note that the average thermal energy of a particle
in the solar core, around 1.35 keV, is 0.3~\% of the
electron rest-mass energy.}
Although this is a fairly trivial correction, 
it does illustrate the sensitivity of the helioseismic inferences
to subtle details of the equation of state.%
\footnote{Gong {\etal}\ (2001) recently presented
a version of the MHD equation of state which includes relativistic
effects for the electrons.}

\begin{figure}
\begin{center}
\inclfig{8.0cm}{\fig/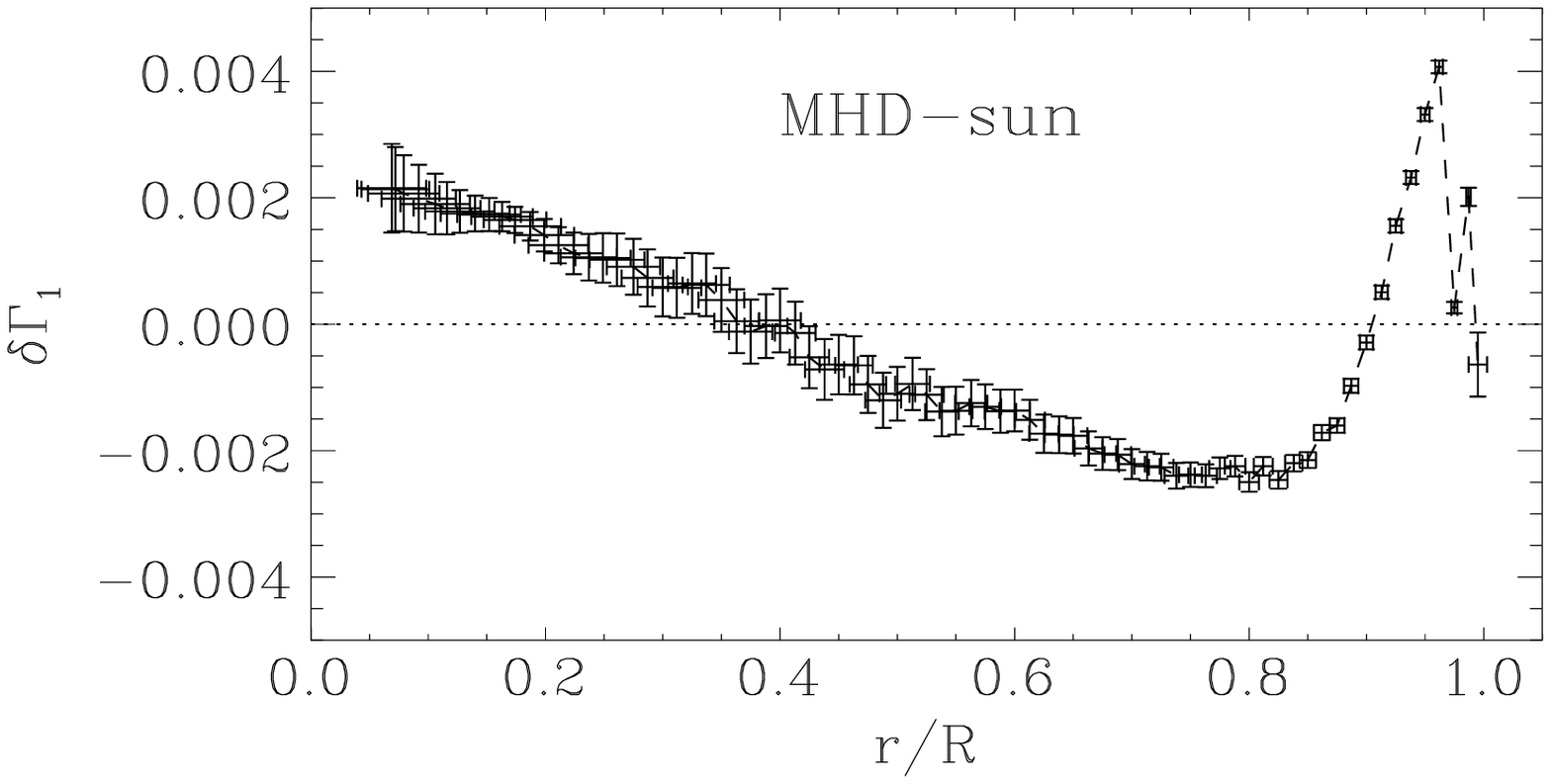}{Gamma1 inversion}
\inclfig{8.0cm}{\fig/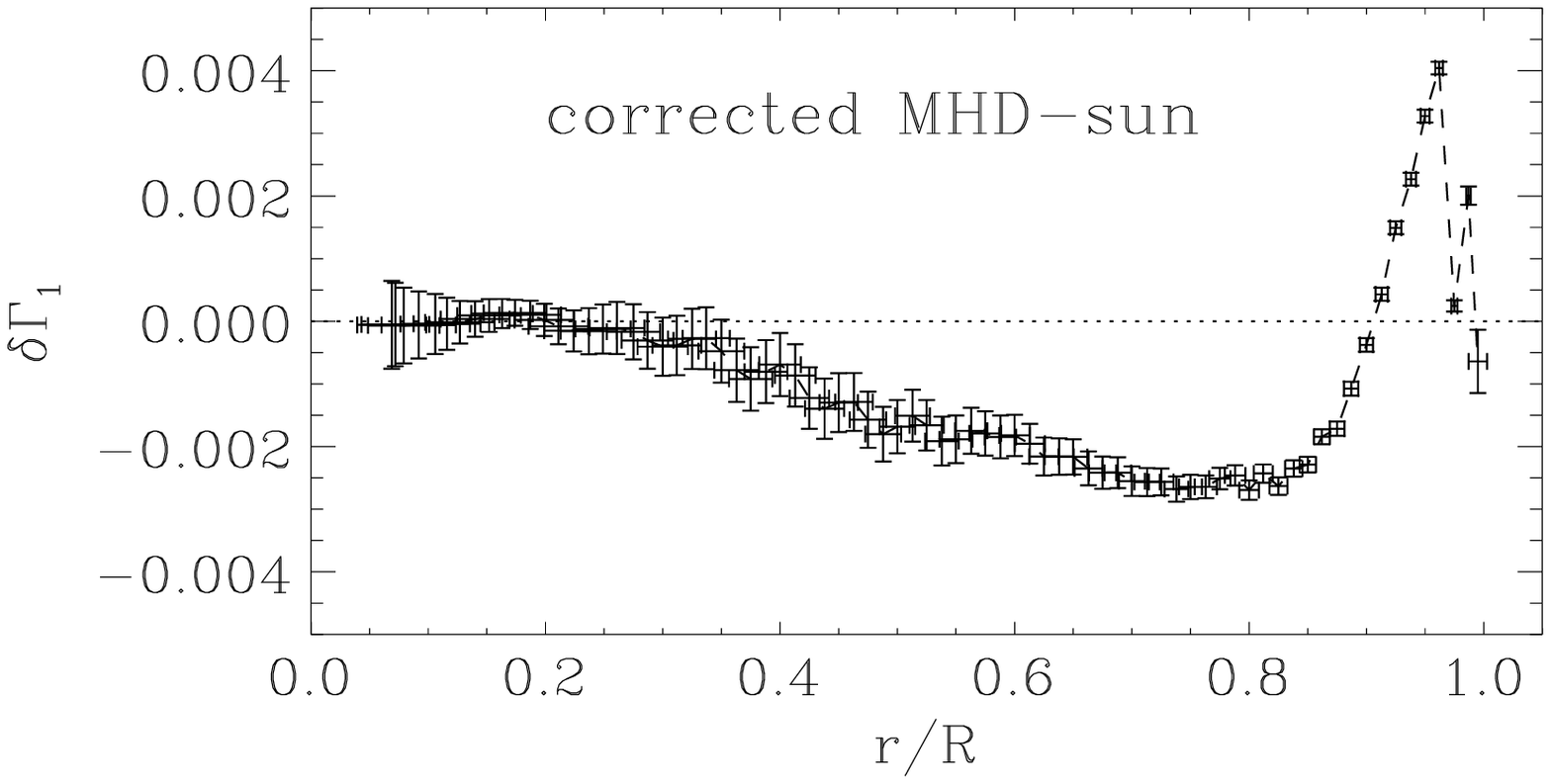}{Gamma1 inversion}
\end{center}
\caption{
\figlab{fig:gaminv1}
The top figure shows the 
difference between $\Gamma_1$ of a solar model with
the MHD equation of state and observation; the bottom figure
shows the result of including a relativistic correction to MHD.
The figures would be qualitatively
similar if OPAL had been used. 
(From Elliot and Kosovichev, 1998.)
}
\end{figure}

Basu, D\"appen, and Nayfonov (1999) made a careful investigation
of the equation of state in the convection zone, determining
intrinsic differences in $\Gamma_1$ for several different models
using the OPAL or MHD equations of state;
this allows tests of these complex and conceptually
very different treatments of the thermodynamic state of solar matter
({\cf}\ Section~\ref{sec:standardsol}).
Some results are illustrated in Fig.~\ref{fig:gaminv2}.
Both equations of state clearly have significant errors,
particularly in the hydrogen and helium ionization zones,
for $r \gwig 0.9 R$;
it appears that the OPAL formulation is closer to the Sun in most
of the region considered, although the situation may be reversed
in the outer 2 -- 3~\% of the radius.
Investigations such as these clearly have great potential for studying
the complex thermodynamic processes in the solar interior, of substantial
value also for other applications of the properties of high-temperature
plasmas.

Several recent determinations of the convection-zone helium abundance $\Ye$
have been made from helioseismic analysis, using both the MHD and the OPAL
equations of state.
The values tend to be in the range 0.24 -- 0.25, with some dependence
on the equation of state, the data set and the analysis method
({\eg}, Basu and Antia, 1995; Richard {\etal}, 1998; Basu, 1998),
although an OLA inversion by Kosovichev (1997) yielded 
rather more disparate values: $\Ye = 0.23$ using MHD and 
$\Ye = 0.25$ using OPAL.
It is striking, in all these cases, that the values obtained are 
substantially below the initial value $Y_0 = 0.27 - 0.28$ 
required to calibrate the models to have the present solar luminosity.
This confirms the importance of settling of helium which reduces
the envelope helium abundance during evolution;
in fact, in Model S the present value, $Y_{\rm e} = 0.245$,
is in reasonably good agreement with the helioseismic determinations.
However, it is evident that the uncertainty resulting from the
possible errors in the equation of state requires further work;
improved results on the helium abundance and the properties of
the equation of state may be expected when reliable data on
high-degree modes become available
({\eg}, Rabello-Soares {\etal}, 2000; Di Mauro {\etal}, 2002).

\begin{figure}
\begin{center}
\inclfig{8.6cm}{\fig/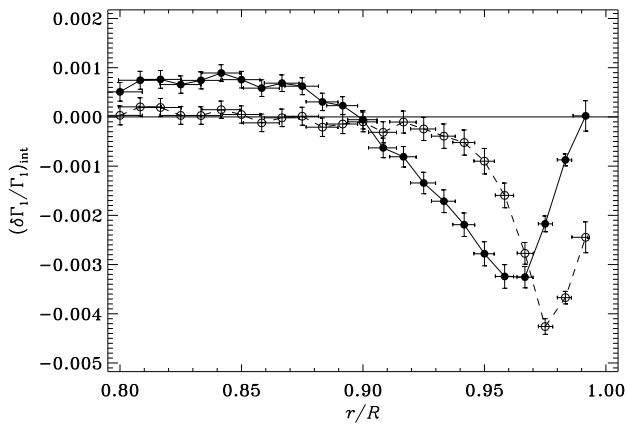}{Gamma1 differences, envelope}
\end{center}
\caption{
\figlab{fig:gaminv2}
Relative difference between $\Gamma_1$ obtained from an inversion of
helioseismological data and $\Gamma_1$ for two solar models.
in the sense ``Sun -- model''. Only the ``intrinsic''
difference in $\Gamma_1$ is shown, that is, the part of the difference
due to the equation of state (see text). 
Lines have been drawn through the results to guide the eye.
The closed circles connected by a solid line are results
obtained with an MHD model, the open circles connected with
a dashed line are results with an OPAL model.
(Adapted from Basu, D\"appen, and Nayfonov, 1999.)
}
\end{figure}

Beneath the convection zone, solar structure depends on the
equation of state, opacity, composition profile and, in the
core, the nuclear energy generation rates.
Here, furthermore, $\Gamma_1$ has very little sensitivity to composition,
at the level of the present accuracy of the inversions;
the determination of the composition depends mainly on its effects
on the mean molecular weight $\mu$ and hence the sound speed
({\cf}\ Eq.\ \ref{approx-sound}), assuming that the temperature is
essentially known.
Thus further constraints, based on the equations of stellar structure
and the assumption of the relevant physical properties, are required
to infer the composition profile.
Gough and Kosovichev (1990) reformulated the inverse problem
in terms of corrections to composition, using the equations of
stellar structure, to determine the
hydrogen abundance in the solar core.
This procedure was also adopted by Kosovichev (1997).
Alternatively, Eqs (\ref{eqn:solstruc}) can be solved, under the
constraint that the model sound speed match the helioseismic inference,
but with no assumption about the hydrogen-abundance profile $X(r)$,
which is then determined as a result of the analysis
(Shibahashi and Takata, 1996; Antia and Chitre, 1998;
Takata and Shibahashi, 1998).
The results of these analyses show considerable scatter,
but they generally confirm the gradient in the hydrogen abundance
just below the convection zone found in solar models, resulting from settling
({\cf} Fig.~\ref{fig:X-mod}).
However, there is a tendency for the gradient to be less steep,
indicating the presence of processes that might partly counteract the settling.
(For a summary of these results, see Christensen-Dalsgaard, 1998.)
As discussed in Section~\ref{sec:helsun}, weak mixing is indeed
a possible explanation for the bump in $\deltar c^2/c^2$ just
beneath the convection zone.

If the composition profile is assumed to be known, on the other hand,
other aspects of the solar interior may be studied.
%Korzennik \& Ulrich (1989) investigated whether changes to the
%opacity might improve the agreement between the observed and
%computed frequencies.
Tripathy and Christensen-Dalsgaard (1998) made a detailed investigation
of the effects of opacity modifications on solar structure and
on this basis Tripathy {\etal}\ (1998)
attempted to determine changes to the opacity that could 
account for the inferred sound-speed difference
illustrated in Fig.~\ref{fig:csq-inv}b.
The required changes, of only a few per cent,
were probably within the general uncertainty in current opacity calculations,
although it is less clear whether their detailed behavior was physically
realistic.
There is little doubt, in any case, that the explanation of the
inferred sound-speed difference will require modifications both
to the composition profile and to the opacity.

\subsection{Helioseismology and the solar neutrino problem}
\plabel{sec:helneutrino}

As discussed in Section~\ref{sec:solneutrino}, the discrepancy
between the predicted and measured flux of solar neutrinos has
cast some doubt on calculations of solar models.
The solar neutrino flux is very sensitive to the temperature of the
solar core.
Thus only relatively modest changes to the structure of the solar
core, reducing the central temperature, are required to bring
the computed neutrino flux into better agreement with the observations.
This is the background for the large number of attempts that have
been made to construct models with a reduced neutrino flux.
It is clear, however, that the close agreement between solar structure
and a standard solar model suggests that such modifications are unlikely
to be consistent with the helioseismic inferences.
The required reduction by roughly a factor of two of the flux
of high-energy neutrinos corresponds approximately to
a reduction in the central temperature of the Sun of about 3 per cent;
if it is assumed that other aspects of the model are roughly unchanged,
this corresponds to a similar decrease in $c^2$, which is 
in obvious conflict with the helioseismically inferred sound-speed difference
({\eg}, Bahcall {\etal}, 1997).
Similar conclusions have been reached by a number of other investigations.%
\footnote{{\eg}, Dziembowski {\etal}\ (1994), 
Ricci {\etal}\ (1997),
%Bahcall, Basu \& Pinnsonneault (1998),
Turck-Chi\`eze {\etal}\ (1998),
Bahcall {\etal}\ (2001).}
More careful analyses, determining limits on the neutrino flux
given the helioseismic constraints, generally confirm this conclusion 
({\eg}, Antia and Chitre, 1997; Takata and Shibahashi, 1998);
Watanabe and Shibahashi (2001) showed that, even assuming a
reduced core abundance of heavy elements, models could not
be constructed which were consistent with both the neutrino and the
helioseismic data.
Also, Turck-Chi\`eze {\etal}\ (2001a) recently constructed a model
essentially consistent with the seismic data and demonstrated that
the neutrino emission from this model was very close to that of
a standard solar model.

It should be noted, none the less, that
conclusions based on helioseismology
concerning the solar neutrino production must be regarded with a little caution.
Since helioseismology essentially provides inferences of 
$T/\mu$, not of $T$ and $\mu$ separately,
a model might in principle be
constructed where $T$ and $\mu$ are both modified in such a way
that their ratio is unchanged, while the neutrino flux is
reduced substantially.
Some reduction in the computed neutrino flux is also possible, 
without increasing the discrepancy in sound speed,
by simply changing the assumed nuclear reaction parameters
suitably within their error limits 
({\eg}, Brun {\etal}, 1998).
Even so, the helioseismic success of the normal solar models
strongly suggests that the solution to the neutrino problem
should be sought not in the physics of the solar interior
but rather in the physics of the neutrino.

This conclusion was dramatically confirmed by the recent
results from the Sudbury Neutrino Observatory which, when combined
with data from the Super-Kamiokande experiment, showed direct 
evidence for solar-neutrino oscillations and yielded a total
rate consistent within errors with standard models
(Ahmad {\etal}, 2001; see Section~\ref{sec:solneutrino}).
Given these results there seems little doubt of the existence
of neutrino oscillations;
also, the results provide independent confirmation of the standard solar model.
With this, the role of helioseismology in the investigations of solar
neutrinos has changed. 
Previously the main issue was to provide evidence
for or against the standard solar models.
Now, the goal is to use helioseismology, together with other relevant
information about the solar core, to constrain as far as possible the
rate of neutrino generation in the Sun%
\footnote{It was noted by Gough (2001bc) that this will require
careful attention to the details of helioseismic inferences about the
solar core; in particular, departures from spherical symmetry
may have to be constrained.};
together with the measurements on Earth of the detection rate of
various types of neutrinos this may offer the best hope for constraining
the properties of the neutrinos, such as masses and interaction parameters.
The importance of this to the further development of physics is obvious.

\section{Inferences of solar internal rotation} \plabel{sec:infrot}

The early inferences of solar internal rotation by Duvall {\etal}\ (1984) 
were based on predominantly sectoral modes, with $m \simeq \pm l$,
and hence provided information about the radial
variation of rotation in a region around the solar equator.
In particular, they established that the interior
of the Sun rotates at approximately
the same speed as the surface, with no evidence for a rapidly rotating core.
To determine the angular velocity $\Omega(r, \theta)$ as a function of
both radius and latitude, through inversion of \Eq{rot-split},
observations of rotational splitting as a function of the azimuthal order
$m$ are required.
These became available with the advent of fully two-dimensional 
observations of solar oscillations
(Brown, 1985; Rhodes {\etal}, 1987;
Libbrecht, 1988, 1989).
Already the initial analyses of these data showed a striking variation
of rotation in the solar interior:
the convection zone largely shared the latitude variation observed
on the surface ({\cf} Eq.~\ref{eqn:surfrot}),
with little variation with depth, 
whereas the radiative interior seemed to rotate like a solid body.%
\footnote{{\eg}, Brown and Morrow (1987),
Christensen-Dalsgaard and Schou (1988),
Kosovichev (1988),
Brown {\etal}\ (1989),
Dziembowski {\etal}\ (1989),
Rhodes {\etal}\ (1990),
Thompson (1990),
Goode {\etal}\ (1991).}
This was at variance with earlier models of the dynamics of the
convection zone ({\cf}\ Section~\ref{sec:solrot}),
and created problems for the dynamo models of the solar magnetic
activity ({\eg}, Gilman {\etal}, 1989).

\begin{figure}
\begin{center}
\inclfig{7.0cm}{\fig/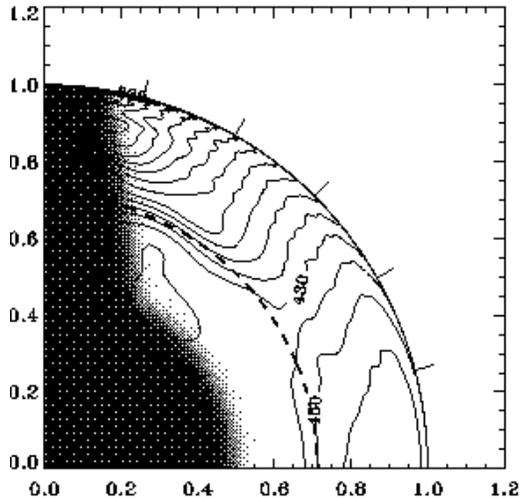}{Contour rotation}
\end{center}
\caption{
\figlab{fig:mdi-rot1}
Inferred rotation rate $\Omega/2 \pi$ in a quadrant of the Sun,
obtained by means of SOLA inversion of 144 days of MDI data.
The equator is at the horizontal axis and the pole is at the vertical
axis, both axes being labelled by fractional radius.
Some contours are labelled in nHz, and, for clarity, selected
contours are shown as bold. The dashed circle is at the base of
the convection zone and the tick marks at the edge of the outer
circle are at latitudes $15^\circ$, $30^\circ$, $45^\circ$, $60^\circ$, and
$75^\circ$.
The shaded area indicates the region in the Sun where no reliable inference
can be made with the current data.
(Adapted from Schou {\etal}, 1998.)
} 
\end{figure}

Very extensive results on rotational splitting have been obtained 
in the last few years.%
\footnote{Examples of recent inferences of solar rotation are
provided by
Thompson {\etal}\ (1996), Wilson {\etal}\ (1997),
and Corbard {\etal}\ (1997).}
These include data from
the GONG network in the form of individual frequency splittings,
and from the SOI/MDI instrument on SOHO in the form of $a$
coefficients extending as high as $a_{35}$.
As discussed in Section~\ref{sec:rotinv}, these observational developments
have been accompanied by the development of efficient inversion algorithms.
Schou {\etal}\ (1998) carried out
analyses of the data from the first 144 days of data from SOI/MDI
using a variety of inversion techniques.
Figure \ref{fig:mdi-rot1} shows the inferred angular velocity,
obtained by means of SOLA inversion.
To illustrate some of the features of the solution more clearly,
Fig.~\ref{fig:mdi-rot2} shows cuts at fixed latitudes.
In accordance with the earlier results, the angular velocity
depends predominantly on latitude in the convection zone,
while there is little significant variation in the radiative interior.
The transition between these two regions, denoted {\it the tachocline}
by Spiegel and Zahn (1992), appears to be quite sharp, and to
coincide approximately with the base of the convection zone.

\begin{figure}
\begin{center}
\inclfig{7.5cm}{\fig/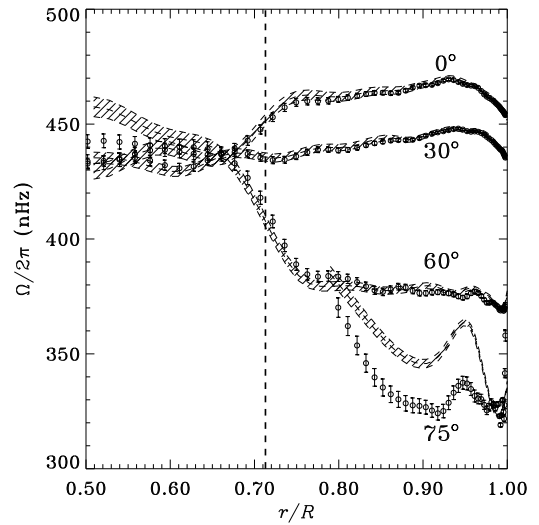}{Cuts of rotation}
\end{center}
\caption{
\figlab{fig:mdi-rot2}
Inferred rotation rate $\Omega/2 \pi$ as a function of radius at
the latitudes indicated,
obtained from inversion of 144 days of MDI data.
The circles with 1-$\sigma$ error bars show results of a SOLA
inversion, while the dashed lines with 1-$\sigma$ error band
were obtained with RLS inversion.
The heavy vertical dashed line marks the base of the convection zone.
(Adapted from Schou {\etal}, 1998.)
} 
\end{figure}

The quality of the MDI data is such that finer details in the
rotation become very apparent.
As was found in earlier analyses, the angular velocity increases
with depth beneath the surface, at least at low latitude,
the maximum angular velocity occurring on the equator at a
depth of around $0.05 \,R$.
Korzennik {\etal}\ (1990), noting the same feature in the
equatorial rotation rate,
pointed out that this variation could be related to the 
different rotation rates inferred from tracking of surface features,
assuming that these features were anchored at different depths.

The tachocline is of very considerable dynamical interest, as
providing the coupling between the latitude-varying rotation
in the convection zone and the nearly solid rotation below it.
Furthermore, it seems likely that the solar dynamo must operate
in this region, with properties that depend sensitively
on the details of the variation in angular velocity
({\eg}, Parker, 1993).
The apparent width of the tachocline in Fig.~\ref{fig:mdi-rot2}
in part reflects the finite resolution of the inversion,
as determined by the radial extent of the averaging kernels.
This must be taken into account in estimating the true width of the tachocline.
Estimates of the width and other properties were made by
Kosovichev (1996a) and Corbard {\etal}\ (1998a, 1999).
Charbonneau {\etal}\ (1999) applied several analysis techniques to LOWL data;
they obtained a tachocline width%
\footnote{The width is defined
as the parameter $w$ in a representation of the
transition of the form $0.5\{1 + {\rm erf}[2(r - r_{\rm c})/w]\}$,
where $r_{\rm c}$ is the central location of the transition.}
of $(0.039 \pm 0.013) \, R$
and an equatorial central radius $r_{\rm c} = (0.693 \pm 0.002) \, R$,
essentially placing the transition beneath the convection zone.
As noted previously by Antia {\etal}\ (1998) and
Di Mauro and Dziembowski (1998),
Charbonneau {\etal}\ found
some indication that $r_{\rm c}$ increased towards the pole,
with an equator-to-pole variation of
$\Delta r_{\rm c} = (0.024 \pm 0.004) \, R$, in the sense
that the tachocline is prolate.

Although the overall features of rotation, as presented above,
have been found using several different data sets and analysis methods,
it should be mentioned that there are problems at the level of finer
details, particularly at higher latitudes.
These have become apparent in comparisons between results based on
data from the GONG and SOI/MDI projects, 
in both cases analyzed with the procedures used by both projects
({\eg}, Howe {\etal}, 2001a; Schou\ {\etal}, 2002).
The origin of the differences in the inferred rotation rates
can be traced back mainly to differences in the analysis procedures
used to determine the oscillation frequencies from the 
spherical-harmonic-filtered time series ({\cf}\ Section~\ref{sec:tempanal}).
Also, as illustrated by the comparison of the SOLA and RLS results 
in Fig.~\ref{fig:mdi-rot2}, different inversion
methods may give different results at high latitude.
Clearly, the underlying causes for these various differences, 
and how to correct for them, need to be identified.

As discussed in Section~\ref{sec:solrot}, models of solar evolution
have suggested the possible existence in the present Sun of a
rapidly rotating core.
Thus it is of obvious interest to infer the properties of rotation 
close to the solar center.
Unfortunately, this is extremely difficult and the
results obtained so far are somewhat contradictory.
Only for modes of the lowest degrees do the kernels extend
into the core and even for these the contribution from the core
to the rotational splitting is small.%
\footnote{The problem is more severe than 
for structure inversion, which also includes modes of degree
$l = 0$; these obviously have no rotational splitting.
Furthermore, the kernels for rotation are suppressed by 
geometrical effects for small $r$.}
In addition, the observational determination of the splitting
is difficult at low degree: here only a few values of
$m$ are available, the total splitting may be comparable
to the natural widths of the peaks in the oscillation power spectra,
and the common procedures for frequency determination may
introduce a systematic bias (Appourchaux {\etal}, 2000b).
A review of the problems in determining the core rotation,
and of the results, was given by Eff-Darwich and Korzennik (1998).
As a result of the small contribution from the core to the splitting,
even fairly modest differences in the observed splittings of low-degree
modes can give disparate results for the core rotation.
Indeed, recent published values range from somewhat higher than
the surface rotation rate ({\eg}, Gizon {\etal}, 1997;
Corbard {\etal}, 1998b), over rates consistent with the bulk of the
envelope ({\eg}, Lazrek {\etal}, 1996) to rotation substantially
below the surface rate ({\eg}, Elsworth {\etal}, 1995b;
Tomczyk, Schou, and Thompson, 1995).
Charbonneau {\etal}\ (1998) showed, based on LOWL data, that
a core of radius $0.1 R$ could rotate at no more than twice the surface rate.
Chaplin {\etal}\ (1999b) attempted 
to obtain averages of rotation localized to the core,
from a combination of BiSON and LOWL splittings.
The results were consistent with constant rotation of the radiative interior,
although with a possible suggestion of a down-turn in the core;
analysis of the averaging kernels showed that constraining the measure
of rotation to the inner 20~\% of the solar radius was only possible
at the expense of very substantial errors in the inferred rotation rate.
Results consistent with uniform rotation of the deep interior were
also obtained by Chaplin {\etal}\ (2001a) who made a careful simulation
of possible systematic errors in the determination of the low-degree
frequency splittings.

\section{The changing, aspherical Sun} \plabel{sec:change}

The Sun is not a static object.
The slow evolutionary changes are likely too small to be detectable
within a human lifetime;
however, the changes associated with the 22-year solar magnetic cycle
({\cf}\ Section~\ref{sec:activity}) may be expected to influence the
structure and dynamics of the solar interior with measurable
consequences for the oscillation frequencies.
One may hope that this can provide information about the inner workings
of the magnetic cycle, including the possible dynamo mechanisms responsible
for it.
In particular, dynamo action just below the convection zone might
produce organized magnetic fields of sufficient strength to be detectable
in the oscillation frequencies.

The first evidence for frequency changes was obtained by
Woodard and Noyes (1985) who found an average decrease in frequencies
of low-degree modes of around $0.42 \muHz$ from 1980 (close to solar
maximum) to 1984 (near solar minimum).
More extensive data by Libbrecht and Woodard (1990, 1991),
covering a substantial range in frequency and degree,
confirmed the general trend and provided information
about the dependence of the frequency change on mode parameters.
It was found that the change largely scaled as the 
inverse mode inertia, much as do the effects of near-surface
errors ({\cf} Section~\ref{sec:oscilprop}).
From 1986 to 1989 ({\ie}, essentially from minimum to close to maximum)
the frequencies increased by up to around $0.8 \muHz$;
the change varied strongly with frequency, from being negligible
below $1.5 \mHz$ to a maximum at $4 \mHz$.
On this basis it was concluded that the dominant source
of the frequency variation was localized very close to the
solar surface.
This was confirmed in a careful comparison of results from several
different data sets by Chaplin {\etal}\ (2001b).
Furthermore, the frequency variations have been shown to be
closely correlated with surface activity, even on time scales of months 
({\eg}, Woodard {\etal}, 1991;
Bachmann and Brown, 1993; Rhodes {\etal}, 1993; Elsworth {\etal}, 1994;
Chaplin {\etal}, 2001c).

A closely related issue are the effects of departures from spherical symmetry.
The resulting variations in wave speed with latitude make a contribution to
the frequency splitting in azimuthal order that is independent of
the sign of $m$;
thus, in terms of the expansion given in \Eq{acoeff} these effects
give rise to even $a$ coefficients.%
\footnote{Quadratic effects of rotation ({\cf} Section~\ref{sec:rotsplit})
also contribute to the even $a$ coefficients; 
these contributions can be calculated from the helioseismically inferred
angular velocity; see, for example, Dziembowski {\etal}\ (1998) and
Antia {\etal}\ (2000b).}
Early measurements of these coefficients were reported by
Duvall {\etal}\ (1986) and Brown and Morrow (1987).
These coefficients, and their variation during the solar cycle,
behaved in a manner corresponding to the time-varying latitude dependence
of the solar surface temperature and magnetic field
({\eg}, Kuhn, 1988; Goode and Kuhn, 1990; Woodard and Libbrecht, 1993).
Extensive data during the rising phase of the present solar cycle
have been obtained from the GONG and SOI/MDI experiments,
greatly strengthening the evidence for a close correlation between
the variations in the oscillation frequencies and the surface
magnetic field
(Dziembowski {\etal}, 1998, 2000; Howe {\etal}, 1999).
Antia {\etal}\ (2001) considered data covering the period 1995--2000
from both GONG and SOI/MDI, and extending to $a_{14}$.
They again found a very close correlation between the variations
in the $a$ coefficients and in the corresponding components of a
Legendre-polynomial expansion of the surface magnetic flux;
this strongly suggests that the behavior of the oscillation frequencies is
directly related to the near-surface magnetic field.
Furthermore, from an inverse analysis of the changes 
they confirmed the superficial
nature of the changes in the wave-propagation speed.

From these results it may appear that the measurements of the frequency changes
and the even $a$ coefficients
have so far added little to our knowledge about solar variability.
Nonetheless, it is still of considerable interest to investigate
the causes for these effects.
Gough and Thompson (1988) concluded that 
the asphericities causing the even $a$ coefficients
in the expansion of frequency splittings were likely of magnetic origin.
Goldreich {\etal}\ (1991) carried out an analysis of the effects
of changes in the near-surface magnetic field and the entropy
of the convection zone and similarly concluded that the dominant 
cause of the frequency change with time was magnetic.
A subsequent analysis by Balmforth {\etal}\ (1996)
confirmed that entropy perturbations alone were unlikely
to be able to account for the observed frequency changes.

The frequency changes for low-degree modes generally follow
the same behavior as seen at high degrees
({\eg}, Elsworth {\etal}, 1994).
However, closer inspection reveals striking differences:
Jim\'enez-Reyes {\etal}\ (1998) found that, when plotted 
against magnetic flux, the frequency changes exhibited
hysteresis, with the frequency at a given flux being larger
during the rising phase of the solar cycle than during the declining phase.
Moreno-Insertis and Solanki (2000) showed that this behavior
could be understood in terms of the variation 
with phase of the solar cycle of the distribution
of the magnetic field over the solar surface, as could
variations in the frequency change with degree.
This behavior is clearly closely related to the changes in the
even $a$ coefficients during the solar cycle, discussed above.

%The models discussed so far generally relied on treating the 
%magnetic effects on the oscillations as small perturbations.
%This is clearly inappropriate within active regions, where
%the field is strong.
%Cunha, Br\"uggen and Gough (1998) carried out an analysis of the
%effect of regions of localized strong field on the modes,
%based on the interaction of the waves with the field as
%inferred from time-distance analyses.
%They concluded that a significant part of the observed
%frequency shift must be due to the influence of sunspots.

According to \Eq{eqn:fmode} the f-mode frequencies are 
determined essentially by the solar surface radius.
This has been used to estimate corrections to the commonly used
value by comparing the observed frequencies to those of solar models
(Schou {\etal}, 1997; Antia, 1998).
Dziembowski {\etal}\ (1998) and Antia {\etal}\ (2000a)
noted that the inferred radius changed with time, reflecting
possible solar-cycle changes in the solar surface radius.
However, it was pointed out by Dziembowski {\etal}\ (2001) that,
as already noted by Gough (1993), 
\Eq{eqn:fmode} should be corrected for the finite radial extent
of the f-mode eigenfunctions;
thus the inferred radius change may in fact take place
in subsurface layers, resulting from changes in magnetic fields
or temperature stratification, with little effect on the photospheric radius
$R_{\rm ph}$.
Dziembowski {\etal}\ (2001) concluded that the change in $R_{\rm ph}$
associated with the solar cycle is only a few kilometres,
of uncertain sign, and hence certainly too small to have a significant
effect on the solar irradiance.

Although the evidence discussed so far points to a superficial nature
of the effects of solar activity on solar structure 
and oscillation frequencies,
it is possible that magnetic fields, or other aspherical perturbations,
sufficiently strong to have an observable effect
may exist deeper within the Sun.
Gough {\etal}\ (1996b) carried out inversion of even $a$ coefficients
to search for radial variations of the asphericity, concluding that
it was confined to a shallow layer close to the surface.
Antia {\etal}\ (2000b) found evidence for an aspherical
perturbation at $r \simeq 0.96 R$; by analyzing frequencies of modes
penetrating beyond the base of the convection zone they also placed an
upper limit of around 30 Tesla
on a possible toroidal magnetic field located in this region.
Evidence for asphericity in the wave speed over a range of depths in
the convection zone was also found by Dziembowski {\etal}\ (2000).
Finally, from analysis of SOI/MDI data Antia {\etal}\ (2001)
found a significant peak, at $r = 0.92 R$ and a latitude of $60^\circ$,
in the time-averaged asphericity, with a similar though weaker signal
in GONG data.
The physical nature of these perturbations is so far unknown;
in particular, as shown by Zweibel and Gough (1995)
it is very difficult to distinguish between direct magnetic effects
and effects of variations in the sound speed.

\begin{figure}
\begin{center}
\inclfig{8.6cm}{\fig/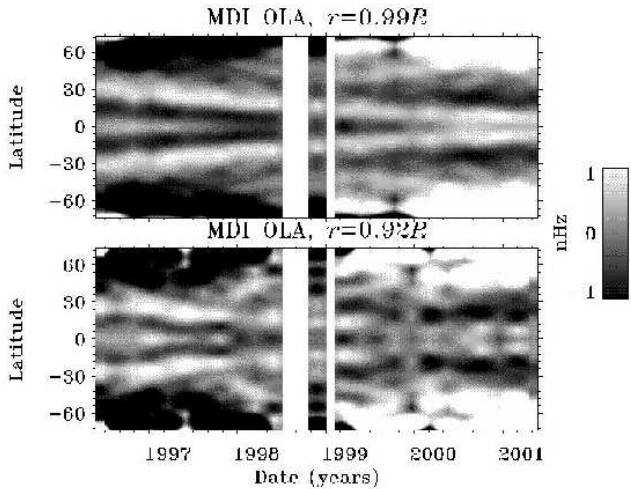}{Zonal flows}
\end{center}
\caption{
\figlab{fig:zonal}
The evolution with time in the zonal flows, inferred from SOLA
inversions of data from SOI/MDI, after subtraction of the time-averaged
rotation rate.
The results are presented as a function of time and latitude,
the grey scale at the right giving the scale in nHz.
The top panel is at a radius of $0.99 R$ and the bottom panel is at $0.92 R$.
Note that the plot has been symmetrized around the equator,
since global rotational inversion only senses the symmetrical component
of the rotation rate ({\cf}\ Section~\ref{sec:rotsplit}).
The white vertical stripes correspond to periods where the SOHO spacecraft
was temporarily non-functional.
(Adapted from Howe {\etal}, 2001b.)
} 
\end{figure}

Solar activity also affects the dynamics of the solar convection zone.
In Doppler observations of the solar surface
Howard and LaBonte (1980) found bands of slightly faster and slower rotation,
which they called torsional oscillations,
shifting towards lower latitudes as the solar cycle progressed
(for more recent results, see Ulrich, 1998, 2001).
Kosovichev and Schou (1997) and Schou {\etal}\ (1998) found similar
variations with latitude in the rotation rate inferred from helioseismic
inversion, extending over perhaps 5~\% of the solar radius.
By analyzing f-mode frequency splittings, Schou (1999) showed
that these bands shifted towards the equator with time, in a manner
very similar to the surface torsional oscillations.
Howe {\etal}\ (2000a, 2001b) studied the depth variation
and time evolution of these so-called zonal flows,
as illustrated in Fig.~\ref{fig:zonal}.
Here data from SOI/MDI have been analyzed in 72-day 
segments, to infer the rotation rate during each of these periods;
an average over all segments over time, at each latitude and radial
location, has been subtracted, and the resulting residuals are displayed.
The bands of faster rotation converging towards the equator are evident.
Remarkably, these can be followed below the surface 
to a depth of at least 8~\% of the solar radius;
on the solar surface, they correspond closely to the variations first
seen by Howard and LaBonte.
Thus these variations involve a substantial fraction of the solar
convection zone.
Similar results were obtained by Antia and Basu (2000, 2001).
Vorontsov {\etal}\ (2002) analyzed SOI/MDI data using an adaptive
regularization technique and found indications that the flows
involve the entire convection zone.
The physical origin of these zonal flows is as yet not clear;
it is interesting, however, that 
Covas {\etal}\ (2000) found a similar spatial and temporal behavior
of rotation in a mean-field dynamo model of the solar magnetic variations.

Birch and Kosovichev (1998) and
Schou {\etal}\ (1998) found that the near-polar rotation was substantially
slower than expected from the directly observed surface rotation rate
({\cf} Eq.~\ref{eqn:surfrot})
or from a simple extrapolation from results at lower latitude.
Similarly, Fig.~\ref{fig:zonal} shows substantial variations at higher
latitudes, not obviously related to the zonal flows at lower latitude.
These variations can be followed to latitudes of at least $80^\circ$ 
(Schou, 1999; Antia and Basu, 2001).

\begin{figure}
\begin{center}
\inclfig{8.6cm}{\fig/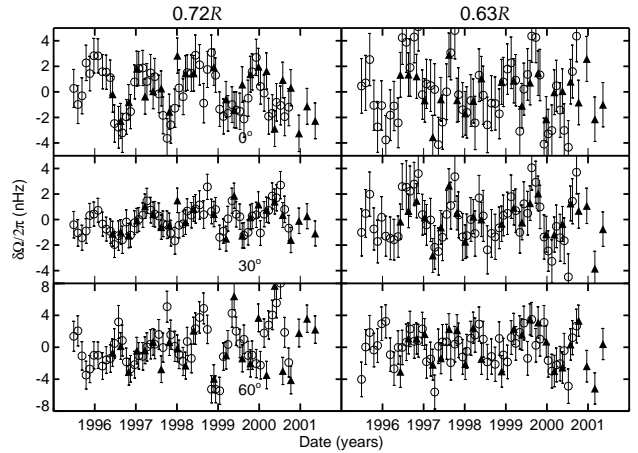}{Tachocline oscillations}
\end{center}
\caption{
\figlab{fig:tacho-osc}
Deviation from the mean rotation rate inferred from inversions
at various locations near the base of the convection zone, as a
function of time.
Open circles represent results from the GONG network, and filled
triangles are from the SOI/MDI experiment.
(From Howe {\etal}, 2001b.)
} 
\end{figure}

Variations in the rotation rate at even greater depth were detected
by Howe {\etal}\ (2000b; see also Howe {\etal}, 2001b).
These are illustrated in Fig.~\ref{fig:tacho-osc}, which again
is based on subtracting time-averaged rotation rates from the results
for time segments.
The most striking variation, seen both in data from GONG and SOI/MDI,
is an oscillation with a period of around 1.3 y in the equatorial
region at the base of the convection zone.%
\footnote{Interestingly, evidence for
variations with a similar period has been detected
in solar activity and the solar wind; {\eg}, Ichimoto {\etal}\ (1985).}
Careful analyses have shown that this cannot simply be an effect
of systematic errors in the observations with an annual period.%
\footnote{In an independent analysis, Basu and Antia (2001) failed
to confirm these findings; although some of their results showed
variations reminiscent of those illustrated in Fig.~\ref{fig:tacho-osc},
the authors did not consider them to be significant.}
At the depth of $0.63 R$, well below the convection zone, there
are indications of an oscillation with the same period but the opposite phase.
Significant variations are also found at higher latitude, although
with less regular periodicities.
Possible physical mechanisms that may be responsible for this behavior
were discussed by Thompson (2001);
a particularly interesting model results from dynamo calculations
which exhibit period-halving bifurcations 
(Covas {\etal}, 2001).

\section{Local helioseismology} \plabel{sec:lochel}

So far, I have considered global modes of solar oscillation,
resulting from the interference of acoustic or surface-gravity waves
travelling through the Sun.
The frequencies of these modes reflect the properties,
such as structure and rotation, of that part of the Sun through which
travel the waves making up a given mode.
By suitably combining the frequencies of these modes, information about
structure and rotation can be localized to limited regions in radius and
latitude, providing inferences about the variation of these properties
with position within the Sun.

Powerful though they are, such analyses have obvious limitations.
The global modes extend over all longitudes; thus analysis of their
frequencies provide essentially no information about longitude
variation of solar properties;
furthermore, as discussed in Section~\ref{sec:rotsplit},
they depend only on that component of, {\eg}, rotation which is symmetric
around the equator.
Also, the properties of global modes have little sensitivity to
meridional or more complex flows,
such as large-scale convective eddies,
which may be present in the solar convection zone.
Further, although the modes are undoubtedly affected by sunspots
or other manifestations of strong localized magnetic fields,
these effects do not lend themselves to detailed inferences of,
say the three-dimensional subsurface structure of a sunspot.

However, it is possible to analyse the observations in ways
that provide more general information.
The wave field in a given region of the solar surface is affected
by the properties of that region, including the subsurface layers
down to the depth of penetration, determined by the lower turning
point ({\cf}\ Eq.\ \ref{turn-point}) of the waves that are observed.
By analyzing the properties of such local waves, it is possible
to infer local three-dimensional structures and flows beneath
the solar surface.

Early investigations of this nature considered the wave field around sunspots.
By carrying out a so-called {\it Hankel transform} of the waves
in cylindrical coordinates, centered on the spot,
Braun, Duvall \& LaBonte (1987) demonstrated that wave energy was 
absorbed or scattered by the spot.
This provided the potential for studying the subsurface structure of
active regions.
Brown (1990) presented a technique for inverting such data to obtain
a map of active-region structure;
he noted that, unlike `classical' helioseismology using oscillations
frequencies, this is based on observations of amplitudes and phases
of the waves.
A detailed review of the seismology of active regions was given
by Bogdan \& Braun (1995).

Studies of local properties of the solar interior,
known as {\it local helioseismology},
are developing very rapidly,
although they have not yet reached the level of maturity of
global helioseismology.
A basic difficulty which has not yet
been fully solved is the treatment of the {\it forward problem},
{\ie}, the calculation of the wave field and the resulting observables
for a given subsurface structure and flow.
(In contrast, in global helioseismology it is straightforward to compute
oscillation frequencies for a solar model with an assumed rotation law.)
As a result, the inferences made through local analysis are 
somewhat difficult to interpret, in terms of their resolution and the extent
to which they reflect the true properties of the solar interior,
although substantial progress has recently been made in this area.

\begin{figure}
\begin{center}
\inclfig{8.6cm}{\fig/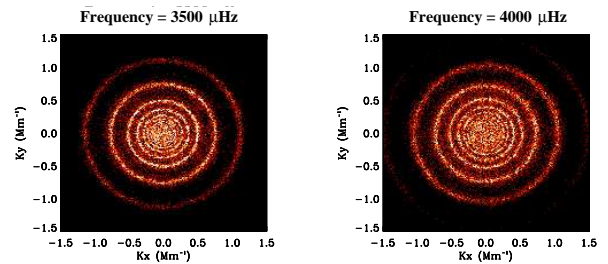}{Rings}
\end{center}
\caption{
\figlab{fig:2rings}
Ring diagrams obtained as cuts through tri-dimensional power spectra
at the frequencies indicated;
the data used are SOI/MDI full-disc Dopplergrams.
Each ring corresponds to a value of the radial order $n$.
(Adapted from Gonz\'alez Hern\'andez {\etal}\ 1998a.)
}
\end{figure}

\subsection{Ring-diagram analysis}

Possibly the first analysis of effects on oscillation frequencies
of local perturbations was presented by Gough and Toomre (1983).
They pointed out that the frequencies would be changed
by a local velocity field, through the advection of the wave pattern;
furthermore, they established the frequency perturbation resulting
from a local perturbation to the sound speed.
This suggestion was developed into a practical procedure by Hill (1988).
He considered the power spectrum, based on the oscillation field
over a restricted area of the solar surface, as a function of
frequency $\omega$ and the components $k_x$ and $k_y$ of the
horizontal wave vector in the longitude and latitude directions.
In the $(\omega, k_x, k_y)$ space the results are `trumpet-like'
surfaces, obtained by rotating the ridges in Fig.~\ref{fig:mdi-lnu}
around the frequency axis.
The analysis is carried out by considering cuts through these
surfaces at fixed frequency: the result is a set of rings,
each corresponding to a ridge in the $l -\nu$ diagram.
({\cf} Fig.~\ref{fig:2rings}).
As shown by Hill (1988) these rings are shifted by the underlying
horizontal flow field, the shift of a given ring being given by
an average of the velocity weighted by the relevant radial eigenfunction.
Similarly, variations in the subsurface sound speed cause
a distortion of the rings.
Thus, by considering different rings and different frequencies,
a set of data is obtained from which the depth variation of the flow
or the sound speed
can be inferred by means of inversion techniques such as those
described in Section~\ref{sec:helinv}.
These results are then assumed to represent horizontal averages
over the region for which the ring diagrams have been determined.
By repeating this for several regions on the solar surface,
a map of the flow and subsurface sound speed can be built up.

\begin{figure}
\begin{center}
\inclfig{7.0cm}{\fig/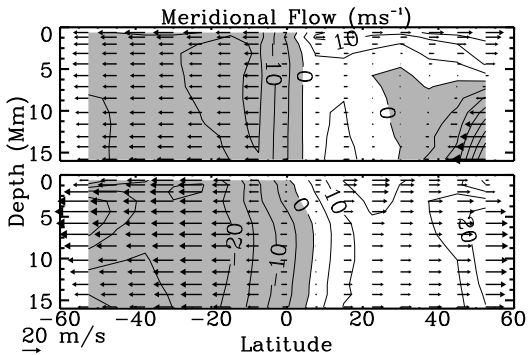}{Meridional circulation}
\end{center}
\caption{
\figlab{fig:mercirc}
Meridional flows in the solar convection zone, as inferred from
ring-diagram analysis, plotted against latitude (abscissa) and
depth beneath the solar surface (ordinate).
The length of the arrows indicate the speed, the scale being indicated
at the lower left; grey regions mark southward flow.
The results in the lower panel were obtained in 1997,
at relatively low solar activity,
whereas the upper panel is from 2001, close to solar maximum activity.
(Adapted from Haber {\etal}, 2002.)
} 
\end{figure}

Detailed analyses have been carried out of the flows in the solar
convection zone by means of this technique. 
Clear evidence has been found for meridional flows, which tend to
be poleward at periods of low solar activity
({\eg}, Gonz\'alez Hern\'andez {\etal}, 1998b; Haber {\etal}, 1998, 2000;
Schou and Bogart, 1998; Basu, Antia, and Tripathy, 1999).
At higher activity, the situation appears to be more complicated.
Some recent results, from an extensive analysis of MDI data by
Haber {\etal}\ (2002), are illustrated in Fig.~\ref{fig:mercirc}.
In the lower panel, obtained near solar minimum, there is a regular
flow from the equator towards the poles at all depths.%
\footnote{The slight North-South asymmetry may be due to a modest
misalignment of the orientation of the solar polar axis
which was assumed in the analysis.}
In the upper panel, however, obtained near solar maximum activity,
the flows in the Northern hemisphere are substantially more complicated,
a countercell with an equator-ward flow having developed at depth at
higher latitudes.

The ring-diagram analysis also allows separate determination of the
rotation rate in the northern and southern hemispheres.
Haber {\etal}\ (2000, 2001) found zonal flows converging towards the
equator, similar to those inferred from global helioseismic inversions
({\cf} Section~\ref{sec:change}), although with a substantial
North-South asymmetry, as illustrated in Fig.~\ref{fig:zonring}.
When symmetrized around the equator, these results were in
reasonable agreement with those obtained from global inversions, however.

\begin{figure}
\begin{center}
\inclfig{6.0cm}{\fig/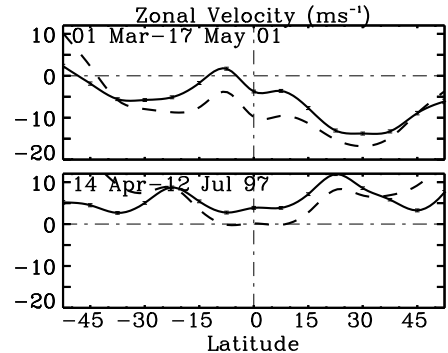}{Zonal flows, from rings}
\end{center}
\caption{
\figlab{fig:zonring}
Longitudinally averaged zonal flows, obtained from ring-diagram analysis.
The dashed curves show results at a depth of 0.9~Mm, and the solid
curves are at a depth of 7~Mm.
The results in the lower panel were obtained in 1997,
at relatively low solar activity,
whereas the upper panel is from 2001, close to solar maximum activity.
This should be compared to the zonal flows obtained from global analysis
({\cf}\ Fig.~\ref{fig:zonal});
note that in the latter figure only the component symmetrical around the
equator is obtained.
(Adapted from Haber {\etal}, 2001.)
} 
\end{figure}

Hindman {\etal}\ (2001) used ring diagrams to determine what essentially
corresponds to the mean multiplet frequency, as a function of position
on the solar disk,
and in this way obtained local frequency shifts
associated with active regions; 
when averaged over the solar disk and time, the results are not inconsistent
with the frequency changes observed for global modes over the
solar cycle ({\cf}\ Section~\ref{sec:change}).
This may provide insight into the physical origins of these frequency changes.
A related theoretical investigation of frequency shifts caused by
localized strong magnetic fields, such as are present in active regions,
was carried out by Cunha {\etal}\ (1998).

\subsection{Time-distance analysis and helioseismic holography}
\plabel{sec:timedist}

In geoseismology the most commonly used procedure is to measure
the travel time for waves between a known source and a detector.
The sources range from distant earthquakes, in investigations
of the global structure of the Earth, to 
vibrators in measurements of local subsurface structures.
The travel time provides an integral of the wave speed along the path
of the wave; many such travel times can be combined to produce a
coherent model of the region under study.
In this way it has been possible to obtain a three-dimensional
model of the interior of the Earth ({\eg}, Dziewonski and Woodhouse, 1987).

In the solar case there are no similarly sharply defined sources
of the waves: as discussed in Section~\ref{sec:solcause},
the waves are continuously excited by the random effects of
the near-surface convection.
Nevertheless, it was argued by Duvall {\etal}\ (1993b) that a similar signal
could be obtained from a suitable correlation analysis of the wave field
observed at the solar surface;
the time delay maximizing the correlation between two points
provides a measure of the travel time along the ray connecting these
two points.
The technique was developed further by D'Silva (1996) and
D'Silva {\etal}\ (1996).
Within the approximation of geometrical acoustics the travel time
along the ray $\Gamma_i$ can be written as 
\be
\tau_i(t) = \int_{\Gamma_i} {\dd s \over 
\cw(\boldr, t) + \boldv(\boldr, t) \cdot \boldn } \; ,
\eel{traveltime}
where $s$ is distance along the ray,
$\boldr$ is the spatial coordinate, $\cw$
is the local wave speed, $\boldv$ is the local flow
velocity and $\boldn$ is a unit vector along the ray;
the appearance of time $t$ indicates that both the wave speed and
flow velocity may depend on time.
The wave speed is predominantly given by the sound speed but may
be perturbed by magnetic fields in active regions.
Given measurements along a sufficient number of rays,
these relations may be inverted to infer $\cw(\boldr, t)$ and
$\boldv(\boldr, t)$ (Kosovichev, 1996b).
Reviews of time-distance techniques were given by 
Kosovichev and Duvall (1997),
and Kosovichev {\etal}\ (2000, 2001).

In practice, the correlation analysis is carried out between
regions of the solar surface, typically a small central area
and a surrounding ring or parts of a ring.
Also, \Eq{traveltime} assumes that the waves can be treated in
the ray approximation.
It was noted by Bogdan (1997) that this approximation is questionable
in the solar case, since the wavelength in general is not small
compared to the scale of the features that are investigated.
Birch and Kosovichev (2000, 2001) studied the effects of wave-speed
perturbations in the first Born approximation to derive travel-time
sensitivity kernels, relating the wave-speed perturbation
to the change in the travel time, as a replacement for the ray approximation.
Jensen {\etal}\ (2000) proposed simple
analytical approximations to such kernels and showed that they
were in reasonable agreement with sensitivity computations based
on solutions to the wave equation.
These kernels were used for inversion to infer wave-speed perturbations
by Jensen {\etal}\ (2001), 
who also determined averaging kernels reflecting the resolution properties
of the inversion.
Birch {\etal}\ (2001) made a careful analysis of the 
accuracy of the Born and ray approximations, by comparing them
with direct calculations of the scattering of acoustic waves
in a uniform medium.
Finally, 
Jensen and Pijpers (2002) derived sensitivity kernels for wave-speed
perturbations and flow velocity in the Rytov approximation,
and compared various approximations to these kernels.
It is very encouraging that the theoretical basis for the time-distance
technique is getting more solidly established through these analyses;
information transfer from similar work in geophysics has been very
fruitful in this regard.

Time-distance analyses have been used to investigate the near-surface
flow fields associated with supergranular convection ({\eg}, Kosovichev, 1996b).
In an interesting analysis based on f modes,
Duvall and Gizon (2000) evaluated the vertical vorticity associated
with the flow and showed that this was in agreement with theoretical
expectations for convection in a rotating system.
Giles {\etal}\ (1997) determined properties of
the meridional flows in the solar convection zone,
from the equator towards the poles, also seen with the ring-diagram
analyses ({\cf} Fig.~\ref{fig:mercirc}).
Inferences of meridional flows over an an extended range of
depths within the convection zone were reported by Duvall and Kosovichev (2001);
interestingly, no evidence was found for a return flow.
Variations with time in the meridional flow, inferred from
time-distance analysis, were discussed by Chou and Dai (2001).
As in the ring-diagram results, the flow showed increasing complexity 
with increasing solar activity; 
however, as Chou and Dai did not carry out an inversion in the radial
direction, a more detailed comparison of the results is not possible.

Investigations have also been made of wave-speed perturbations
associated with emerging active regions ({\eg}, Kosovichev {\etal}, 2000;
Jensen {\etal}, 2001).
An example is shown in Fig.~\ref{fig:wavespeed}.
It is evident that the emerging magnetic field is associated with
a complex structure of generally increased wave speed below the solar surface.
Zhao {\etal}\ (2001) recently inferred the velocity
field beneath a large sunspot;
they found a strong mass flow across the spot at depth of 9--12~Mm,
indicating that the magnetic field responsible for the spot has a rather
loose structure at these depths.

\begin{figure}
\begin{center}
\inclfig{7.5cm}{\fig/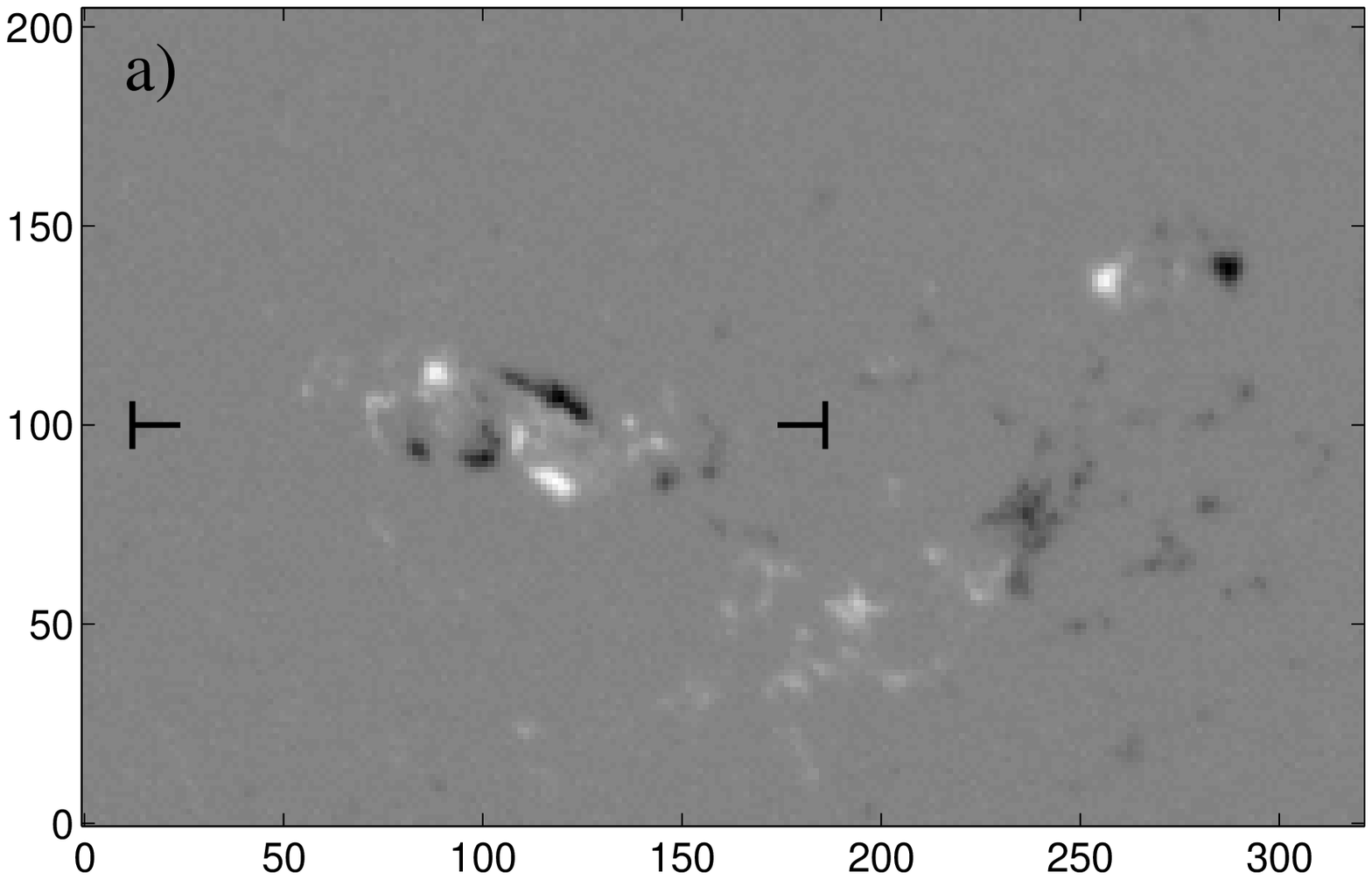}{Magnetic map}
\inclfig{7.5cm}{\fig/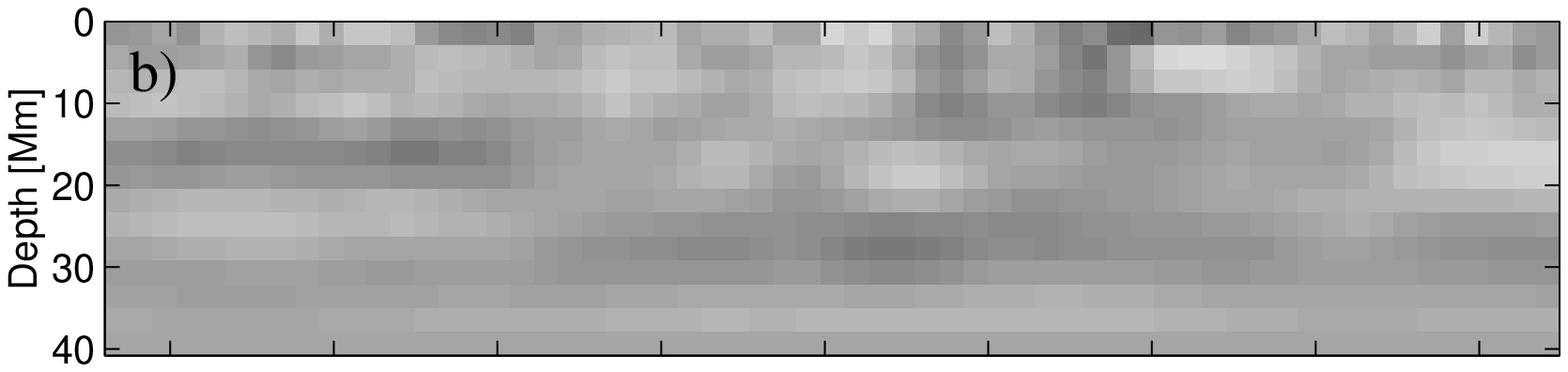}{Wavespeed perturbations}
\inclfig{7.5cm}{\fig/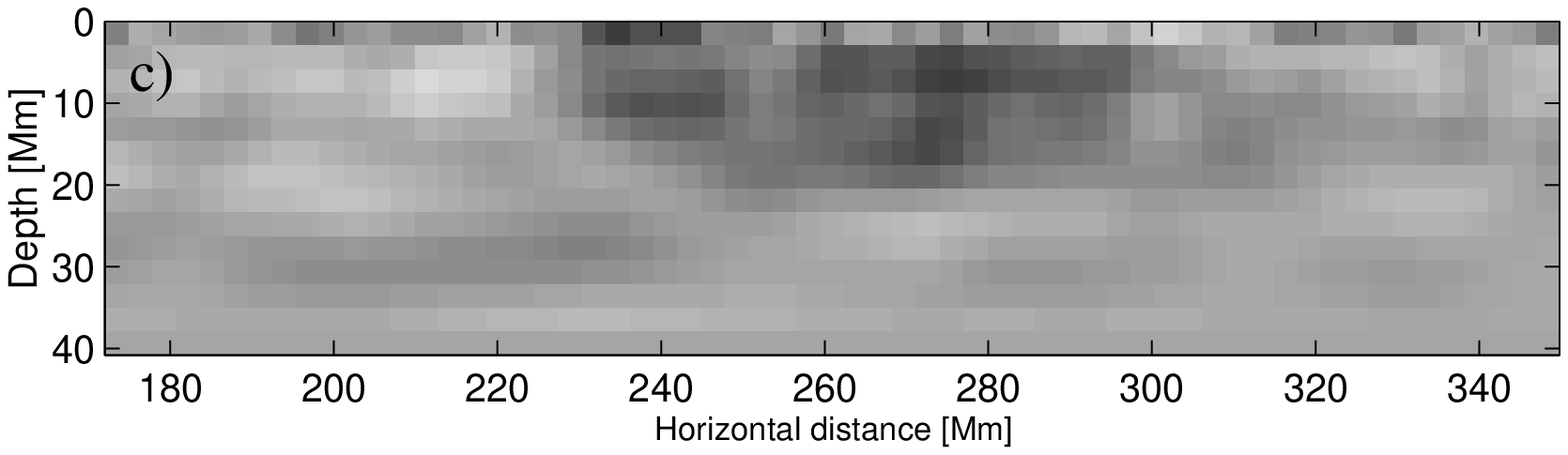}{Wavespeed perturbations}
\end{center}
\caption{
\figlab{fig:wavespeed}
Panel (a) shows an MDI magnetogram of an emerging active region;
distances are in Mm, and the black and white regions indicate
magnetic fields of opposite polarity,
over a range between $-0.1$ and $0.1$ Tesla.
Panels (b) and (c) show wave-speed perturbations below the surface, 
at the cross section marked by $\vdash$ and $\dashv$ in panel (a);
the grey scale ranges from perturbations of $-0.2 \km \s^{-1}$
(white) to $0.5 \km \s^{-1}$ (black).
Panel (c) was obtained at the same time as the magnetogram in panel (a),
while panel (b) was taken 16 hours earlier.
(Adapted from Jensen {\etal}, 2001.)
} 
\end{figure}

A technique closely related to time-distance helioseismology
is known as {\it helioseismic holography}.
It goes back to a proposal by Roddier (1975) to use holographic
methods to visualize acoustic sources below the solar surface,
followed by a suggestion by Lindsey and Braun (1990)
that it might be possible to form an acoustic image of sunspots on
the back of the Sun.%
\footnote{Peri and Libbrecht (1991) searched for, but failed to find,
a deficit of acoustic power at the antipodes of far-side active regions.}
However, the first practical application of the technique
seems to have been by Lindsey and Braun (1997) and the parallel
development of the so-called technique of acoustic imaging
by Chang {\etal}\ (1997).
In these techniques, the acoustic wave field on the solar surface
is combined coherently, taking into account the phase information,
to reconstruct the presence of acoustic absorbers or scatterers
in the subsurface layers.
The methods have predominantly been used to investigate
the subsurface structure and acoustic properties of
active regions
({\eg}, Chen {\etal}, 1998;
Braun {\etal}, 1998; Lindsey and Braun, 1998; 
Braun and Lindsey, 2000).
A tutorial review of helioseismic holography was given by
Lindsey and Braun (2000a),  while Chou (2000) reviewed the work done
on acoustic imaging.
A technique for inversion of the holographic data was presented
by Skartlien (2002).

\begin{figure}
\begin{center}
%\inclfig{6cm}{\fig/dbraun-holo1.eps}{Backside imaging}
\inclfig{5.5cm}{\fig/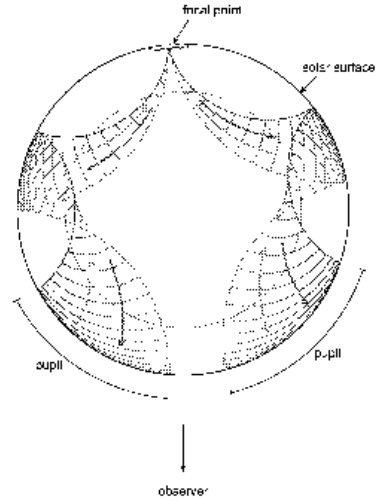}{Backside imaging}
\end{center}
\caption{
\figlab{fig:backimage}
Schematic illustration of the principle in imaging active regions
on the far side of the Sun.
The figure shows the propagation of waves in a cross section of the Sun,
starting from a focal point in an active region on the far side
and observed in the pupil on the near side.
See text for details.
(Adapted from Lindsey and Braun, 2000b.)
} 
\end{figure}

The ability of the holographic analysis to detect active regions
on the far side of the Sun was convincingly demonstrated by
Lindsey and Braun (2000b), through analysis of data from SOI/MDI.
The principle is illustrated schematically in Fig.~\ref{fig:backimage}.
Waves emerging from the far side of the Sun can be measured on the near
side, in the region denoted `pupil', after one or more reflections
at the solar surface; 
through appropriate analysis of the measured wave field it is possible
to focus on specific regions on the far side.
Relative to the neighboring quiet photosphere waves from the active
region suffer a phase shift which can be detected.
This is illustrated in Fig.~\ref{fig:backspot} where the
phase shift (expressed as a change in travel time) determined on
the far side is compared to a magnetogram of the same region
after it has moved to the near side of the Sun as a result of solar rotation.
There is clearly a striking agreement between the features in the
acoustic and direct image.
Further developments of this technique has allowed imaging of the
entire far side of the Sun, extending also to the region near the
solar limb and over the poles (Braun and Lindsey 2001).

I finally mention that Woodard (2002) has developed a new analysis method
where the intermediate steps between data and inferences are to some
extent bypassed, hence approaching the ideal case presented in the
introduction to Section~\ref{sec:anal}.
Specifically, he obtained a relation between inhomogeneity-induced
correlations in the observed wave field and the underlying
supergranular flow.
Results of analysis of data from the MDI instrument showed a very promising
correlation with the directly measured surface flow field.
A detailed comparison of this technique with other techniques of
local helioseismology, evaluating its advantages and possible disadvantages,
still remains to be carried out, however.

\begin{figure}
\begin{center}
\inclfig{8.6cm}{\fig/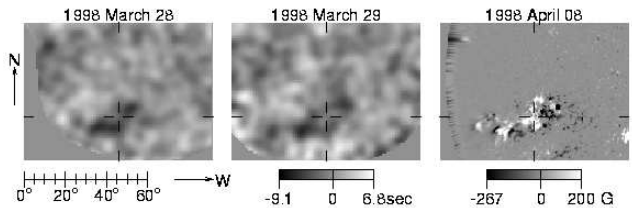}{Backside images}
\end{center}
\caption{
\figlab{fig:backspot}
The leftmost two panels show travel-time perturbations $\Delta t$,
in the vicinity of an active region on the far side of the Sun.
The right-hand panel shows a magnetogram of the same region 10 days later,
after the region has become visible on the near side.
The rule indicates angular distance on the solar surface.
(Adapted from Lindsey and Braun, 2000b.)
} 
\end{figure}

\section{The helioseismic Sun} \plabel{sec:helsun}

It seems unlikely that even the most optimistic predictions 
in the early phases of helioseismology, around 1975, 
could have foreseen the extent to which the solar interior can now be probed.
Inferences of solar structure have shown that standard calculations
of solar models reproduce the actual structure to a precision
better than 0.5~\% in sound speed.
This is a remarkable demonstration of the ability of physics,
including our current understanding of the microscopic properties
of matter under stellar conditions, to predict properties
of such a relatively complex object as the Sun.
It also provides strong evidence that the discrepancy between
the predicted and measured capture rates of neutrinos results from
the properties of the neutrinos, rather than from errors in the
modeling of the solar interior.
Indeed, the strong constraints on solar structure from
helioseismology provide a basis for using the solar core as
a well-calibrated neutrino source for the study of neutrino physics.
The solar rotation rate has been determined in much of the solar
interior, revealing striking variations with position, and changes in time.
Further, information is emerging about the flows in the solar convection
zone and the subsurface structure of magnetically active regions.

These are remarkable achievements, in providing observational 
information about the internal properties of a star.
However, an important goal is now to understand the results
in physical terms, and evaluate their broader consequences for the
modeling of stellar structure and evolution,
as well as their implications for our understanding of physics
of matter in stars.

Investigations of the thermodynamic properties in the convection
zone have shown that even the present complex descriptions are
inadequate, at the level of precision reached by the helioseismic inferences;
this demonstrates the possibility of using the Sun as a laboratory for the
study of the equation of state of partially ionized matter, 
in very great detail.
Although the effects are subtle in the solar case, 
they could have substantial importance under other astrophysical
circumstances, such as in lower-mass stars or giant planets where
the interactions between the constituents of the plasma are much stronger.

The successes in overall solar modeling should not overshadow the failures:
the differences between the inferred solar sound speed and the predictions
of the models are far larger than the observational uncertainty.
A particularly striking feature is the localized region just below the
convection zone where the solar sound speed is substantially
higher than that of the models.
This is a region where the models predict a strong gradient in
the hydrogen abundance, as a result of settling of helium from
the convection zone towards the interior.
It also approximately coincides with the tachocline, {\ie},
the transition between the latitudinally varying rotation in the
convection zone and the almost uniform rotation in the radiative interior.
It was suggested by Gough and McIntyre (1998) that the uniform
rotation of the interior is maintained by a weak magnetic field,
the tachocline being established as a boundary layer;
within this region circulation is established which leads to mixing.%
\footnote{Detailed numerical modeling of this mechanism has been
started by Garaud (2002).}
Mixing would also result from the strongly anisotropic turbulence
originally suggested by Spiegel and Zahn (1992) to explain the tachocline.
In either case, mixing of the region just beneath the convection zone
would tend to reduce the composition gradients,
locally increasing the hydrogen abundance and hence the sound speed,
as required by the helioseismic results
(Brun {\etal}, 1999; Elliott and Gough, 1999).
Such smoothing of the gradient was also suggested
by the inversions, discussed in Section~\ref{sec:helphys},
for the hydrogen abundance.

Independent evidence for mixing beneath the convection zone comes
from the reduction in the solar photospheric lithium abundance,
relative to the primordial value 
({\cf}\ Section~\ref{sec:sunprop}).
Lithium is destroyed by nuclear reactions at temperatures above
$2.5 \times 10^6 \K$,
substantially higher than the temperature at the base of the convection zone.
In fact, models with mixing have been computed which match 
the lithium abundance, suppressing also the peak in the sound-speed
difference just below the convection zone
({\eg}, Richard {\etal}, 1996; Chaboyer, 1998; Brun {\etal}, 1999).
On the other hand, the fact that the photospheric beryllium abundance 
is close to the primordial value indicates that significant mixing 
does not extend to temperatures as high as $3.5 \times 10^6 \K$
where beryllium is destroyed.

The inferences of solar internal rotation show that the rotation rate
is almost constant in the radiative interior:
unlike simple models of the Sun's rotational evolution from
an assumed state of rapid initial rotation, there is no indication
of a rapidly rotating core.
An important consequence is that the solar oblateness,
which can be calculated precisely from the inferred rotation rate,
has no significant effects on tests, based on planetary motion,
of Einstein's theory of general relativity ({\eg}, Pijpers, 1998;
Roxburgh, 2001).
The nearly uniform rotation of the radiative interior
indicates the presence of efficient transport of angular momentum,
coupling the radiative interior to the convection zone, from which
angular-momentum loss has taken place through the solar wind.
It was proposed by Kumar and Quataert (1997) and Talon and Zahn (1998)
that angular-momentum transport might take place by means of
gravity waves generated at the base of the solar convection zone.
However, Gough and McIntyre (1998), with reference to analogous
phenomena in the Earth's atmosphere, argued that gravity waves would
be unlikely to have the required effect;
they identified magnetic effects as the only plausible transport 
mechanism, a weak primordial field being sufficient to ensure the
required coupling.

The variation of the rotation rate in the convection zone,
reflected also in the latitude dependence
observed on the solar surface, is presumably maintained by
angular-momentum redistribution within the convection zone,
through interaction between rotation, convection and possibly other flows.
The observed variation is inconsistent with relatively simple models
which tend to predict a rotation rate depending on the distance to the
rotation axis ({\cf}\ Section~\ref{sec:solrot}).
It was pointed out by Gough (1976)
that the interaction between rotation and small-scale
convection might lead to an anisotropic turbulent viscosity which
could affect angular-momentum transport;
an estimate of the anisotropic Reynolds stress tensor was made
on the basis of three-dimensional hydrodynamical simulations by
Pulkkinen {\etal}\ (1993).
Pidatella {\etal}\ (1986) used simple models of this nature to 
interpret early helioseismic inferences of rotation in the convection zone.
Recently, the numerical resolution in full hydrodynamical simulations
of the solar convection zone has become sufficient to capture
at least some aspects of the smaller-scale turbulence
({\eg}, Miesch, 2000; Miesch {\etal}, 2000; Brun and Toomre, 2002);
the results of these simulations show an encouraging
similarity to the helioseismically inferred rotation profile.

It is likely that interaction between convection and rotation is
responsible for the formation of the large-scale solar magnetic field
and its 22-year variation in the solar magnetic cycle, through some
kind of dynamo mechanism.
Dynamo models have in fact been constructed which are based on
the helioseismically inferred rotation rate
({\eg}, Parker, 1993; Charbonneau and MacGregor, 1997).

Analyses of data during the period leading to the present maximum
in solar activity have shown striking variations in solar rotation.
Zonal flows converging towards the solar equator, previously
detected in surface observations, have been shown to extend
over a substantial fraction of the convection zone.
These bands of somewhat faster and slower rotation appear to
be related to the equator-ward drift of locations of sunspots
as the solar cycle progresses
({\eg}, Howard and LaBonte, 1980; Snodgrass, 1987; Ulrich, 1998, 2001);
however, the physical connection is as yet not understood.
Even more surprising has been the detection of oscillations with a period
of 1.3~y in the rotation rate near and below the base of the convection zone;
one may hope that they can provide additional information 
about conditions in this region and possibly about the mechanism of
the solar dynamo.
It is evident that such temporal variations provide strong arguments
for further detailed observations of solar oscillations, ideally through
at least one full 22-year magnetic cycle.

Further information about the detailed structure and dynamics 
of the convection zone has been obtained from local helioseismology.
Large-scale convective flow patterns have been detected,
as well as meridional flows with complex structure that appears to
depend on the level of magnetic activity.
This may lead to a detailed understanding of
the mechanisms controlling convection and rotation, including the
angular-momentum transport,
when the observations are
combined with the increasingly realistic modeling of the
dynamics of the solar convection zone.
Also, detailed information is becoming available about the 
subsurface structure and time evolution of active regions, which
will likely lead to a better understanding of the processes underlying
their formation.
Particularly interesting is the detection of active regions on
the far side of the Sun;
by giving advance warning before they reach the near side 1 -- 2 weeks
later and hence have the potential to unleash eruptions in
the direction of the Earth, such observations may be helpful in 
reducing the risk of harmful effects from such eruptions.

The causes of the solar oscillations are not central to the use of the
frequencies for helioseismic investigations, although the processes
responsible for the excitation and damping undoubtedly contribute to
the frequency shifts, suppressed in inverse analyses,
which are induced by the superficial layers of the Sun.
The statistical properties of the observed modes seem largely to
be consistent with stochastic excitation of damped oscillations,
as discussed in Section~\ref{sec:solcause}
({\eg}, Chaplin {\etal}, 1997; Chang and Gough, 1998).
Furthermore, Stein and Nordlund (1998b, 2001) 
showed that hydrodynamical simulations
of solar near-surface convection predicted the excitation of oscillations,
with an energy input approximately consistent with the observations;
a detailed comparison by Georgobiani {\etal}\ (2000) between observations
and hydrodynamical simulations of solar oscillations also showed 
overall agreement, including indications of asymmetry.
On this basis, we can be reasonably confident that we understand the
overall aspects of the excitation of the solar modes;
thus it may be possible to use the observed properties of the oscillations,
including the statistics of the amplitudes, to obtain information
about convection beneath the solar surface.

\section{Per aspera ad astra} \plabel{sec:astra}

Although impressive advances have been made on the helioseismic study of the
solar interior, this provides information about only an individual,
relatively simple star.
Complete testing of the theory of stellar structure and evolution would
require studies of the broad range of stellar types, spanning
very different physical properties and processes, that are observed.
These include effects, such as rapid rotation or convective cores,
that cannot be investigated in the solar case.
Fortunately, it has been found that stars of very different types,
covering most stellar masses and evolutionary states, show pulsations;
often, these stars are multi-mode pulsators and hence in principle 
offer relatively detailed information about their interiors.
For example, such stars include the $\gamma$ Doradus and $\delta$ Scuti 
stars, the slowly pulsating B stars and the $\beta$ Cephei stars, which
span the main sequence from masses of $1.5$ to more than 10 solar masses,
and various classes of white dwarfs.%
\footnote{For extensive discussions of general stellar pulsations, see,
for example, Unno {\etal}\ (1989), as well as the proceedings
edited by Breger and Montgomery (2000) and Aerts {\etal}\ (2002).}
Thus there would appear to be an excellent potential for 
{\it asteroseismology},%
\footnote{This terminology has given rise to some discussion.
Motivated by Trimble (1995) who questioned the appropriateness of the
term, Gough (1996b) gave what in my view is its definitive etymological
justification.}
probing the stellar interiors on the basis of the observed frequencies.

In most cases, the oscillations are caused by intrinsic driving 
resulting from various radiative or perhaps convective mechanisms.
Although modes may be unstable in a substantial range of frequencies,
the modes observed are typically only a relatively small subset of
the unstable modes, and the selection of modes which reaches observable
amplitudes is complex and poorly understood.
As a result, it is difficult to identify the observed frequencies with
specific modes, characterized by their degree, radial and azimuthal order;
this has severely limited the possibilities for using the modes for
investigating the stellar interiors.

In contrast, oscillations excited in a manner similar to what is 
observed in the Sun are expected to show a broad spectrum of observable
modes, most modes in this range being present since no subtle
selections are at work in the determination of the mode amplitudes.
This makes solar-like oscillations very attractive for asteroseismology.
Furthermore, the relation between stellar structure and the oscillation
frequencies is relatively well understood.
In the foreseeable future stellar observations will be restricted
to disk-averaged data, and hence to low-degree modes;%
\footnote{However, a space-based interferometric mission has been
considered which would allow resolution of modes on distant stars
with degrees as high as 50 ({\eg}, Carpenter and Schrijver, 2000).
This would, for example, allow some resolution of the structure and
rotation near the base of the convection zone in a star similar to the Sun.}
however, as discussed in Section~\ref{sec:pasymp} these are precisely
the modes that give information about the properties of stellar cores.

On the basis of our understanding of the source
of the solar oscillations (see Section~\ref{sec:helsun}),
one may expect similar oscillations in other stars
with outer convection zones, although the predictions of their
amplitudes are still somewhat uncertain
({\eg}, Christensen-Dalsgaard and Frandsen, 1983; Houdek {\etal}, 1999).
However, in any case the expected amplitudes in main-sequence stars
is extremely low, as is also observed in the solar case:
the predicted velocity amplitudes are typically below $1 \m \s^{-1}$
and the relative luminosity amplitudes are below around 10 parts per million,
severely stretching the capabilities of ground-based observations faced
with instrumental problems and fluctuations in the Earth's atmosphere.%
\footnote{On the other hand, quite substantial amplitudes are predicted
for red-giant stars.
In fact, Christensen-Dalsgaard {\etal}\ (2001) found evidence,
based on observations by the American Association of Variable Star
Observers, that semi-regular variability in red giants might
be caused by the same stochastic mechanism that is responsible
for the solar oscillations.}
Thus it is hardly surprising that the observational results until recently
have been at best tentative.
An extensive co-ordinated campaign with most of the World's
then largest telescopes (Gilliland {\etal}, 1993) failed to find
oscillations in stars in the open cluster M67, in some cases
with upper limits below the theoretical predictions.
Promising results were obtained by Brown {\etal}\ (1991) and
Marti\'c {\etal}\ (1999) 
for Procyon, again with amplitudes somewhat below predictions.
Detailed results for the sub-giant $\eta$~Bootis were obtained
by Kjeldsen {\etal}\ (1995); 
interestingly, modeling by Christensen-Dalsgaard {\etal}\ (1995)
and Guenther and Demarque (1996) 
showed that there
might be evidence for g-mode-like behavior in the observed frequencies.
However, it must be noted that the observations were 
questioned by Brown {\etal}\ (1997). 

\begin{figure}
\begin{center}
\inclfig{7cm}{\fig/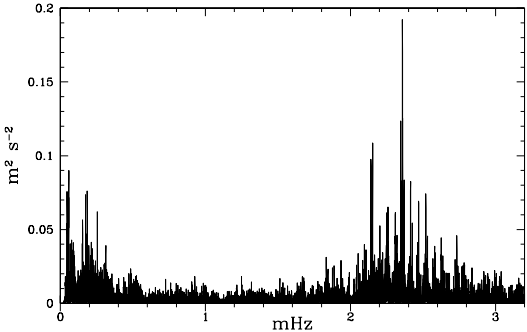}{alpha Cen}
\end{center}
\caption{
\figlab{fig:alphacen}
Power spectrum of oscillations of $\alpha$ Cen A,
from radial-velocity observations with the CORALIE 
fiber-fed echelle spectrograph
on the 1.2 m Swiss telescope at the 
La Silla site of the European Southern Observatory.
(From Bouchy and Carrier, 2001.)
} 
\end{figure}

In the last few years the observational situation has undergone
a dramatic improvement.
Interesting results have been obtained from photometric
observations with the star tracker
on the otherwise failed WIRE satellite ({\eg}, Buzasi {\etal}, 2000;
Schou and Buzasi, 2001).
Furthermore, techniques have been developed for very stable
radial-velocity measurements in connection with the search
for extra-solar planets.
This has resulted in the detection of
evidence for solar-like oscillations 
in the star $\beta$ Hydri (Bedding {\etal}, 2001);
also, as shown in Fig.~\ref{fig:alphacen},
a very clear detection has been made in 
the `solar twin' $\alpha$ Centauri A (Bouchy and Carrier, 2001).

Further developments are expected of ground-based observing facilities,
including the HARPS spectrograph (Queloz {\etal}, 2001)
to be installed on the 3.6 m telescope
of the European Southern Observatory at La Silla.
A major breakthrough of asteroseismology of solar-like stars will
result from observations from space.
The Canadian MOST satellite (for {\bf M}icrovariability and {\bf O}scillations
of {\bf ST}ars; Matthews 1998) will be launched late in 2002.
The French COROT satellite (for {\bf CO}nvection, {\bf RO}tation
and planetary {\bf T}ransits; Baglin {\etal}\ 1998, 2002)
is scheduled for launch in 2004, and will obtain very extended
time series, with correspondingly high frequency resolution,
for a handful of stars.
Two other projects are under development.
The Danish R{\o}mer satellite is being developed, with the
MONS project (for {\bf M}easuring {\bf O}scillations in {\bf N}earby
{\bf S}tars; Christensen-Dalsgaard 2002), with launch planned for 2005;
this will be in an orbit that allows access to stars in essentially
the entire sky.
Finally, the Eddington mission (Favata, 2002) currently has the status
as reserve mission in the programme of the European Space Agency;
it will carry out precise measurements of oscillations of a large number
of stars of a variety of types.

Data for distant stars will obviously never be as detailed as those
that have been obtained for the Sun;
however, there is no doubt that asteroseismic investigations of
a broad range of stars, with very different properties, 
will contribute greatly to our understanding of stellar
structure and evolution over the coming decades.

\section*{Acknowledgments}

I am very grateful to many colleagues, too numerous to list,
for discussions and collaborations within the field of helioseismology.
I thank S. Basu, F. Bouchy, 
D. C. Braun, D. A. Haber, R. Howe, J. M. Jensen, C. Lindsey
and J. Schou for assistance with figures.
The work reported here was supported in part by the Danish National
Research Foundation through the establishment of the 
Theoretical Astrophysics Center.
The National Center for Atmospheric Research is sponsored by the
National Science Foundation.

\end{document}